\documentclass{umich-thesisnick}
\usepackage{amsfonts}
\usepackage{amsmath}
\usepackage{amssymb}
\usepackage{graphicx}

\input{tcilatex}

\begin{document}

\frontmatter

\title{Theory and modeling of particles with DNA-mediated interactions}
\author{Nicholas A. Licata}
\date{2008}
\degree{Doctor of Philosophy} \department{Physics} 
\chairperson{Assistant Professor Alexei V.
Tkachenko} 
\committee{Professor Sharon C. Glotzer \\
Professor Bradford G. Orr \\ Professor Leonard M. Sander \\ Associate
Professor Jens-Christian D. Meiners}
\maketitle
\makecopyright

\acknowledgements{First and foremost I would like to thank my advisor, Alexei Tkachenko, without 
whose support and guidance this thesis would not have been possible. \ Your patience, enthusiasm, and never ending 
flow of ideas made my work environment both pleasant and productive.  \ I am greatly indebted to my parents for 
putting me in a position where this whole endeavor seemed feasible.  \ Thank you for answering all of my questions 
and your never ending support.  \ Thanks to Ant, who blazed a trail, and to Joe, for reminding me I didn't have to 
follow it. \ To Keli, only you could make the last three years the best three years. 
\ Your love and companionship has made my time in Ann Arbor a period I will undoubtedly look back on with much fondness 
for the rest of my life. 
  \ I would like to thank Rice University for providing a wonderful 
setting to begin this journey.  \  In particular to Paul Stevenson, for his inspiring lectures, and to Nathan Harshman, 
for supervising my senior thesis.  \ To the faculty who provided guidance 
during various stages of the work, Leonard Sander and Brad Orr, thank you.  \ To our wonderful graduate coordinator 
Kimberly Smith, thank you.  \ To my fellow students 
at the University of Michigan: Clement Wong, Glenn Strycker, Clark Cully, Ross O'Connell, Gourab Ghoshal, Elizabeth Leicht, 
Brian Karrer, Pascale Leroueil, Chris Kelly, from making the slog through problem sets a bit less dreary, to hearing 
my grievances about malfunctioning code, thank you. \ To the gents at the 407, Dan Stick, 
Dave Moehring, Matt Schwantes, Mark Gordon, Brian Cline, your friendship has made my time in Ann Arbor all the more 
pleasant.  \ Thanks for making Thursday night the best night of the week. \  To the Lindquist clan, thanks for 
making Tuesday night a close second.  \ }

\abstract{In recent years significant attention has been attracted to proposals which utilize DNA for
nanotechnological applications.  Potential applications of these ideas range from the programmable self-assembly 
of colloidal crystals, to biosensors and nanoparticle based drug delivery platforms.  In Chapter I we introduce the system, which generically consists 
of colloidal particles functionalized with specially designed DNA markers.  The sequence of
bases on the DNA markers determines the particle type.  Due to the hybridization between complementary 
single-stranded DNA, specific, 
type-dependent interactions can be introduced between particles by choosing the appropriate DNA marker sequences.  
In Chapter II we develop a statistical mechanical description of the aggregation and melting behavior of particles with DNA-mediated interactions.  A quantitative comparison between the theory and experiments is made by calculating the 
experimentally observed melting profile.  In Chapter III a model is proposed to describe the dynamical 
departure and diffusion of particles which form reversible key-lock connections.  The model predicts a crossover 
from localized to diffusive behavior.  The random walk statistics for 
the particles' in plane diffusion is discussed.  The lateral motion is analogous to dispersive transport in disordered 
semiconductors, ranging from standard diffusion with a renormalized diffusion coefficient to anomalous, subdiffusive behavior.  
In Chapter IV we propose a method to self-assemble nanoparticle clusters using DNA scaffolds.  An optimal concentration ratio is determined 
for the experimental implementation of our self-assembly proposal.  A natural extension is discussed in Chapter V, the programmable self-assembly 
of nanoparticle clusters where the desired cluster geometry is encoded using DNA-mediated interactions.  We determine the probability 
that the system self-assembles the desired cluster geometry, and discuss the connections to jamming in granular and colloidal systems.  
In Chapter VI we consider a nanoparticle based drug delivery platform for targeted, cell specific chemotherapy.  A key-lock model is proposed to describe
the results of in-vitro experiments, and the situation in-vivo is discussed.  The cooperative binding, and hence the specificity to cancerous 
cells, is kinetically limited.  The implications for optimizing the design of nanoparticle based drug delivery platforms is discussed.  
In Chapter VII we present prospects for future research: the connection between DNA-mediated colloidal crystallization 
and jamming, and the inverse problem in self-assembly.  } 

\makeacknowledgements

\tableofcontents
\listoffigures
\listoftables
\listofappendices
\makeabstract
\mainmatter

\chapter{A DNA-colloidal primer}

\section{\protect\bigskip Miniaturization}

\qquad Advances in science have made possible the manipulation of matter on
a smaller and smaller scale. \ Controlling the spatial arrangement of atoms
and molecules enables the control of bulk material properties. \
Miniaturization has attracted significant attention in its own right,
particularly with respect to integrated circuit design for computer
hardware. \ This trend, known as Moore's Law, states that the number of
transistors which can be placed on an integrated circuit has been increasing
exponentially, approximately doubling every two years \cite{moore}. \
Independent of our ambition to quench the thirst for increased computing
power, miniaturization will likely play an important role in the future of
medical science. \ The ability to engineer nanodevices which interact with
individual cellular components has a number of potentially exciting
applications, ranging from smart drug delivery vehicles \cite{folate1}, \cite%
{pamam}, \cite{target1}, \cite{target2}, \cite{target3}, \cite{target4} to
biosensors which can detect an astonishingly low concentration of pathogens 
\cite{biotech}, \cite{pathogen}, \cite{pathogen2}. \ The realization of
these goals depends fundamentally on our ability to control the structure
and arrangement of individual components on the nanoscale. \ On these
lengths we encounter problems with traditional top-down assembly approaches
to miniaturization, for example lithography \cite{lithography}. \ One
proposed resolution is to proceed from the bottom-up, harnessing the
incredible molecular recognition properties of DNA \cite{natreview}, \cite%
{blocks}, \cite{falls}, \cite{nucleic}. \ 

\section{DNA}

\qquad Deoxyribonucleic acid, hereafter simply DNA, is a biopolymer which
contains the genetic information for the function of all living organisms. \
The macromolecule consists of a sugar-phosphate backbone chain with the
saccharide unit carrying a nucleotide of four possible types \cite{nanotech}%
. \ The primary structure of DNA refers to the sequence of these nucleotides
from the four letter DNA alphabet consisting of cytosine (C), thymine (T),
adenine (A), and guanine (G). \ The secondary structure of DNA refers to the
short range order which manifests itself as a result of interactions between
monomers which are in close proximity \cite{statprop}. \ Hydrogen bonding
between complementary DNA base pairs results in a DNA\ double helix, in
which two DNA molecules wind around each other. \ The complementarity rule
states that adenine bonds with thymine, and cytosine bonds with guanine. \
The double helix is approximately $2nm$ in diameter, and the repeat in the
direction of the helix axis is every $3.4\mathring{A}$ which is about every $%
10$ base pairs. \ The energy gain associated with forming a base pair in the
double helix is comparable to the hydrogen bond energy $\Delta E\sim 0.1eV$.
\ The formation of the T-A (C-G) pair is a result of two (three) hydrogen
bonds. \ As a result the characteristic energy required for double stranded
DNA\ to denature and form two single strands is comparable to the thermal
energy at room temperature $T\sim 300K$. \ In many respects DNA appears to
be an excellent candidate to control matter on the nanoscale. \ The
interactions between nucleotides are highly specific. \ In addition, the
number of potential sequences $4^{N}=\exp (N\log 4)$\ grows exponentially
with the number of nucleotides $N$. \ 

\section{Polymer Physics}

\qquad Part of the usefulness of DNA in controlling matter on the nanoscale
stems not from its chemical specifics, but general conformational properties
of long chain-like molecules \cite{scaling}. \ Here we introduce some of the
basic ideas in studying the conformations of polymers which will be of use
later on. \ 

A very idealized model of a polymer is a sequence of $N$ rigid links of
length $l$, where the direction between consecutive links is independent. \
In this freely-jointed model the end to end distance of the polymer chain $%
\mathbf{r}$ can be expressed in terms of the bond vectors $\mathbf{u}_{i}=%
\mathbf{x}_{i+1}-\mathbf{x}_{i}$ where $\mathbf{x}_{i}$ is the radius vector
of the $i^{th}$ segment. \ \ 
\begin{equation}
\mathbf{r=}\sum\limits_{i=1}^{N}\mathbf{u}_{i}
\end{equation}%
The radius of gyration of the chain $R_{g}$ is defined in terms of the
average mean-squared displacement. \ 
\begin{equation}
R_{g}^{2}=\left\langle \mathbf{r\cdot r}\right\rangle =\left\langle
\sum\limits_{i=1}^{N}\sum\limits_{j=1}^{N}\mathbf{u}_{i}\mathbf{\cdot u}%
_{j}\right\rangle =\sum\limits_{i=1}^{N}\left\langle \mathbf{u}%
_{i}^{2}\right\rangle +\sum\limits_{i\neq j}\left\langle \mathbf{u}_{i}%
\mathbf{\cdot u}_{j}\right\rangle =Nl^{2}  \label{Rg}
\end{equation}%
The cross terms vanish when averaged since we assumed the angular
orientation of the links was uncorrelated. \ Note that the characteristic
size of the ideal polymer $R_{g}=N^{\frac{1}{2}}l$ is significantly smaller
than that of the fully extended chain. \ In the limit of large $N$ the
probability distribution function $P_{N}(\mathbf{r})$ for a particular end
to end distance $\mathbf{r}$ is Gaussian. \ \ 
\begin{equation}
P_{N}(\mathbf{r})=\left( \frac{2\pi Nl^{2}}{3}\right) ^{-\frac{3}{2}}\exp
\left( -\frac{3\mathbf{r}^{2}}{2Nl^{2}}\right)
\end{equation}%
This statement follows from the central limit theorem, since the end to end
distance can be expressed as a sum of independent bond vectors. \
Alternatively one can consider the polymer configuration as a random walk,
in which case $P_{N}(\mathbf{r})$ satisfies the diffusion equation \cite%
{statprop}, \cite{scmbook}. \ Hence at fixed $\mathbf{r}$ the entropy of the
polymer chain is%
\begin{equation}
S(\mathbf{r})=\log (P_{N}(\mathbf{r}))=const-\frac{3\mathbf{r}^{2}}{2Nl^{2}}
\end{equation}%
Note that thourhgout this thesis we refrain from writing the Boltzmann
constant, choosing natural units with $k_{B}=1$. \ Since there is no
interaction energy in this model the free energy can be written as follows:%
\begin{equation}
F(\mathbf{r})=E-TS(\mathbf{r})=const+\frac{3T}{2Nl^{2}}\mathbf{r}^{2}
\end{equation}%
\qquad \qquad

If we stretch the chain by applying a stretching force $\mathbf{f}$ on both
ends of the polymer the free energy increases. \ In equilibrium the
corresponding elastic restoring force $\mathbf{f}_{el}=-\mathbf{f}$. \ The
extension of the polymer chain $\mathbf{R}$ as a result of applying a
stretching force is \ 
\begin{equation}
\mathbf{f=}\left. \mathbf{\nabla }F(\mathbf{r})\right\vert _{\mathbf{r=R}}=%
\frac{3T}{Nl^{2}}\mathbf{R}
\end{equation}%
which is valid provided the chain is not stretched too much $\left\vert 
\mathbf{R}\right\vert \ll Nl$. \ Hence we see that the ideal polymer behaves
like a mechanical spring with spring constant $k=3T/\left( Nl^{2}\right) $.
\ The chain stretches along the direction of the applied force, and the
corresponding restoring force is of purely entropic origin (i.e. since there
are fewer configurations of the stretched chain). \ 

On long enough length scales the single chain will be ideal. \ When excluded
volume interactions are included between the monomers, we expect to see
deviations from the ideal chain behavior. \ Flory presented the following
argument \cite{Flory} to determine how the size of the polymer chain depends
on the number of monomers $N$. \ We expect that the excluded volume between
monomers will favor swelling of the chain. \ If chain is confined to a
volume $R^{3}$ the average monomer concentration $c\sim N/R^{3}$. \ As a
result the total repulsive energy associated with monomer-monomer
interactions is proportional to $F_{rep}\sim Tvc^{2}R^{3}$ where we have
introduced the excluded volume parameter $v$ which in general may be
temperature dependent. \ However, stretching the chain costs entropy, so
there is a contribution to the free energy $F_{el}\sim \frac{TR^{2}}{Nl^{2}}$%
. \ Minimizing the total free energy $F=F_{rep}+F_{el}$ to determine the
preferred chain size $R_{F}$ we have%
\begin{eqnarray}
\frac{\partial }{\partial R}(F_{rep}+F_{el}) &=&-\frac{3TvN^{2}}{R^{4}}+%
\frac{2TR}{Nl^{2}}=0 \\
R_{F} &\simeq &N^{\frac{3}{5}}l^{\frac{2}{5}}v^{\frac{1}{5}}\simeq N^{\frac{3%
}{5}}l
\end{eqnarray}%
The Flory exponent $\nu =3/5$ gives the dependence of the chain size $%
R_{F}\sim N^{\nu }$ on the monomer number $N$. \ In writing the last
equality I have estimated the excluded volume parameter $v\simeq l^{3}$. \
Note that the chain is stretched as compared to the ideal chain which has $%
\nu =1/2$. \ 

With this in mind, we can return to the question of determining the
stretching response of a chain with excluded volume interactions. \ Here we
present a scaling argument due to Pincus \cite{Pincus}. \ The characteristic
length which enters the problem is the flory radius $R_{F}=N^{\nu }l$. \ The
other parameters of the problem are the magnitude of the stretching force $f$
and the thermal energy $T$. \ A scaling function $\varphi (x)$ with
dimensionless argument $x=(R_{F}f)/T$ is introduced to determine the
elongation of the polymer $R$ in response to the stretching force. \ 
\begin{equation}
R\simeq R_{F}\varphi (x)
\end{equation}%
When the stretching force is small the chain is weakly perturbed and the
response should be proportional to $f$. \ Hence for $x\ll 1$ we have $%
\varphi (x)\sim x$ and 
\begin{equation}
R\simeq R_{F}x=lN^{\frac{6}{5}}\left( \frac{lf}{T}\right) .
\end{equation}%
So we see that the spring constant of the chain with excluded volume
interaction $k\sim N^{-6/5}$ is reduced as compared to that of the ideal
chain $k\sim N^{-1}$. \ In the opposite regime of strong stretching we
require that the extension $R$ be linear in $N$. \ Hence for $x\gg 1$ we
assume $\varphi (x)\sim x^{m}$ and determine $m=1-(1/\nu )=2/3$ which
satisfies this condition. \ 
\begin{equation}
R\simeq R_{F}x^{m}=lN\left( \frac{lf}{T}\right) ^{\frac{2}{3}}
\end{equation}%
For strong stretching, the chain with excluded volume interaction has a
nonlinear force-extension relation $f\sim R^{3/2}$ which deviates from the
linear Hooke's Law for the ideal polymer chain. \ These results will be of
interest in later chapters when we model the interaction of colloids which
are connected by polymer springs. \ 

In a real polymer system there will be correlations between adjacent links,
in which case our assumption $\left\langle \mathbf{u}_{i}\mathbf{\cdot u}%
_{j}\right\rangle =0$ in Eq. \ref{Rg} is no longer valid. \ The persistence
length $l_{p}$ of the polymer chain provides a measure of the chain
flexibility, and is roughly the maximum length for which the polymer chain
remains straight. \ Let $\theta (s)$ be the angle between two segments of
the chain separated by a distance $s$. \ In these terms the persistence
length is defined as \cite{statprop} 
\begin{equation}
\left\langle \mathbf{\hat{u}}(0)\mathbf{\cdot \hat{u}}(s)\right\rangle
=\left\langle \cos \theta (s)\right\rangle =\exp \left( \frac{-s}{l_{p}}%
\right)
\end{equation}%
The persistence length of single-stranded DNA ($l_{p}\simeq 1nm$) is
significantly shorter than that of double-stranded DNA ($l_{p}\simeq 50nm$).
\ The double helix structure is quite rigid, whereas the single strand is
more flexible. \ For lengths $L<l_{p}$ the chain can be treated effectively
as a rigid rod. \ 

\section{DNA Grafted Colloids}

\qquad Here we present one approach whereby DNA can be used to organize
particles on the nanoscale. \ The general idea is to graft many DNA strands
onto the surface of a colloidal particle \cite{rational}. \ The size and
chemical composition of the colloid depends on the application. \ In some
experiments polystyrene beads with diameter $d\sim 1\mu m$ are utilized for
this purpose \cite{chaikin}, \cite{crocker}. \ Another common experimental
approach \cite{theory}, \cite{mirkincrystal}, \cite{colloidalcrystal} is to
use gold nanoparticles with $d\sim 10nm$. \ In this case the grafting is
made possible by attaching a thiol group to one end of the DNA strand which
binds to the surface of the gold nanoparticle. \ The result is a system of
monodisperse "octopus-like" particles where each particle has many DNA arms.
\ One end of each DNA chain is attached to the surface of the particle, and
the other end is free. \ In preparing such a system the experimenter can
control both the average DNA grafting density, and the particular nucleotide
sequence of the DNA\ arms. \ Note that preparing these "octopus-like"
particles relies on diffusion of DNA\ chains which adsorb to the particle
surface. \ This adsorption process is random or stochastic, as a result one
cannot control the exact number $N$ of DNA\ chains attached to a given
particle. \ Instead one controls the average number of DNA\ chains per
particle $\left\langle N\right\rangle $ by choosing the appropriate ratio of
the total DNA concentration $C_{DNA}$ to the total particle concentration $%
C_{particle}$ during preparation. \ 
\begin{equation}
\left\langle N\right\rangle =\frac{C_{DNA}}{C_{particle}}
\end{equation}%
This parameter $\left\langle N\right\rangle $ completely defines the
probability distribution for the number of DNA arms $N$ attached to a given
particle, which due to the random character of the preparation process must
have the Poisson form. \ 
\begin{equation}
P(N)=\frac{\left\langle N\right\rangle ^{N}\exp (-\left\langle
N\right\rangle )}{N!}
\end{equation}%
Here $P(N)$ is the probability that a particle has exactly $N$ DNA arms,
with $\left\langle N\right\rangle $ the average number of DNA arms on the
particles. \ 

The ability to control the sequence of DNA nucleotides attached to the
particles leads to interactions between particles of different types in
solution. \ We say that the "type" or "color" of the particle is determined
by the sequence of DNA nucleotides attached to the particle. \ For example,
consider particles grafted with many single-stranded DNA with sequence
ACTGAG. \ We call these "red" particles. \ We could also prepare "green"
particles with sequence CTCAGT. \ Here I label the sequences with the
following rule. \ The first letter is the base which is closest to the
grafting point, out to the last letter which is the base at the free end of
the DNA chain. \ Note that I have chosen the green sequence complementary to
the red one as dictated by the rule for complementary hydrogen bonding. \ In
solution when DNA arms of the red particles encounter DNA arms of the green
particles these two single strands of DNA can hybridize to form a double
strand. \ Provided that we are working under appropriate experimental
conditions (temperature, salt concentration, etc.) the formation of the
double strand results in a lower free energy state than if the two strands
were denatured. \ This provides a practical method to link particles through
the formation of a DNA bridge. \ The bond that results between particles
connected by DNA bridges is reversible, since we can change the temperature
or pH of the solution to denature the two DNA strands composing the bridge.
\ The binding energy for the formation of a DNA bridge will depend on a
number of factors, including the length of the complementary DNA sequence
and properties of the DNA chains attached to the particles \cite{statmech}.
\ For now we simply note that the interaction is highly-specific and
tunable. \ 

\section{Interactions}

\qquad The DNA-colloidal system we are considering is quite complex. \ In
general the interaction potential between colloids in solution combines
specific (or type-dependent) interactions with non-specific
(type-independent) interactions. \ The specific interactions pertain to the
formation of DNA bridges between colloids as a result of DNA-DNA
hybridization. \ The specificity is determined by the sequence of DNA
nucleotides attached to the particles and the complementary rule for DNA
base pairing. \ The non-specific interactions include all the interactions
which are independent of the particular DNA sequence. \ For example, this
includes the van der Waals attraction and electrostatic repulsion as
described by the DLVO theory \cite{israelachvili}, \cite{statmechsurface}. \
In addition we must take into account the steric repulsion between colloids
that arises from grafting many DNA chains to the surface of the particles. \
At first glance a quantitative treatment of the system appears discouraging
given the complexity and diversity of the interactions. \ However, by
comparing the characteristic energy and length scales we will see that the
most important interactions for our purposes are those directly related to
DNA, specifically DNA-DNA hybridization and steric repulsion. \ 

We first consider the non-specific interactions between colloids in
solution. \ The electrostatic interactions between charged colloids in ionic
solution are described by the Poisson-Boltzmann equation. \ Because the
equation is nonlinear an analytic treatment is generally only possible with
simple geometries in the context of some approximation scheme. \ In the
context of the Debye-H\"{u}ckle approximation the equation can be linearized
to obtain \cite{derjag}, \cite{overbeek}, \cite{bell} the pair potential
between two spherical colloids of radius $a$ carrying fixed charge $-Ze$. \ 
\begin{equation}
\frac{U_{el}(r)}{T}=Z^{2}\left( \frac{\exp (\kappa a)}{1+\kappa a}\right)
^{2}\frac{l_{B}}{r}\exp (-\kappa r)
\end{equation}%
Here $l_{B}=e^{2}/(\epsilon T)$ is the Bjerrum length and $\epsilon $ is the
dielectric constant of water \cite{statprop}. \ For water at room
temperature $\epsilon \approx 80$ which gives $l_{B}\approx 0.7nm$. \ The
presence of counterions in solution leads to screening of the electrostatic
potential. \ For monovalent counterions of concentration $n$ the Debye
screening length $\kappa ^{-1}=1/\sqrt{4\pi l_{B}n}$. \ The Debye length is
the length at which the counterions screen out electric fields. \ For
example, in a NaCl solution with concentration $0.2M$ the Debye length $%
\kappa ^{-1}\approx 0.68nm$. \ This ion concentration is typical of many
animal fluids. \ This estimate indicates that stabilizing colloidal
suspensions against non-specific aggregation electrostatically is not a
particularly appealing method due to the incredibly short range of the
resulting repulsive potential. \ This is especially true in many biological
applications where temperature and ion concentration are not set by the
experimenter. \ In general we will consider situations where the colloids
themselves are not charged, and electrostatic interactions can be neglected.
\ 

Even if the colloids are not charged, we still need to consider the DNA. \
In solution the phosphates which constitute the DNA backbone dissociate and
each carries a negative charge. \ Because each of the links carries charge,
we might expect that the repulsive interactions will lead to highly
stretched conformations of the chain $R_{g}\sim N$. \ Here $N$ is the number
of monomers in a single DNA chain. \ However, in ionic solution the charges
are screened. \ In fact for a strongly charged polyelectrolyte in ionic
solution the counter ions condense on the chain, effectively neutralizing
its charge \cite{manning}. \ Roughly speaking, the counterions condense once
the linear charge density of the chain $\rho $ exceeds the critical value $%
\rho _{crit}=e/l_{B}$. \ Electrostatic effects play a role in determining
certain properties of the DNA chains, for example they increase the
persistence length as compared to a neutral chain. \ However, from our
perspective the fact that the DNA backbone is charged will not be of great
importance. \ 

We now consider the van der Waals interaction between colloids. \ Consider
an atom which on average has a spherically symmetric charge distribution. \
Quantum mechanical fluctuations of the valence charge give rise to an
instantaneous dipole moment. \ The instantaneous dipole results in an
electric field at a distance $r$ from the atom $\overrightarrow{E}\sim
1/r^{3}$. \ This field induces a dipole moment $\overrightarrow{p}\sim 
\overrightarrow{E}$ in a nearby atom. \ The resulting interaction energy $%
U\sim -\overrightarrow{p}\cdot \overrightarrow{E}\sim -1/r^{6}$. \ Assuming
that the interaction between a collection of atoms is pairwise additive and
nonretarded one can write \cite{israelachvili} the following expression (in
the Derjaguin approximation) for the interaction potential between two
spheres of radius $a$. \ Here the spheres are separated by a surface to
surface distance $D=r-2a$ and the expression is valid for $D\ll a$. \ \ 
\begin{equation}
\frac{U_{vdW}(r)}{T}=-\frac{\widetilde{A}}{12}\frac{a}{D}  \label{vdw}
\end{equation}%
Here $\widetilde{A}=A/T$ is the reduced Hamaker constant. \ At $T=300K$ the
reduced Hamaker constant $\widetilde{A}\simeq 3$ for polystyrene in water,
and $\widetilde{A}\simeq 76$ for gold in water. \ For quantitative
comparisons Eq. \ref{vdw} is not particularly useful. \ A more detailed
treatment is required which takes into account the effects of retardation. \ 

The resulting attraction is insignificant when compared to the specific
attraction generated by DNA hybridization \cite{gangDNAcontrol}. \ For
example, in a recent study with micron sized polystyrene spheres, the van
der Waals attraction was estimated to be $U_{vdW}\simeq -3T$ at surface to
surface separations of $D=14nm$ and $U_{vdW}\simeq -10T$ at $D=10nm$. \ This
is to be compared with the energy scale for the DNA hybdriziation, which
will depend on the length of the complementary hybridization sequence. \ For
a $15$ base pair linker at room temperature $U_{DNA}\simeq -30T$ per DNA\
bridge! \ 

We now consider the steric repulsion of the DNA chains which prevents the
non-specific aggregation of colloids. \ Understanding the behavior of
polymer brushes is an active field of research. \ Treatments of increasing
complexity are available, from scaling arguments to self-consistent field
theories \cite{brush}, \cite{grafted}, \cite{wittenbrush}, \cite{milnerbrush}%
. \ Here we present a simple argument to outline the qualitative behavior of
grafted polymer brushes. \ As the surface grafting density $\sigma $ of the
DNA chains increases, there is a competition between entropic and excluded
volume effects. \ There is an energy penalty associated with monomer-monomer
contacts which favors stretching of the chain. \ However stretching the
chain costs entropy as discussed earlier. \ The result is the formation of a
DNA brush on the surface of the colloid. \ These brushes interact giving
rise to a repulsive potential between particles grafted with polymer chains.
\ 

Writing the competition between the excluded volume and entropic
interactions the free energy per chain $F$ in the brush of height $h$ is 
\begin{equation}
F=\frac{3T}{2Nl^{2}}h^{2}+T\frac{v}{2}N\left( \frac{N\sigma }{h}\right) 
\text{.}
\end{equation}%
Here $(N\sigma )/h$ is the average monomer concentration in the brush with
surface grafting density $\sigma $ and $v$ is the excluded volume parameter.
\ Minimization with respect to $h$ gives the free energy per chain $F_{\ast
} $ and the equilibrium brush height $h_{\ast }$ where we have estimated $%
v\simeq l^{3}$. \ 
\begin{eqnarray}
F_{\ast } &\simeq &TN(l^{2}\sigma )^{\frac{2}{3}} \\
h_{\ast } &\simeq &Nl(l^{2}\sigma )^{\frac{1}{3}}
\end{eqnarray}%
The resulting DNA brush is characterized by highly extended conformations of
the DNA chains, in particular the equilibrium height of the brush $h_{\ast }$
is proportional to the number of monomers in a chain $N$. \ 

There is an energy penalty associated with compressing the brushes once the
particle separation $D\lesssim 2h_{\ast }$. \ A more detailed treatment of
the problem takes into account the distribution of chain ends within the
brush. \ By making the analogy to an associated quantum mechanical problem
the authors of \cite{wittenbrush}\ have calculated the free energy penalty
associated with compressing the brush to a height $h<h_{\ast }$. \ Here we
quote the result for the free energy per chain $F(u)$ associated with
compressing the brush to a height $h<h_{\ast }$. \ The dimensionless
parameter $u=h/h_{\ast }$. \ 
\begin{equation}
F(u)=\frac{5F_{\ast }}{9}\left[ \frac{1}{u}+u^{2}-\frac{u^{5}}{5}\right]
\end{equation}%
An order of magnitude estimate \cite{colloidalcrystal} for compressing the
DNA\ brush below its equilibrium height repulsive gives several $T$ per DNA
chain. \ Therefore by tuning the brush height steric repulsion prevents
particles from ever approach at separations close enough to feel a
significant effect of the van der Waals attraction. \ Grafting polymers to
the particle surface is a controlled technique one can utilize to prevent
non-specific aggregation of particles in solution. \ Note that during this
discussion we considered a DNA\ brush, but the mechanism is largely
independent of the particular monomer chemistry. \ Another water soluble
polymer could play a similar role, one common choice in experiments is
polyethylene glycol (PEG). \ 

The result of this discussion indicates that in the DNA-colloidal system we
will consider, the pertinent interactions are those directly relating to the
DNA. \ There is a specific attraction associated with DNA\ hybridization,
and a non-specific steric repulsion which arises as a result of grafting
many DNA chains on the surface of the colloids. \ 

\section{Literature Review}

\qquad In this section we will highlight some of the literature which
addresses problems related to the topics of this thesis. \
In the past two decades, there have been a number of experimental advances
in DNA based self-assembly. \ These ideas originally stem from work in the lab
of Ned Seeman, who introduced the first schemes for building nanostructured
objects using specifically designed DNA \cite{natreview}, \cite{nucleic}. \
This approach has been adapted to demonstrate the ability of DNA\ to
rationally assemble aggregates of colloidal particles. \ There have been and
number of important contributions, including research in the groups of Mirkin \cite%
{rational}, Alivisatos \cite{alivisatos}, Soto \cite{blocks}, and many
others \cite{micelle}, \cite{nanocrystals}, \cite{designcrystals}, \cite%
{template}, \cite{angstrom}, \cite{nanotube}, \cite{chaikin}, \cite{crocker}%
, \cite{synthesis}, \cite{supra}, \cite{braun}, \cite{biotech}, \cite{kiang}%
, \cite{colloidalgold}, \cite{crockerdnaletter}, \cite{gangDNAcontrol}. \
One particularly interesting recent advance is the self-assembly of
colloidal crystals using DNA-mediated interactions by the groups of Gang 
\cite{colloidalcrystal} and Mirkin \cite{mirkincrystal}. \ These systems
have also attracted attention from a theoretical perspective. \ In one of
the first theoretical works on the subject \cite{morphology}, Tkachenko
studied the equilibrium phase behavior for a binary system of particles
decorated with DNA. \ The system exhibits a diverse spectrum of crystalline
phases, including the diamond phase which is of interest for the
self-assembly of photonic crystals. \ 

Some previous theoretical work on the aggregation and melting behavior of DNA-colloidal
assemblies include the work of Jin et al. \cite{meltingprop}, Park and
Stroud \cite{theory}, and Lukatsky and Frenkel \cite{phasebehav}. \ These
authors studied the aggregation behavior and optical properties of
DNA-mediated colloidal assemblies. \ One drawback to the previous work is
that the results were based on phenomenological or lattice based models
which give limited insight into the physics underlying the aggregation
phenomena. \ In Chapter II \cite{statmech} we develop an off-lattice,
statistical mechanical description of aggregation and melting in these
systems. \ The results of the theory are compared quantitatively to recent
experiments by the groups of Chaikin \cite{chaikin} and Crocker \cite%
{crocker}. \ There are connections between this aggregation behavior and the
sol-gel tranasition in branched polymers \cite{scaling}. \ Other soft matter
systems exhibit similar phenomena, for example a system of microemulsion
droplets connected by telechelic polymers \cite{microemulsion}. \ 

In addition to the work on bulk systems, DNA is a promising candidate to
self-assemble small clusters of particles, or nanoblocks. \ Independent of
the DNA based studies, Manoharan et al. \cite{packing} devised a scheme to
self-assemble small clusters of microspheres. \ The microspheres are
attached to the surface of liquid emulsion droplets, and the clusters
self-assemble by removing fluid from the droplet. \ The clusters are
packings of spheres that minimize the second moment of the mass
distribution. \ This packing sequence is somewhat ubiquitous in soft matter
systems. \ Glotzer et al. \cite{Glotzer}, \cite{Cone} have demonstrated that
cone-shaped clusters with $N\leqslant 10$ particles self-assemble into
clusters with the same packing sequence as \cite{packing}. \ This result is
not necessarily expected, since the self-assembly processes are driven by
different mechanisms. \ In the experiments capillary forces are responsible
for the assembly process, whereas in the simulations the interactions
between cone-particles are anisotropic and highly specific. \ Similar ideas
can be used to explain the structure of prolate virus capsids \cite{virus}.
\ In Chapter IV \cite{nanoclusters} we propose a method to self-assemble
clusters of particles with the same packing sequence, where the
self-assembly process is mediated by DNA. \ Other recent studies of the DNA
based assembly of nanoscale building blocks include \cite{biomolecule} and 
\cite{dnablock}. \ 

\chapter{DNA-mediated colloidal aggregation}

\section{Introduction}

\qquad In the past ten years, there have been a number of advances in
experimental assembly of nanoparticles with DNA-mediated interactions \cite%
{rational}, \cite{storhoff}, \cite{synthesis}, \cite{supra}, \cite%
{alivisatos}, \cite{braun}. \ While this approach has a potential of
generating highly organized and sophisticated structures \cite{morphology}, 
\cite{licata}, most of the studies report random aggregation of colloidal
particles \cite{chaikin}, \cite{crocker}. \ Despite these shortcomings, the
aggregation and melting properties may provide important information for
future development of DNA-based self--assembly techniques. \ These results
also have more immediate implications. For instance, the observed sharp
melting transition is of particular interest for biosensor applications \cite%
{biotech}. \ For these reasons the development of a statistical mechanical
description of these types of systems is of great importance. \ It should be
noted that the previous models of aggregation in colloidal-DNA systems were
either phenomenological or oversimplified lattice models \cite{meltingprop}, 
\cite{phasebehav}, \cite{theory}, which gave only limited insight into the
physics of the phenomena. \ 

In this chapter \cite{statmech}, we develop a theory of reversible
aggregation and melting in colloidal-DNA systems, starting from the known
thermodynamic parameters of DNA (i.e. hybridization free energy $\Delta G$),
and geometric properties of DNA-particle complexes. \ The output of our
theory is the relative abundance of the various colloidal structures formed
(dimers, trimers, etc.) as a function of temperature, as well as the
temperature at which a transition to an infinite aggregate occurs. \ The
theory provides a direct link between DNA\ microscopics and experimentally
observed morphological and thermal properties of the system. \ It should be
noted that the hybridization free energy $\Delta G$ depends not only on the
DNA\ nucleotide sequence, but also on the salt concentration and the
concentration of DNA\ linker strands tethered on the particle surface \cite%
{tetherhyb}. \ In this paper $\Delta G$ values refer to hybridization
between DNA free in solution. \ 

\begin{figure}[h]
{\includegraphics[width=3.9332in,height=2.9603in]{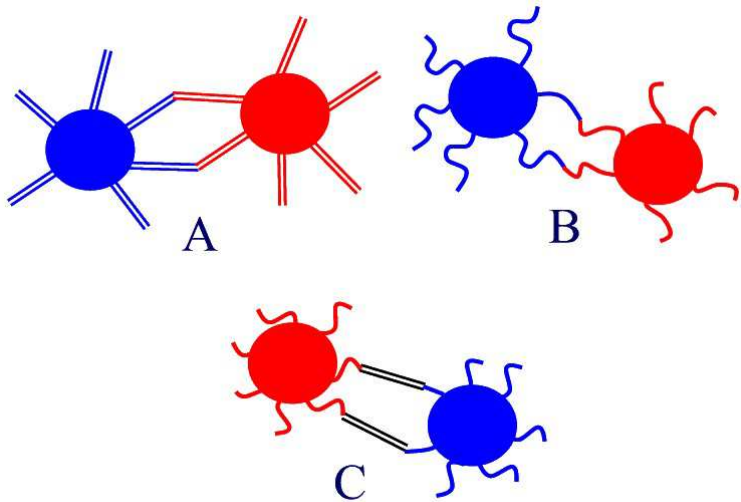}}
\caption{Graphical depiction of various schemes for DNA\ bridging. \ A) A
freely-jointed, rigid bridge constructed from complementary linker DNA. \ B)
A flexible bridge can be constructed using complementary linker DNA. \ C) A
rigid bridge constructed from short, flexible linker DNA and a long, rigid
linker. \ }
\label{hybridschemes}
\end{figure}

In a generic experimental setup, \ particles are grafted with DNA\ linker
sequences which determine the particle type($A$ or $B$). \ In this chapter
we will restrict our attention to a binary system\footnote{%
This restriction to binary systems is consistent with the current
experimental approach. \ In a later chapter we will demonstrate that if each
particle has a \textit{unique} linker sequence, one might be able to
programmably self-assemble nanoparticle clusters of desired geometry\cite%
{licata}.}. \ These linkers may be flexible or rigid. \ A selective,
attractive potential between particles of type $A$ and $B$ can then be
turned on by joining the linkers to form a DNA\ bridge. \ This DNA bridge
can be constructed directly if the particle linker sequences are chosen to
have complementary ends. \ Alternatively, the DNA bridge can be constructed
with the help of an additional linker DNA. \ This additional linker is
designed to have one end sequence complementary to the linker sequence of
type $A$ particles, and the other end complementary to type $B$. \ The
properties of the DNA bridge formed will depend on the hybridization
scheme(see figure \ref{hybridschemes}). \ 

The plan for the chapter is as follows. \ In section 2.2 we provide a
description of the problem. \ In section 2.3 we determine the bridging
probability for the formation of a DNA\ bridge between two colloids,
assuming the known thermodynamic parameters of DNA(hybridization free energy 
$\Delta G$). \ Using this bridging probability as input, in section 2.4 we
calculate the effective binding free energy $\epsilon _{AB}$ for the
formation of a dimer. \ Section 2.5 establishes the connection between the
theory and the experimentally determined melting profile $f(T)$, the
fraction of unbound particles as a function of temperature. \ In particular,
we demonstrate how knowledge of $\epsilon _{AB}$ can be used to determine
this profile, including the effects of particle aggregation. \ In section
2.6 the theory is compared with two recent experiments detailing the
reversible aggregation of colloids with DNA-mediated attraction \cite%
{meltingprop}, \cite{chaikin}. \ Section 2.7 presents a detailed description
of how the results can be applied to fit the experimental melting curves for
a binary system of DNA-grafted colloids. \ The main results of the model are
summarized in section 2.8. \ 

\section{Description of the Problem}

\qquad We consider particles of type $A$ and $B$ which form reversible $AB$
bonds as a result of DNA hybridization. \ The task at hand is to determine
the relative abundance of the various colloidal structures that form as a
function of temperature. \ From this information we can determine which
factors affect the melting and aggregation properties in DNA-colloidal
assemblies. \ To do so we must determine the binding free energy for all of
the possible phases(monomer, dimer, ..., infinite aggregate), and then apply
the rules for thermodynamic equilibrium. \ As we will see, these binding
free energies can all be simply related to $\epsilon _{AB}$, the binding
free energy for the formation of a dimer. \ Our task is thus reduced to
determining $\epsilon _{AB}$ from the thermodynamic parameters of DNA and
structural properties of the DNA\ linkers. \ In our statistical mechanical
framework, $\epsilon _{AB}$ is calculated from the model partition function,
taking into account the appropriate ensemble averaging for the non-ergodic
degrees of freedom. \ The result is related to the bridging probability for
a pair of linkers. \ By considering the specific properties of the DNA
bridge that forms, the bridging probability can be related to the
hybridization free energy $\Delta G$ of the DNA. \ In this way, we obtain a
direct link between DNA\ thermodynamics and the global aggregation and
melting properties in colloidal-DNA systems. \ 

\section{Bridging Probability}

\qquad To begin we relate the hybridization free energy $\Delta G$ for the
DNA in solution to the bridging probability for a pair of linkers. \ This
bridging probability is defined as the ratio $\frac{P_{bound}}{P_{free}}$,
with $P_{bound}$ the probability that the pair of linkers have hybridized to
form a DNA bridge, and $P_{free}$ the probability that they are unbound. \
This ratio is directly related to the free energy difference of the bound
and unbound states of the linkers $\Delta \widetilde{G}$ (throughout this
thesis we will use units with $k_{B}=1$): \ 
\begin{align}
\frac{P_{bound}}{P_{free}}& =\exp \left[ \frac{-\Delta \widetilde{G}}{T}%
\right] =\frac{c_{eff}}{c_{o}}\exp \left[ \frac{-\Delta G}{T}\right]
\label{bindprob} \\
c_{eff}& =\frac{\int P(\mathbf{r}_{1},\mathbf{r})P(\mathbf{r}_{2},\mathbf{r}%
)d^{3}\mathbf{r}}{\left( \int P(\mathbf{r},\mathbf{r}^{\prime })d^{3}\mathbf{%
r}\right) ^{2}}  \label{ceff}
\end{align}%
Here $c_{o}=1M$ is a reference concentration. \ $P(\mathbf{r},\mathbf{r}%
^{\prime })$ is the probability distribution function for the linker chain
which starts at $\mathbf{r}^{\prime }$ and ends at $\mathbf{r}$. The overlap
density $c_{eff}$ is a measure of the change in conformational entropy of
the linker DNA as a result of hybridization. \ It will depend on the
properties of the linker DNA(ex: flexible vs. rigid), and the scheme for
DNA\ bridging(ex: hybridization of complementary ends vs. hybridization
mediated by an additional linker). $\ c_{eff}$ is the concentration of free
DNA which would have the same hybridization probability as the grafted
linkers in our problem. \ As discussed in section 2.6, the DNA linker
grafting density also plays an important role in determining the possible
linker configurations and hence $c_{eff}$. \ 
\begin{figure}[h]
{\includegraphics[width=4.2774in,height=3.2188in]{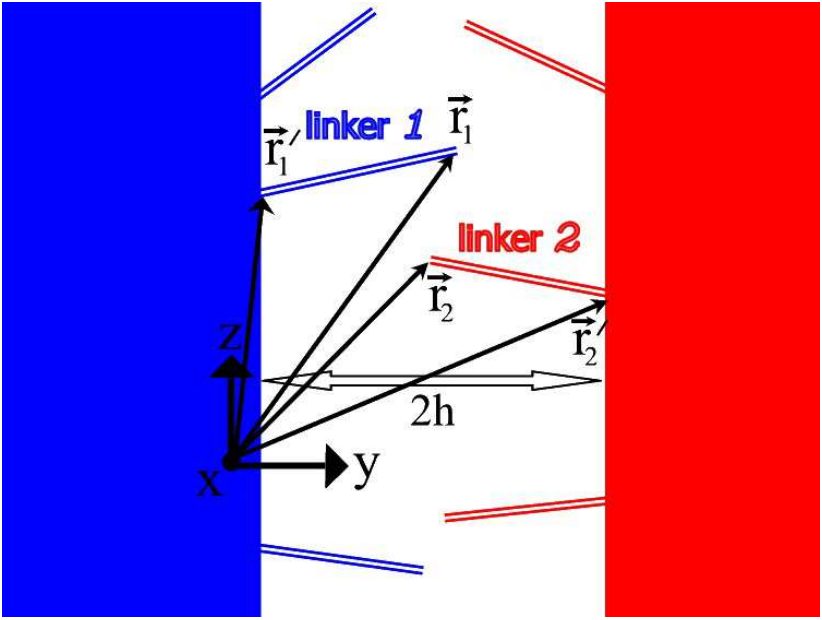}}
\caption{The statistical weight of a bound state is calculated by
determining the number of hybridized configurations for two complementary
linker chains relative to the number of unhybridized configurations. \ }
\label{setup}
\end{figure}

Assuming that the size of the linkers is much smaller than the particle
radius $R$, we first consider the problem in a planar geometry. Let the two
linkers be attached to two parallel planar surfaces separated by a distance $%
2h$. \ Referring to figure \ref{setup} we see that $\mathbf{r}^{\prime }$ is
the location where the linker DNA is grafted onto the particle surface, and $%
\mathbf{r}$ is the position of the free end. \ \ 

\subsection{Hybridization Scheme A: Freely-Jointed Rigid Linkers}

\qquad In this section we consider hybridization by complementary, rigid
linker DNA (scheme A in Figure \ref{hybridschemes}). \ This scheme is
particularly interesting since it is directly related to several recent
experiments \cite{chaikin}, \cite{meltingprop}. \ We assume that $L<l_{p}$
and $L\ll R$, where $l_{p}\simeq 50nm$ is the persistence length of ds DNA
and $L$ is the ds linker DNA length. In this regime, the linker chains can
be treated as rigid rods tethered on a planar surface. \ The interaction is
assumed to be point-like, in which a small fraction $\Delta /L$\ of the
linker bases hybridize.

\begin{figure}[h]
{\includegraphics[width=4.2774in,height=3.2188in]{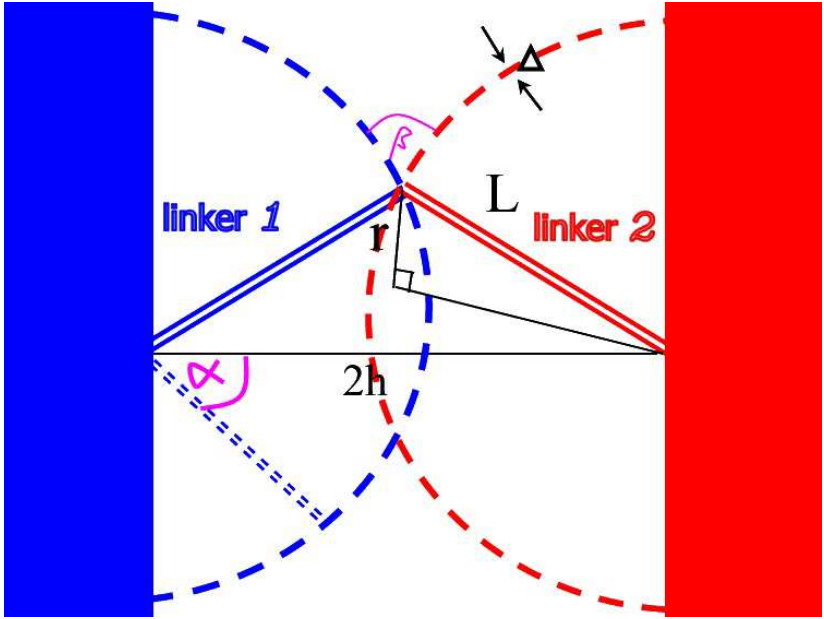}}
\caption{Cross-sectional view of the hybridization of two complementary
rigid linker DNA. \ The overlap density is calculated in a planar
approximation to the particle surface. \ }
\label{hybridization}
\end{figure}

We can calculate the overlap density by noting that the integral in Eq. (\ref%
{ceff}) is proportional to the volume of intersection of two spherical
shells (red and blue circles in Figure \ref{hybridization}) : \ 
\begin{equation}
c_{eff}=\frac{2\pi rA}{\left( 2\pi L^{2}\Delta \right) ^{2}}=\frac{\Theta
\left( L-\left\vert \mathbf{r}_{1}^{\prime }-\mathbf{r}_{2}^{\prime
}\right\vert /2\right) }{2\pi L^{2}\left\vert \mathbf{r}_{1}^{\prime }-%
\mathbf{r}_{2}^{\prime }\right\vert },
\end{equation}%
here $A=\Delta ^{2}/\sin \beta \ $and $r=\sqrt{L^{2}-\left\vert \mathbf{r}%
_{1}^{\prime }-\mathbf{r}_{2}^{\prime }\right\vert ^{2}/4}$ (see notations
in Figure \ref{hybridization}). We have used the fact that $\cos \beta
/2=\left\vert \mathbf{r}_{1}^{\prime }-\mathbf{r}_{2}^{\prime }\right\vert
/2L$. $\ c_{eff}$ and the binding probability are largest when the linkers
are grafted right in front of each other, i.e. when $\left\vert \mathbf{r}%
_{1}^{\prime }-\mathbf{r}_{2}^{\prime }\right\vert \sim 2h$. \ By taking the
limit $h\approx L$\ we arrive at the following result for the corresponding
"bridging" free energy \ 
\begin{equation}
\Delta \widetilde{G}_{A}\approx \Delta G_{A}+T\log \left[ 4\pi L^{3}c_{o}%
\right] .  \label{rigidbridge}
\end{equation}%
This free energy remains nearly constant for any pair of linkers, as long as
they can be connected in principle, i.e. $\left\vert \mathbf{r}_{1}^{\prime
}-\mathbf{r}_{2}^{\prime }\right\vert <2L$. This limits the maximum lateral
displacement of the linkers: \textbf{\ }$r_{\bot }<2\sqrt{L^{2}-h^{2}}$, and
therefore sets the \textit{effective cross-section} of the interaction: 
\begin{equation}
a=\mathbf{\ }\pi r_{\bot }^{2}=4\pi \left( L^{2}-h^{2}\right)
\label{rigidarea1}
\end{equation}%

\subsection{Hybridization Scheme B: Complementary Flexible Linkers}

\qquad We will now consider scenario B of figure \ref{hybridschemes},
hybridization of complementary, flexible linker DNA. \ This situation can be
realized in experiment by choosing linker DNA(ss or ds), provided the chain
length $L>>l_{p}$, the persistence length. \ We perform the calculation in a
planar approximation to the particle surface, which implies the particle
radius $R>>R_{g}$, the radius of gyration of the linker chain. \ In scenario
B, we must also take into account the entropic repulsion $G_{rep}$ of the
linker DNA which arises as a result of confining the chains between planar
surfaces. \ Since we are working with Gaussian chains, we can use the result
of Dolan and Edwards \cite{edwards}. \ Making the appropriate modification
to equation \ref{bindprob} the binding probability for a pair of linkers is
the following\footnote{$G_{rep}$ gives the free energy for a \textit{single }%
linker with one end grafted on the planar surface, and the other end free. \
The binding probability contains a factor of $2G_{rep}$ since each DNA
bridge is made by joining $2$ linkers. \ }. \ 
\begin{equation}
\exp \left[ \frac{-\Delta \widetilde{G}_{B}}{T}\right] =\frac{c_{eff}}{c_{o}}%
\exp \left[ \frac{-(\Delta G_{B}+2G_{rep})}{T}\right]
\end{equation}%
Defining $x\equiv \frac{h}{R_{g}}$ with the planar surfaces separated by a
distance $2h$, the free energy of repulsion has the following asymptotic
behavior. \ 
\begin{eqnarray}
G_{rep} &=&-T\log \left[ \frac{1}{2x}\sqrt{\frac{8\pi }{3}}\sum_{k=1\text{, }%
k\text{\ }odd}^{\infty }\exp \left[ -\frac{\pi ^{2}k^{2}}{24x^{2}}\right] %
\right] \\
&\simeq &T\left[ \log \left( 2x\sqrt{\frac{3}{8\pi }}\right) +\frac{\pi ^{2}%
}{24x^{2}}\right] \text{ \ \ }x\ll 1 \\
G_{rep} &=&-T\log \left[ 1-2\sum_{k=1}^{\infty }(-1)^{k+1}\exp \left[
-6x^{2}k^{2}\right] \right] \\
&\simeq &-T\log \left[ 1-2\exp \left( -6x^{2}\right) \right] \text{ \ \ }%
x\gg 1
\end{eqnarray}%
The details of the calculation are given in Appendix 1. \ The final result
gives the behavior of the binding probability $\exp \left[ \frac{-\Delta 
\widetilde{G}_{B}}{T}\right] $ between complementary, flexible linkers as a
function of $x$ and the separation between grafting points $\mathbf{\Delta r}%
^{\prime }=\mathbf{r}_{1}^{\prime }-\mathbf{r}_{2}^{\prime }$. \ 
\begin{eqnarray}
\Delta \widetilde{G}_{B} &\simeq &\Delta G_{B}+T\left[ 
\begin{array}{c}
\frac{3}{4}\left( \frac{\Delta r^{^{\prime }}}{R_{g}}\right) ^{2}+\log
\left( R_{g}^{3}c_{o}\right) +\log \left( \frac{32}{\pi ^{2}}\right) \\ 
+3\log (x)+\frac{\pi ^{2}}{12x^{2}}%
\end{array}%
\right] \text{ for\ \ }x\ll 1  \label{flexiblebridge} \\
\Delta \widetilde{G}_{B} &\simeq &\Delta G_{B}+T\left[ 
\begin{array}{c}
\frac{3}{4}\left( \frac{\Delta r^{^{\prime }}}{R_{g}}\right) ^{2}+\log
\left( R_{g}^{3}c_{o}\right) -\log \left( \frac{9}{4}\sqrt{\frac{3}{\pi }}%
\right) \\ 
-2\log (x)+3x^{2}%
\end{array}%
\right] \text{ \ for\ }x\gg 1
\end{eqnarray}%
Interpolating between the two regimes, we can see from the figure that the
minimum free energy is at $x\lesssim 1$.

\begin{figure}[h]
{\includegraphics[width=4.19in,height=3.15313in]{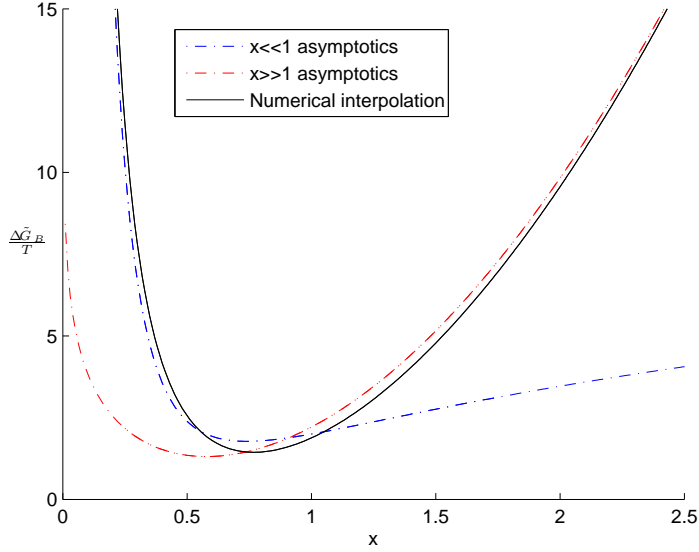}}
\caption{ $\frac{\Delta \widetilde{G}_{B}(\mathbf{\Delta r}^{^{\prime }}=0)}{%
T}$ as a function of $x$ for flexible DNA bridges. \ The figure is
normalized by choosing $\frac{\Delta G_{B}}{T}+\log (R_{g}^{3}c_{o})=0$. \ \ 
}
\label{ssdnamin}
\end{figure}

\subsection{Hybridization Scheme C: Short Flexible Markers with a Long Rigid
Linker}

\qquad We now turn our attention to scenario C of figure \ref{hybridschemes}%
, hybridization of short, flexible marker DNA with radius of gyration $R_{g}$
to a long, rigid linker DNA of length $L$. \ We will consider the case $%
\zeta \equiv \frac{L}{R_{g}}\gg 1$. \ For this reason we can neglect the
entropic repulsion $G_{rep}$ of the short linkers, since they only feel the
presence of the surface to which they are attached. \ However, in this
scenario we must take into account the loss of entropy of the long, rigid
DNA\ linker. \ After hybridization this linker strand does not have the full 
$4\pi $ steradians of rotational freedom it does when free in solution. \
The appropriate modification to the binding probability is:%
\begin{eqnarray}
\exp \left[ \frac{-\Delta \widetilde{G}_{C}}{k_{B}T}\right] &=&\frac{c_{eff}%
}{c_{o}}\exp \left[ \frac{-\Delta G_{C}}{k_{B}T}\right] \\
c_{eff} &=&\frac{1}{4\pi L^{2}}\frac{\int P(\mathbf{r}_{1},\mathbf{r}%
_{1}^{\prime })P(\mathbf{r}_{2},\mathbf{r}_{2}^{\prime })\delta (|\mathbf{r}%
_{1}-\mathbf{r}_{2}|-L)d^{3}\mathbf{r}_{1}d^{3}\mathbf{r}_{2}}{\left( \int P(%
\mathbf{r},\mathbf{r}^{\prime })d^{3}\mathbf{r}\right) ^{2}}
\end{eqnarray}%
Once again, the reader interested in the details of the calculation is
directed to Appendix B. \ For completeness we quote the result here. \ \ 
\begin{eqnarray}
\Delta \widetilde{G}_{C}(\mathbf{\Delta r}^{^{\prime }} &=&0,\epsilon
=\epsilon ^{\ast })\simeq \Delta G_{C}+T\log \left[ \frac{8\sqrt{\frac{\pi }{%
3}}L^{2}R_{g}c_{o}}{\left( e^{-\frac{3}{2}}-e^{-\frac{9}{4}}\right) }\right]
\label{linkedbridge} \\
&=&\Delta G_{C}+4.24T+T\log \left[ L^{2}R_{g}c_{o}\right]
\end{eqnarray}

\section{Effective Binding Free Energy}

\ \qquad We now proceed with the calculation of the effective free energy $%
\epsilon _{AB}$ , which is associated with the formation of a dimer from a
pair of free particles, $A$ and $B$. \ Since the DNA coverage on the
particle surface is not uniform, this free energy, and the corresponding
partition function $Z$, would in principle depend on the orientations of the
particles with respect to the line connecting their centers. The equilibrium
binding free energy would correspond to the canonical ensemble of all
possible orientations, i.e. $\epsilon _{AB}=-T\log 4\pi \left\langle
Z\right\rangle $. However, this equilibrium can only be achieved after a
very long time, when the particle pair samples all possible binding
configurations, or at least their representative subset. The real situation
is different. After the first DNA-mediated bridge is created the particle
pair can still explore the configurational space by rotating about this
contact point. However, after the formation of two or more DNA bridges (at
certain relative orientation of the particles), further exploration requires
multiple breaking and reconnecting of the DNA links, which is a very slow
process. We conclude that the system is ergodic with respect to the various
conformations of the linker DNA for fixed orientations of the particles, but
the orientations themselves are \textit{non-ergodic variables}. The only
exceptions are the single-bridge states: the system quickly relaxes to a
more favorable orientational state (unless the DNA coverage is extremely
low, and finding a second contact is very hard). \ \ If $N$ denotes the
number of DNA\ bridges constituting the $AB$ bond, the appropriate
expression for $\epsilon _{AB}$ in this partially ergodic regime is the
so-called component averaged free energy \cite{glasses}, \cite%
{brokenergodicity}: 
\begin{equation}
\epsilon _{AB}=-T\left\langle \log Z\right\rangle _{N\geq 2}
\end{equation}

Each DNA bridge between particles can be either open or closed. \ 
\begin{equation}
Z_{bridge}=1+\exp \left( -\frac{\Delta \widetilde{G}(h^{\prime },\mathbf{r}%
_{1}^{\prime }-\mathbf{r}_{2}^{\prime })}{T}\right)
\end{equation}%
Here $\mathbf{r}_{i}^{\prime }$ is the 2D position where the bridge is
grafted onto surface $i$. \ We now consider a generic case when the
interaction free energy $\Delta \widetilde{G}$ depends on the separation
between planar surfaces $2h^{\prime }$, and the separation of grafting
points $\mathbf{r}_{1}^{\prime }-\mathbf{r}_{2}^{\prime }$, without
assumption of a particular bridging scheme. \ If the probability for bridge
formation is small, two DNA linkers on the same surface will not compete for
complementary linkers. \ In this regime the free energy can be calculated by
summing over the contribution from each bridge that forms between dimers. \ 
\begin{equation}
F=-T\sum_{i}\sum_{j}\log \left[ 1+\exp \left( -\frac{\Delta \widetilde{G}%
(h^{\prime },\mathbf{r}_{i}^{\prime }-\mathbf{r}_{j}^{\prime })}{T}\right) %
\right]
\end{equation}%
We convert the summation to integration by introducing the linker areal
grafting density $\sigma $. \ 
\begin{equation}
F=-T\int \int \sigma _{1}(\mathbf{r}_{1}^{\prime })\sigma _{2}(\mathbf{r}%
_{2}^{\prime })\log \left[ 1+\exp \left( -\frac{\Delta \widetilde{G}%
(h^{\prime },\mathbf{r}_{1}^{\prime }-\mathbf{r}_{2}^{\prime })}{T}\right) %
\right] d^{2}\mathbf{r}_{1}^{\prime }d^{2}\mathbf{r}_{2}^{\prime }
\end{equation}%
Changing variables to $\mathbf{\Delta r}=\mathbf{r}_{1}^{\prime }-\mathbf{r}%
_{2}^{\prime }$ and $\mathbf{\rho }=\left( \mathbf{r}_{1}^{\prime }+\mathbf{r%
}_{2}^{\prime }\right) /2$, we can reintroduce the notion of a bridging
cross-section $a(h^{\prime })$, this time in a model-independent manner: \ 
\begin{equation}
a(h^{\prime })\log \left[ 1+\exp \left( -\frac{\Delta \widetilde{G}%
_{o}(h^{\prime })}{T}\right) \right] \equiv \int d^{2}\mathbf{\Delta r}\log %
\left[ 1+\exp \left( -\frac{\Delta \widetilde{G}(h^{\prime },\mathbf{\Delta r%
})}{T}\right) \right]  \label{area}
\end{equation}%
\ Here $\Delta \widetilde{G}_{o}(h^{\prime })\equiv \Delta \widetilde{G}%
(h^{\prime },\mathbf{\Delta r}=0)$ is the minimum free energy with respect
to the separation between grafting points $\mathbf{\Delta r}$. We can now
write the free energy: 
\begin{equation}
F=-T\int \sigma _{1}(\overrightarrow{\mathbf{\rho }})\sigma _{2}(%
\overrightarrow{\mathbf{\rho }})a(h^{\prime })\log \left[ 1+\exp \left( -%
\frac{\Delta \widetilde{G}_{o}(h^{\prime })}{T}\right) \right] d^{2}\mathbf{%
\rho }
\end{equation}%
We now convert from the planar geometry to the spherical particle geometry
using the Derjaguin approximation \cite{intermolecular}. \ 
\begin{align}
d^{2}\mathbf{\rho }& =\rho d\rho d\phi \\
h^{\prime }& =h+\frac{\rho ^{2}}{2R}
\end{align}%
Let $\Delta \widetilde{G}_{\ast }$ \ be the minimal value of the bridging
free energy. \ \ Then the result for $F$ can be rewritten as:

\bigskip 
\begin{equation}
F=-TN\log \left[ 1+\exp \left( -\frac{\Delta \widetilde{G}_{\ast }}{T}%
\right) \right]  \label{nbarFrelation}
\end{equation}%
Here $N$ has a physical meaning as the number of potential bridges for given
relative positions and orientations of the particles: 
\begin{equation}
N\equiv \int \sigma _{1}(\overrightarrow{\mathbf{\rho }})\sigma _{2}(%
\overrightarrow{\mathbf{\rho }})a(h^{\prime })\left( \frac{\log \left[
1+\exp \left( -\Delta \widetilde{G}_{o}(h^{\prime })/T\right) \right] }{\log %
\left[ 1+\exp \left( -\Delta \widetilde{G}_{\ast }/T\right) \right] }\right)
d^{2}\mathbf{\rho }
\end{equation}

One can calculate the average value of $N$ in terms of the average grafting
density, $\sigma =\left\langle \sigma _{1}\right\rangle =\left\langle \sigma
_{2}\right\rangle :$ \ 
\begin{equation}
\left\langle N\right\rangle \equiv 2\pi R\sigma ^{2}\int a(h^{\prime
})\left( \frac{\log \left[ 1+\exp \left( -\Delta \widetilde{G}_{o}(h^{\prime
})/T\right) \right] }{\log \left[ 1+\exp \left( -\Delta \widetilde{G}_{\ast
}/T\right) \right] }\right) dh^{\prime }  \label{nbar}
\end{equation}%
In a generic case of randomly grafted linkers, $\left\langle N\right\rangle $
completely defines the overall distribution function of $N$, which must have
a Poisson form: $P(N)=\frac{\left\langle N\right\rangle ^{N}e^{-\left\langle
N\right\rangle }}{N!}$. \ The average number of bridges $\left\langle
N\right\rangle $ between two particles depends on both the DNA linker
grafting density $\sigma $ and the bridging probability determined from $%
\Delta \widetilde{G}$. \ 

The free energy for the formation of a dimer $\epsilon _{AB}=\left\langle
F\right\rangle _{2+}-T\log \Omega $. \ The second term is the entropic
contribution to the free energy, which comes from integration over the
orientational and translational degrees of freedom of the second particle. \
Because the system is not ergodic in these degrees of freedom, the
accessible phase space $\Omega $ will be reduced by a factor of $P_{2+}$. $\
P_{2+}$ is the probability that there are at least two DNA bridges between
the particles. \ In terms of the average number of bridges $\left\langle
N\right\rangle $ between particles, we have the following relations: \ 
\begin{align}
P_{2+}& =1-(1+\left\langle N\right\rangle )e^{-\left\langle N\right\rangle }
\\
\left\langle N\right\rangle _{2+}& =\frac{\left\langle N\right\rangle \left(
1-e^{-\left\langle N\right\rangle }\right) }{P_{2+}}
\end{align}%
\begin{equation}
\epsilon _{AB}=-T\left\{ \left\langle N\right\rangle _{2+}\log \left[ 1+\exp
\left( -\frac{\Delta \widetilde{G}_{\ast }}{T}\right) \right] +\log \left[
P_{2+}4\pi \delta (2R)^{2}c_{o}\right] \right\}  \label{epsilonAB}
\end{equation}%
Here $\delta $ is the localization length of the $AB$ bond, which comes from
integrating the partition function over the radial distance between
particles. \qquad

\subsection{Scheme A}

\qquad We now can calculate $\left\langle N\right\rangle $ for the case of
freely-jointed rigid bridging considered earlier (i.e. for scheme A). \ In a
previous section we provided a direct calculation of the interaction free
energy, $\Delta \widetilde{G}_{o}\left( h\right) \approx const=\Delta 
\widetilde{G}_{\ast }$ (eq.\ref{rigidbridge}), and bridging cross-section, $%
a(h^{\prime })=4\pi (L^{2}-h^{\prime 2})$. Applying eq. \ref{nbar} we arrive
immediately at the following result. \ 
\begin{equation}
\left\langle N\right\rangle =8\pi ^{2}\sigma
^{2}R\int\limits_{0}^{L}(L^{2}-h^{\prime 2})dh^{\prime }=\frac{16\pi
^{2}\sigma ^{2}RL^{3}}{3}  \label{nbarA}
\end{equation}

\subsection{Scheme B}

\qquad We note that in this case, since the binding probability for a given
pair of linkers is Gaussian in the separation between grafting points $%
\mathbf{\Delta r}=\mathbf{r}_{1}^{\prime }-\mathbf{r}_{2}^{\prime }$, we can
perform an analytic calculation of the effective cross section. \ In what
follows $x^{\prime }=h^{\prime }/R_{g}$. \ Recall the definition of $%
a(h^{\prime })$:%
\begin{equation}
a(h^{\prime })\log \left[ 1+\exp \left( \frac{-\Delta \widetilde{G}%
_{o}(h^{\prime })}{T}\right) \right] =\dint d^{2}\mathbf{\Delta r}\log \left[
1+\exp \left( \frac{-\Delta \widetilde{G}_{B}(h^{\prime },\mathbf{\Delta r})%
}{T}\right) \right]
\end{equation}%
\begin{equation}
\exp \left[ \frac{-\Delta \widetilde{G}_{B}(\mathbf{\Delta r})}{T}\right]
=\Lambda (x^{\prime })\exp \left[ -\frac{3}{4}\left( \frac{\mathbf{\Delta r}%
}{R_{g}}\right) ^{2}\right]
\end{equation}%
\begin{equation}
\Lambda (x^{\prime })\equiv \frac{c(x^{\prime })}{R_{g}^{3}c_{o}}\exp \left[ 
\frac{-(\Delta G_{B}+2G_{rep})}{T}\right]
\end{equation}%
The explicit form of the dimensionless concentration $c(x^{\prime })$ is
given in Appendix A (see Eq. \ref{dimconc}). \ Changing to polar coordinates 
$(r,\theta )$ we have:%
\begin{equation}
a(x^{\prime })\log \left[ 1+\Lambda (x^{\prime })\right] =\dint_{0}^{2\pi
}d\theta \dint\limits_{0}^{\infty }r\log \left[ 1+\Lambda (x^{\prime })\exp %
\left[ -\frac{3}{4}\left( \frac{r}{R_{g}}\right) ^{2}\right] \right] dr
\end{equation}%
Define a new variable $u(r)\equiv \Lambda (x^{\prime })\exp \left[ -\frac{3}{%
4}\left( \frac{r}{R_{g}}\right) ^{2}\right] $. \ 
\begin{equation}
a(x^{\prime })\log \left[ 1+\Lambda (x^{\prime })\right] =-\frac{4\pi }{3}%
R_{g}^{2}\dint\limits_{\Lambda (x^{\prime })}^{0}du\frac{\log \left[ 1+u%
\right] }{u}=-\frac{4\pi }{3}R_{g}^{2}Li_{2}\left[ -\Lambda (x^{\prime })%
\right]
\end{equation}%
The calculation yields the following result, with $Li_{2}\left[ z\right]
\equiv \dsum\limits_{k=1}^{\infty }\frac{z^{k}}{k^{2}}$ the Dilogarithm. \ 
\begin{equation}
a(x^{\prime })=-\frac{4\pi }{3}\frac{Li_{2}\left[ -\Lambda (x^{\prime })%
\right] }{\log \left[ 1+\Lambda (x^{\prime })\right] }R_{g}^{2}
\end{equation}%
Since $Li_{2}(z)<0$ for $z<0$, $a(x^{\prime })$ is positive as required. \
From this effective cross section we can compute the average free energy $%
\left\langle F\right\rangle $ as a result of DNA bridging between particles,
with $\sigma $ the average areal grafting density of DNA\ linkers. \ Here $%
\mathbf{\rho }=\mathbf{r}_{1}^{^{\prime }}+\mathbf{r}_{2}^{^{\prime }}$,
with $\mathbf{r}_{i}^{^{\prime }}$ the location where linker $i$ is grafted
on the planar surface. \ 
\begin{equation}
\left\langle F\right\rangle =-T\sigma ^{2}\dint a(x^{^{\prime }})\log \left[
1+\Lambda (x^{^{\prime }})\right] d^{2}\mathbf{\rho }
\end{equation}%
Converting from the planar geometry to the spherical nanoparticle geometry
using the Derjaguin approximation we have:%
\begin{equation}
x^{^{\prime }}=x+\frac{\rho ^{2}}{2RR_{g}}
\end{equation}%
\begin{equation}
\left\langle F\right\rangle =T\frac{8\pi ^{2}}{3}\sigma
^{2}RR_{g}^{3}\dint\limits_{0}^{\infty }Li_{2}\left[ -\Lambda (x^{^{\prime
}})\right] dx^{^{\prime }}  \label{fbarB}
\end{equation}%
This integration can be performed numerically. \ As discussed previously,
the free energy $\epsilon _{AB}$ for the formation of a dimer($AB$ pair)
also contains an entropic contribution from integration over the
orientational and translational degrees of freedom of the second particle. \ 

\subsection{Scheme C}

\qquad We can also determine the free energy in this scenario using the
approximation method developed. \ 
\begin{equation}
F=-T\left\langle N\right\rangle \log \left[ 1+\exp \left( -\frac{\Delta 
\widetilde{G}_{C}(\mathbf{\Delta r}^{^{\prime }}=0,\epsilon =\epsilon ^{\ast
})}{T}\right) \right]
\end{equation}%
We provide a simple geometrical argument to determine the average number of
DNA\ bridges $\left\langle N\right\rangle $ between particles. \ We assume
that the rigid linkers are aligned with a small component parallel to the
surface. \ 
\begin{eqnarray}
a(h) &\simeq &\pi y^{2} \\
y &=&(L+\Delta )\tan \theta _{\max }\approx R_{g}(1+\epsilon )
\end{eqnarray}%
Then applying equation \ref{nbar} with $h=\frac{L}{2}(1+\epsilon )$ we have:%
\begin{equation}
\left\langle N\right\rangle =\pi ^{2}\sigma
^{2}RR_{g}^{2}L\dint\limits_{0}^{2\epsilon ^{\ast }}(1+\epsilon
)^{2}d\epsilon \simeq 2\sqrt{2}\pi ^{2}\sigma ^{2}RR_{g}^{3}  \label{nbarC}
\end{equation}

\section{Aggregation and Melting Behavior}

\qquad At this stage we have calculated the binding free energy $\epsilon
_{AB}$ for an $AB$ pair, starting with the thermodynamic parameters of DNA
(hybridization free energy $\Delta G$). \ In this section we establish the
connection between that result and the experimentally observable
morphological behavior of a large system. One of the ways to characterize
the system is to study its melting profile $f(T)$, which is the fraction of
unbound particles as a function of temperature. \ To determine the profile
we calculate the chemical potential for each phase(monomer, dimer, etc.) and
apply the thermodynamic rules for phase equilibrium. \ We will demonstrate
how the single binding free energy $\epsilon _{AB}$ can be used to determine
the contribution of each phase to the melting profile, including the effects
of aggregation. \ 

\subsection{Dimer Formation}

\qquad To begin we discuss the formation of dimers via the reaction $%
A+B\rightleftharpoons AB$. \ We can express the chemical potential of the $%
i^{th}$ species $\mu _{i}$ in terms of the particle concentrations $c_{i}=%
\frac{N_{i}}{V}$. \ 
\begin{align}
\mu _{A}& =T\log \left( c_{A}\right) \\
\mu _{B}& =T\log \left( c_{B}\right) \\
\mu _{AB}& =T\log \left( c_{AB}\right) +\epsilon _{AB}
\end{align}%
Here $\epsilon _{AB}$ is the binding free energy for the formation of a
dimer. \ In terms of the potential $V(r)$ between $A$ and $B$ type particles
we have: 
\begin{equation}
\epsilon _{AB}=-T\log \left[ 4\pi (2R)^{2}c_{o}\int dr\exp \left( -\frac{V(r)%
}{T}\right) \right]  \label{dnapotential}
\end{equation}

In this section we are not particularly concerned with the specific form of
the DNA-induced potential $V(r)$, having already determined $\epsilon _{AB}$%
\ in the previous section. \ We simply note that the prefactor $4\pi
(2R)^{2} $ arises since the interaction is assumed to be isotropic, with $R$
the particle radius. \ Equilibrating the chemical potential of the various
particle species, we obtain the condition for chemical equilibrium. $\ $ 
\begin{equation}
\mu _{A}+\mu _{B}=\mu _{AB}
\end{equation}%
The result is a relationship between the concentration of dimers and
monomers. \ \ 
\begin{equation}
c_{AB}=\frac{c_{A}c_{B}}{c_{o}}\exp \left[ \frac{-\epsilon _{AB}}{T}\right]
\end{equation}%
The overall concentration of particles in monomers and dimers must not
differ from the initial concentration. \ 
\begin{align}
c_{A}^{i}& =c_{A}+c_{AB} \\
c_{B}^{i}& =c_{B}+c_{AB}
\end{align}%
If the system is prepared at equal concentration, $c_{A}^{i}=c_{B}^{i}=\frac{%
1}{2}c_{tot}$, subtracting the two equations we see that $c_{A}=c_{B}\equiv
c $. \ Written in terms of the fraction of unbound particles $f=\frac{c}{%
\frac{1}{2}c_{tot}}$ we have a quadratic equation for the unbound fraction.
\ 
\begin{equation}
1=f+\exp \left[ \frac{-\widetilde{\epsilon }_{AB}}{T}\right] f^{2}
\end{equation}%
To simplify we have defined an effective free energy $\widetilde{\epsilon }%
_{AB}$ for the formation of a dimer. \ 
\begin{equation}
\widetilde{\epsilon }_{AB}=\epsilon _{AB}-T\log \left[ \frac{c_{tot}}{2c_{o}}%
\right]  \label{epsilontransformed}
\end{equation}%
The solution for the fraction of unbound particles as a function of
temperature is simply: 
\begin{equation}
f=\frac{-1+\sqrt{1+4\exp \left[ \frac{-\widetilde{\epsilon }_{AB}}{T}\right] 
}}{2\exp \left[ \frac{-\widetilde{\epsilon }_{AB}}{T}\right] }
\end{equation}%
\qquad \qquad

Previous studies\cite{chaikin} only included the dimer contribution to the
melting properties of DNA colloidal assemblies. \ With the basic formalism
at hand, we can now extend the preceding analysis to include the
contribution of trimers and tetramers. \ 

\subsection{Trimers and Tetramers}

\qquad Now consider the formation of a trimer via $2A+B\rightleftharpoons
ABA $. \ The chemical potential is slightly different in this case. \ 
\begin{equation}
\mu _{ABA}=T\log \left( c_{ABA}\right) +\epsilon _{ABA}
\end{equation}%
Taking into account that there are now two $AB$ bonds in the structure, one
might conclude that $\epsilon _{ABA}=2\epsilon _{AB}$. \ This is not quite
correct, since there is a reduction in solid angle available to the third
particle. \ To form a trimer, an $AB$ bond forms first, which contributes $%
\epsilon _{AB}$ to $\epsilon _{ABA}$. \ Some simple geometry shows that the
remaining $A$ particle only has $3\pi $ steradians of possible bonding sites
to particle $B$. \ Making this change in the prefactor of eq. \ref%
{dnapotential}, one can see that the second bond contributes $\epsilon
_{AB}-T\log \left( \frac{3}{4}\right) $ to $\epsilon _{ABA}$. \ 
\begin{equation}
\epsilon _{ABA}=2\epsilon _{AB}-T\log \left( \frac{3}{4}\right)
\end{equation}%
The equation for chemical equilibrium can once again be expressed in terms
of the particle concentrations. \ 
\begin{align}
2\mu _{A}+\mu _{B}& =\mu _{ABA} \\
c_{ABA}& =\frac{3}{4}\frac{c_{A}^{2}c_{B}}{c_{o}^{2}}\exp \left[ \frac{%
-2\epsilon _{AB}}{T}\right]
\end{align}%
To include the trimer contribution, we note that there are two possible
varieties, with $\epsilon _{ABA}=\epsilon _{BAB}$. \ 

\begin{align}
c_{A}^{i} & =c_{A}+c_{AB}+2c_{ABA}+c_{BAB} \\
c_{B}^{i} & =c_{B}+c_{AB}+c_{ABA}+2c_{BAB}
\end{align}
Following the same line of reasoning as before, the resulting equation for
the unbound fraction $f$ is: 
\begin{equation}
1=f+\exp\left[ \frac{-\widetilde{\epsilon}_{AB}}{T}\right] f^{2}+\frac{9} {4}%
\exp\left[ \frac{-2\widetilde{\epsilon}_{AB}}{T}\right] f^{3}
\end{equation}

For tetramers we will follow the same general reasoning, however in this
case there are two different structure types. \ The reaction $%
2A+2B\rightleftharpoons ABAB$ results in the formation of string like
structures. 
\begin{equation}
\mu_{ABAB}=T\log\left( c_{ABAB}\right) +\epsilon_{ABAB}
\end{equation}
As in the trimer case, the last particle has $3\pi$ steradians of possible
bonding sites, and contributes $\epsilon_{AB}-T\log\left( \frac{3}{4}\right) 
$ to $\epsilon_{ABAB}$. \ 
\begin{align}
\epsilon_{ABAB} & =3\epsilon_{AB}-T\log\left[ \left( \frac{3}{4}\right) ^{2}%
\right] \\
2\mu_{A}+2\mu_{B} & =\mu_{ABAB} \\
c_{ABAB} & =\left( \frac{3}{4}\right) ^{2}\frac{c_{A}^{2}c_{B}^{2}} {%
c_{o}^{3}}\exp\left[ \frac{-3\epsilon_{AB}}{T}\right]
\end{align}
If an $A$ type particle approaches a trimer of variety $ABA$, a branched
structure can result. \ The reaction $3A+B\rightleftharpoons AAAB$ results
in the formation of these branched structures. \ 
\begin{equation}
\mu_{AAAB}=T\log\left( c_{AAAB}\right) +\epsilon_{AAAB}
\end{equation}
For the branched case, the last particle has approximately $2\pi$ steradians
of possible bonding sites, and contributes $\epsilon_{AB}-T\log\left( \frac{1%
}{2}\right) $ to $\epsilon_{AAAB}$. \ 
\begin{align}
\epsilon_{AAAB} & =3\epsilon_{AB}-T\log\left( \frac{3}{8}\right) \\
3\mu_{A}+\mu_{B} & =\mu_{AAAB} \\
c_{AAAB} & =\left( \frac{3}{8}\right) \frac{c_{A}^{3}c_{B}}{c_{o}^{3}} \exp%
\left[ \frac{-3\epsilon_{AB}}{T}\right]
\end{align}
To include all of the tetramer contributions, note that there are two
branched varieties, with $\epsilon_{AAAB}=\epsilon_{BBBA}$. \ Finally we
impose the constraint that the initial particle concentrations do not differ
from the concentration of all the n-mers, for n=1,2,3,4. \ 
\begin{align}
c_{A}^{i} & =c_{A}+c_{AB}+2c_{ABA}+c_{BAB}+2c_{ABAB}+3c_{AAAB}+c_{BBBA} \\
c_{B}^{i} & =c_{B}+c_{AB}+c_{ABA}+2c_{BAB}+2c_{ABAB}+c_{AAAB}+3c_{BBBA}
\end{align}
The final result is an equation for the unbound fraction $f$ expressed
entirely in terms of the effective free energy $\widetilde{\epsilon}_{AB}$
of a dimer. \ 
\begin{equation}
1=f+\exp\left[ \frac{-\widetilde{\epsilon}_{AB}}{T}\right] f^{2}+\frac{9} {4}%
\exp\left[ \frac{-2\widetilde{\epsilon}_{AB}}{T}\right] f^{3}+\frac {21}{8}%
\exp\left[ \frac{-3\widetilde{\epsilon}_{AB}}{T}\right] f^{4}
\label{polynomial}
\end{equation}

For high temperatures, the melting profile is governed by the solution to
this polynomial equation for $f$. \ For temperatures below the melting point
we expect to find particles in large extended clusters. \ We now proceed to
calculate the equilibrium condition between monomers in solution and the
aggregate. \ 

\subsection{Reversible Sol-Gel Transition}

\qquad To understand the basic structure of the aggregate, we simply note
that there are many DNA\ attached to each particle. \ This gives rise to
branching, as in the discussion of possible tetramer structures. \ Since the
DNA\ which mediate the interaction are grafted onto the particle surface,
once two particles are bound, the relative orientation of the pair is
essentially fixed. \ The resulting aggregate is a tree-like structure, and
the transition to an infinite aggregate at low temperatures is analogous to
the sol-gel transition in branched polymers \cite{scaling}. \ 
\begin{figure}[h]
\begin{center}
\includegraphics[
height=3.2206in,
width=4.28in
]{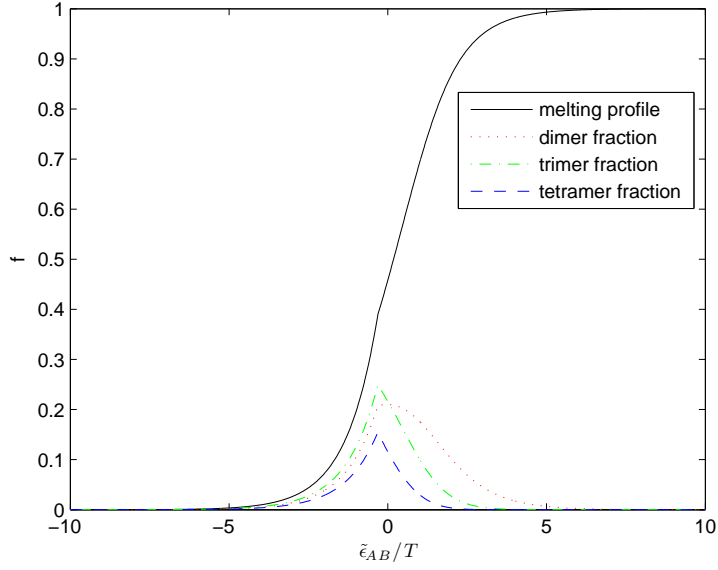}
\end{center}
\caption{The actual unbound fraction $f$\ is the concatenation of the
aggregate profile for $T<T^{\ast }\ $and the n-mer profile for $T>T^{\ast }$%
. \ The fraction of particles in dimers, trimers, and tetramers is also
plotted. \ }
\label{melt2}
\end{figure}

Particles in the aggregate are pinned down by their nearest neighbor bonds,
so we do not consider their translational entropy. $\ $As a result the
chemical potential is simply $\mu_{\infty}=\epsilon_{\infty}$. \
Equilibrating the chemical potential of the monomer in solution and in the
aggregate we have: 
\begin{align}
T\log\left( c\right) & =\epsilon_{\infty} \\
\epsilon_{\infty} & =\epsilon_{AB}-T\log\left( \gamma_{\infty}\right)
\end{align}
Here $\gamma_{\infty}\simeq1$ is the configurational entropy of the branched
aggregate, per particle.

The concentration of particles in the aggregate $c_{\infty}$ is the the
total concentration minus the n-mer concentration. \ Here $c_{1}=c_{A}+c_{B}$
is the total monomer concentration, $c_{2}=c_{AB}$ is the total dimer
concentration, etc. \ 
\begin{equation}
c_{\infty}\approx c_{tot}-c_{1}-c_{2}-c_{3}-c_{4}
\end{equation}
Expressed in terms of $\widetilde{\epsilon}_{AB}$ and the fraction of solid
angle available to particles in the aggregate $\gamma_{\infty}=\frac {%
\Omega_{\infty}}{4\pi}$ we have: 
\begin{equation}
f_{\infty}=\frac{1}{\gamma_{\infty}}\exp\left[ \frac{\widetilde{\epsilon }%
_{AB}}{T}\right]  \label{cluster}
\end{equation}

The transition from dimers, trimers, etc. to the aggregation behavior is the
temperature $T^{\ast}$ at which $f_{\infty}(T^{\ast})$ is a solution to eq. %
\ref{polynomial}. \ In words, $T^{\ast}$ is the temperature at which the
aggregate has a non-zero volume fraction. \ The fraction of unbound
particles for these colloidal assemblies will be governed by eq. \ref%
{cluster} for $T<T^{\ast}$ and eq. \ref{polynomial} for $T>T^{\ast}$. \ As
claimed, we can simply relate the unbound fraction to $\widetilde{\epsilon}%
_{AB}$ for both n-mers and the aggregate. \ 

\section{Comparison to the Experiments}

\qquad Let's consider the experimental scheme of Chaikin et al \cite{chaikin}%
. \ In the experiment, $R=.5\mu m$ polystyrene beads were grafted with ds
DNA linkers of length $L\simeq 20nm$. \ The 11 end bases of the $A$ and $B$
type particles were single stranded and complementary. \ We have already
determined the bridging probability in this scenario(see scheme A). \ In the
experiment \cite{chaikin} a polymer brush is also grafted onto the particle
surface, which will have the effect of preferentially orienting the rods
normal to the surface(See Figure \ref{hybridization}). \ This confinement of
the linker DNA can be incorporated quite easily into our results for $\Delta 
\widetilde{G}$ and $\left\langle N\right\rangle $. \ To modify Eq. \ref%
{rigidbridge}, when integrating over linker conformations we simply confine
each rigid rod to a cone of opening angle $2\alpha $. \ The upper bound for
the polar integration is now $\alpha $ as opposed to $\pi $. \ 
\begin{equation}
\Delta \widetilde{G}_{A}\simeq \Delta G_{A}+T\log \left[ 4\pi
L^{3}c_{o}(1-\cos \alpha )^{2}\right]  \label{gaconfined}
\end{equation}

The alignment effect should also be taken into account when calculating $%
\left\langle N\right\rangle $. \ If the particles are separated by less than 
$2L\cos \alpha $ the end sequences will be unable to hybridize. \ Following
the same steps as before, the lower bound for the $h^{^{\prime }}$
integration is now $L\cos \alpha $ as opposed to $0$. \ 
\begin{align}
\left\langle N\right\rangle & =8\pi ^{2}\sigma ^{2}R\int_{L\cos \alpha
}^{L}(L^{2}-h^{^{\prime }2})dh^{^{\prime }}  \label{nbarconfined} \\
& =\frac{16}{3}\pi ^{2}\sigma ^{2}RL^{3}\left[ 1+\frac{\cos \alpha }{2}(\cos
^{2}\alpha -3)\right]
\end{align}%
In the absence of the brush, and at sufficiently low linker grafting density 
$\sigma $, the alignment effect could be removed by setting $\alpha =\frac{%
\pi }{2}$, in which case we recover our previous results. Since the polymer
brush is stiff, it also imposes a minimum separation of $2h$ between
particles, where $h$ is the height of the brush. \ As a result, in the
expression for $\epsilon _{AB}$ we can approximate the radial flexibility of
the $AB$ bond as $\delta \simeq L-h$. \ 
\begin{figure}[h]
\begin{center}
\includegraphics[
height=3.2206in,
width=4.28in
]
{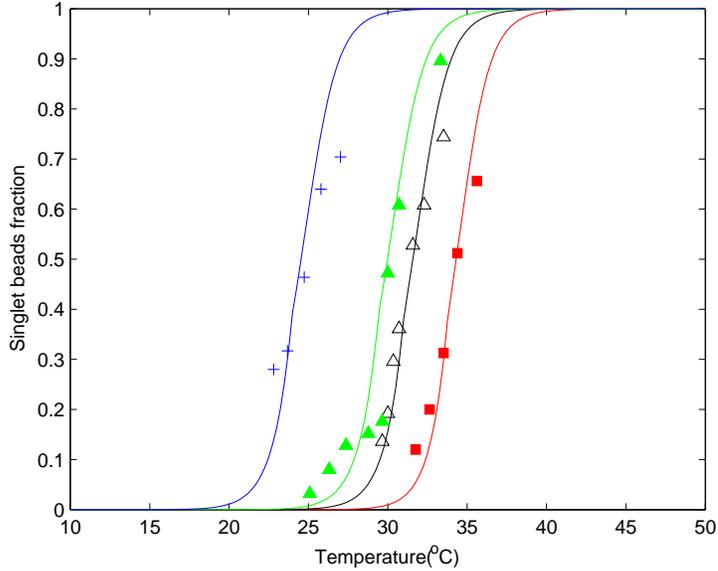}
\end{center}
\caption{Comparison of the melting curves $f(T)$ determined by our model to
the experimental data of Chaikin et al(See Fig.2 in \protect\cite{chaikin}).
\ The four data sets are for the four different polymer brushes used. \ For
the model fits we find that $\left\langle N\right\rangle _{2+}$=2.01 for
crosses, 2.07 for solid triangles, 2.13 for empty triangles, and 2.35 for
squares. \ }
\label{chaikinplot}
\end{figure}

We have now related the free energy $\epsilon _{AB}$ to the known
thermodynamic parameters of DNA($\Delta G=\Delta H-T\Delta S$, $\Delta
H=-77.2\frac{kcal}{mol}$ and $\Delta S=-227.8\frac{cal}{molK}$), and the
properties of linker DNA chains attached to the particles(grafting density $%
\sigma \simeq 3\times 10^{3}\frac{DNA}{\mu m^{2}}$ and linker length $%
L\simeq 20nm$). \ The height of the polymer brush is $h=13\pm 5nm$ \cite%
{chaikin}. \ In fitting the experimental data we have taken the average
value $\left\langle h\right\rangle =13nm$. \ Changing $h$ within these
bounds does not have a major effect on the melting curves. \ As a result
there is one free parameter in the model, the confinement angle $\alpha $. \
This angle determines $\left\langle N\right\rangle $ and $\Delta \widetilde{G%
}$, which in turn determine $\widetilde{\epsilon }_{AB}$, and finally the
melting profile $f$. \ 

With some minor modifications we can also analyze the "tail to tail"
hybridization mode in a recent experiment of Mirkin et al \cite{meltingprop}%
. \ In this experiment, $R=6.5nm$ gold nanoparticles were chemically
functionalized with ss DNA linkers. \ The last $15$ bases on the markers for
particles of type $A$ and $B$ were chosen to be complementary to a $30$ base
ss DNA linker. \ Since the strands are not ligated after hybridization, the
experimental pictures are similar. \ 
\begin{figure}[h]
\begin{center}
\includegraphics[
height=3.218in,
width=4.2774in
]
{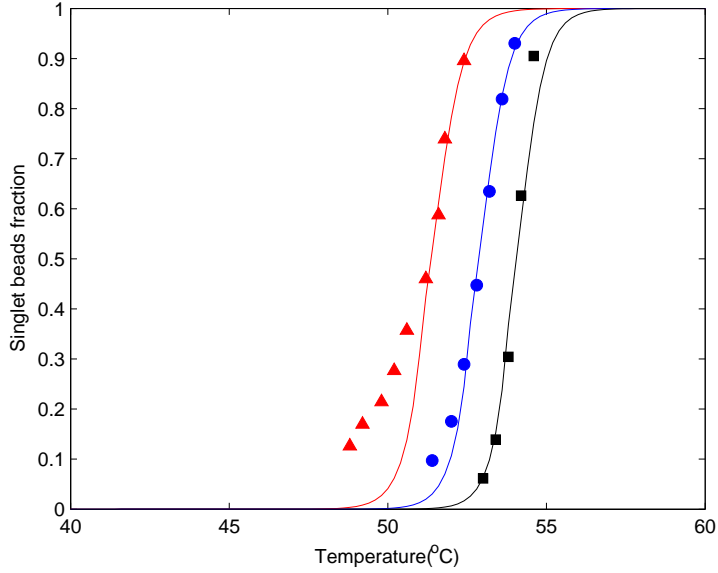}
\end{center}
\caption{The effect of the linker DNA grafting density $\protect\sigma $ on
the melting profile$\ f(T)$. \ The results of the model are compared with
experimental data in \protect\cite{meltingprop}. The three data sets
represent grafting densities of 100\%(squares), 50\%(circles), and
33\%(triangles) for which $\left\langle N\right\rangle _{2+}$=2.32, 2.16,
and 2.05 respectively. \ }
\label{mirkinplot}
\end{figure}

The unhybridized portion of the ss DNA linker simply serves as a spacer, and
the hybridized portions become ds DNA, which we can again treat as rigid
rods. \ This experiment is done without the addition of a polymer brush, but
the grafting density is two orders of magnitude larger than the experiment
of Chaikin et al. \ As a result, there is still an entropic repulsion \cite%
{morphology} associated with compressing the particles below separation $2h$%
. \ Here $h$ could loosely be interpreted as the radius of gyration of the
unhybridized portion of the linker. \ Despite the fact that $L\sim R$, our
planar calculation of $\Delta \widetilde{G}$ provides a good fit to the
experimental data. \ The other major difference is that now the attraction
between particles is mediated by an additional DNA linker. 
\begin{equation}
\Delta G=\Delta G_{1}+\Delta G_{2}-T\log \left[ \frac{c_{link}}{c_{o}}\right]
\label{addlinker}
\end{equation}

The term $\Delta G_{1}(\Delta G_{2})$ is the contribution to the free energy
from the hybridization of the linker on an $A(B)$ type particle to the
complementary portion of the 30 base ss linker. \ The hybridization free
energies $\Delta G_{1}$ and $\Delta G_{2}$ were calculated with the DINAMelt
web server \cite{dinamelt}. \ The last term is the contribution to the free
energy from the translational entropy of the additional linker DNA, with $%
c_{link}$ the additional linker concentration. \ This highlights some
incorrect assumptions of the thermodynamic melting model \cite{meltingprop},
where the two hybridization free energies were not calculated separately,
and the translational entropy of the additional linker DNA\ was ignored. \
By introducing dilutent strands to the system, one can probe the effect of
the linker grafting density $\sigma $ on the melting properties of the
assembly(See Figure 2B in \cite{meltingprop}). \ The agreement between the
experimental data and our theory is good, except at small $f$ values. \ This
is not surprising, since comparing the two requires relating the measurement
of optical extinction to the unbound fraction $f$. \ This is a nontrivial
matter when dealing with aggregation, which corresponds to the small $f$
regime. \ 

\section{Fitting Algorithm}

\qquad In this section we present a step by step method for fitting the
melting curves obtained experimentally for a binary system of DNA-grafted
colloids. \ 

\textbf{Step 1: Determine }$\Delta G$

The first step is to determine the hybridization free energy $\Delta G$ for
the DNA strands free in solution. \ In many cases the value has been
determined experimentally. \ Alternatively, there are a number of web based
applications which calculate hybridization free energies. \ For example, the
DINAMelt server which can be located at
http://www.bioinfo.rpi.edu/applications/hybrid/hybrid2.php and NUpack which
can be located at http://piercelab.caltech.edu/nupack. \ Note that in the
case where the hybridization is mediated by an additional linker, the
translational entropy of that linker must be taken into account (see Eq. \ref%
{addlinker}). \ 

\textbf{Step 2: }$\Delta G\rightarrow \Delta \widetilde{G}$

Since the DNA linkers in our problem are grafted onto the particle surface,
we need to determine how the grafting effects the bridging probability (see
Eqs. \ref{bindprob} and \ref{ceff}). \ This entails calculating the overlap
density $c_{eff}$ which is a measure of the change in conformational entropy
of the DNA strands upon hybridization. \ Determining the appropriate
calculation will depend on the hybridization scheme (see Fig. \ref%
{hybridschemes}). \ In this chapter calculations have been performed for
three different schemes (see Eqs. \ref{rigidbridge}, \ref{flexiblebridge}, %
\ref{linkedbridge}), although the effects of linker confinement have only
been taken into account in scheme $A$ (see Eq. \ref{gaconfined}). \ 

\textbf{Step 3: Calculate }$\left\langle N\right\rangle $

The next step in the procedure is to determine the average number of bridges 
$\left\langle N\right\rangle $ that form between an $AB$ pair. \ The general
starting point is Eq. \ref{nbar}. \ In this chapter calculations have been
performed for three different hybridization schemes (see Eqs. \ref{nbarA}, %
\ref{fbarB}, \ref{nbarC}). \ The effects of linker confinement have been
taken into account in scheme $A$ (see Eq. \ref{nbarconfined}). \ Note that
in our approximation scheme the general relation between $\left\langle
N\right\rangle $ and the free energy $F$ is given by Eq. \ref{nbarFrelation}%
. \ 

\textbf{Step 4: Determine }$\widetilde{\epsilon }_{AB}$

The next step in the procedure is to relate $\left\langle N\right\rangle $
to the binding energy for the formation of a dimer pair $\epsilon _{AB}$
(see Eq. \ref{epsilonAB}). \ The quantity of interest for the fitting $%
\widetilde{\epsilon }_{AB}$ is simply related to $\epsilon _{AB}$ by Eq. \ref%
{epsilontransformed}. \ 

\textbf{Step 5: Determine the melting profile }$f(T)$

We are now in a position to relate the calculation to the experimentally
measured quantity $f(T)$, which is the fraction of monomers as a function of
temperature. \ For high temperatures $f$ is determined by the solution of
the polynomial Eq. \ref{polynomial}. \ As the temperature is lowered at $%
T=T^{\ast }$ we reach the point where $f$ determined from the n-mer profile
(Eq. \ref{polynomial}) is equal to $f$ determined from the aggregate profile
(Eq. \ref{cluster}). \ For $T<T^{\ast }$ the melting profile is determined
by Eq. \ref{cluster}. \ 

\section{Summary}

\qquad We have developed a statistical mechanical description of aggregation
and melting in DNA-mediated colloidal systems. \ First we obtained a general
result for two-particle binding energy in terms of DNA hybridization free
energy $\Delta G$, and two model--dependent parameters: the average number
of available bridges $\left\langle N\right\rangle $ and the overlap density
for the DNA $c_{eff}$. \ We have also shown how these parameters can be
calculated for a particular bridging scheme. In our discussion we have
explicitly taken into account the partial ergodicity of the problem related
to slow binding-unbinding dynamics.

In the second part it was demonstrated that the fractions of dimers, trimers
and other clusters, including the infinite aggregate, are universal
functions of a parameter $\ \widetilde{\epsilon }_{AB}/T=\epsilon
_{AB}/T-\log \left[ c_{tot}/2c_{0}\right] $. \ The theory has been
calculated for three separate hybridization schemes. \ The obtained melting
curves are in excellent agreement with two types of experiments, done with
particles of nanometer and micron sizes. Furthermore, our analysis of the
experimental data give an additional insight into microscopic physics of DNA
bridging in these systems: it was shown that the experiments cannot be
explained without the introduction of angular localization of linker dsDNA.
\ The corresponding localization angle $\alpha $ is the only fitting
parameter of the model, which allows one to fit both the position and width
of the observed melting curves.

There are several manifestations of the greater predictive power of our
statistical mechanics approach, compared to the earlier more
phenomenological models. \ First, once $\alpha $ is determined for a
particular system, our theory allows one to calculate the melting behavior
for an alternative choice of DNA linker sequences. \ Second, if the
resulting clusters are separated, for example in a density gradient tube,
the relative abundance of dimers, trimers, and tetramers can be compared to
the values determined from the theory. \ 

Finally, the theory predicts aging of the colloidal structures, one
experimental signature for which is hysteresis of the melting curves. \ Such
an experiment proceeds by preparing a system above the melting temperature,
and measuring the unbound fraction of colloids as the temperature is
lowered. \ The system is allowed to remain in this cooled state for a very
long time, perhaps months, during which multiple DNA bridges break and
reform. \ During this time the colloids relax into a more favorable
orientation state, including states which are not accessible by simply
rotating about the contact point formed by the first DNA bridge between
particles. \ This favorable orientation state is characterized by an average
number of DNA\ bridges $\left\langle N\right\rangle $ greater than what we
calculate in the partially ergodic regime. \ If the unbound fraction is then
measured as the temperature is increased, the melting curve will shift to a
higher temperature, consistent with a larger value of $\left\langle
N\right\rangle $. \ 

\chapter{Dynamics of "Key-Lock" Interacting Particles}

\section{Introduction}

\qquad In this chapter \cite{keylock}, \cite{keylockletter} we present a
theoretical study of desorption and diffusion of particles which interact
through key-lock binding of attached biomolecules. \ It is becoming common
practice to functionalize colloidal particles with single-stranded DNA
(ssDNA) to achieve specific, controllable interactions \cite{crocker}, \cite%
{chaikin}, \cite{micelle}, \cite{natreview}, \cite{rational}, \cite%
{synthesis}. \ Beyond the conceptual interest as a model system to study
glassiness \cite{licata} and crystallization, there are a number of
practical applications. \ Colloidal self-assembly may provide a fabrication
technique for photonic band gap materials \cite{photonic}, \cite{wiremesh}.
\ One of the major experimental goals in this line of research is the
self-assembly of colloidal crystals using DNA mediated interactions. \ The
difficulty stems in part from the slow relaxation dynamics in these systems.
\ The main goal of this chapter is to understand how the collective
character of key-lock binding influences the particle dynamics. \ In doing
so we gain valuable insight into the relaxation dynamics, and propose a
modified experimental setup whose fast relaxation should facilitate
colloidal crystallization. \ 

Similar systems have also attracted substantial attention in other areas of
nanoscience. \ In particular, by functionalizing nanoparticles with
antibodies to a particular protein, the nanoparticles have potential
applications as smart, cell-specific drug delivery vehicles \cite%
{membranebend}, \cite{dendrimer}. \ These nanodevices take advantage of the
fact that certain cancerous cells overexpress cell membrane proteins, for
example the folate receptor. \ An improved understanding of desorption and
diffusion on the cell membrane surface may have implications for optimizing
the design of these drug delivery vehicles. \ This is the subject of chapter
7. \ 

In what follows we present our results on the dynamics of particles which
interact through reversible key-lock binding. \ The plan for the chapter is
the following. \ In section 3.2 we introduce the key-lock model and explain
the origin of the two model parameters $\Delta $ and $\overline{m}$. \ The
parameter $\Delta $ determines the binding energy for the formation of a
key-lock pair. \ The parameter $\overline{m}$ is the mean of the
distribution for the number of key-lock bridges. \ Depending on $\overline{m}
$, which is related to the coverage of the functional groups (e.g. ssDNA),
there are two distinct regimes. \ At low coverage there is an exponential
distribution of departure times, but no true lateral diffusion. \ As the
coverage increases, we enter a regime where the particle dynamics is a
result of the interplay between desorption and diffusion. \ An estimate is
provided for the value of $\overline{m}$ which determines the crossover from
the localized to diffusive regime in section 3.3. \ In section 3.4 the
localized regime is discussed in detail. \ In this regime the particle is
attached to a finite cluster and remains localized near its original
location until departing. \ We derive the partition function for the finite
clusters, and calculate the departure time distribution. \ In section 3.5 we
determine the departure time distribution in the diffusive regime. \ We
present an effective Arrhenius approximation for the hopping process and a
Fourier transform method which greatly simplifies the calculation. \ In
section 3.6 we discuss the random walk statistics for the particles'
in-plane diffusion. \ A set of parametric equations is derived to relate the
average diffusion time to the mean squared displacement. \ The lateral
motion is analogous to dispersive transport in disordered semiconductors,
ranging from standard diffusion with a renormalized diffusion coefficient to
anomalous, subdiffusive behavior. \ In section 3.7 we connect our results to
recent experiments with DNA-grafted colloids. \ We then discuss the
implications of the work for designing an experiment which facilitates
faster colloidal crystallization. \ In section 3.8 we conclude by
summarizing our main results. \ 

\begin{figure}[tbp]
\includegraphics[width=4.6112in,height=3.4705in]{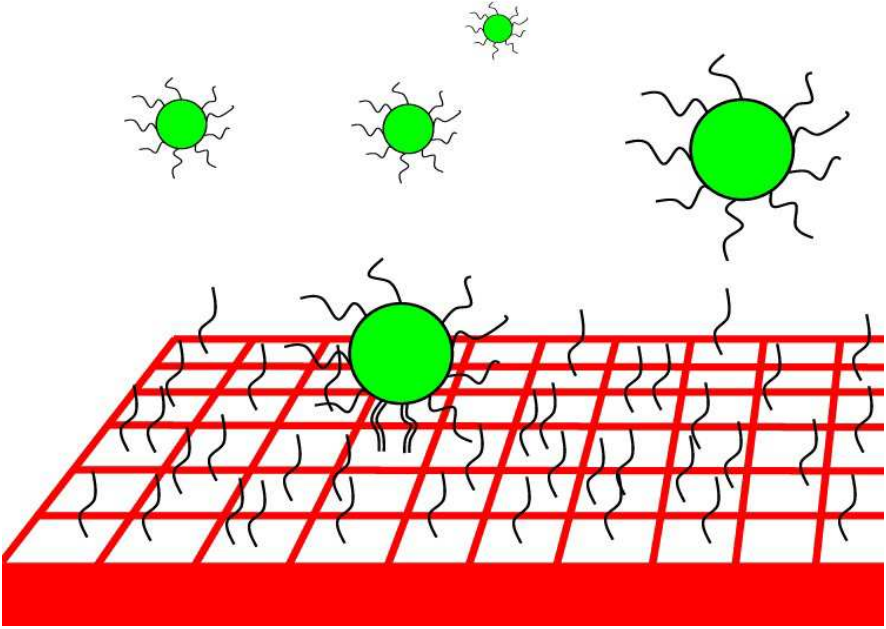}
\caption{Graphical depiction of particles interacting with a flat, 2D
substrate by multiple key-lock binding. }
\label{substrate}
\end{figure}

\section{Model Description}

\qquad We now present the model, where a single particle interacts with a
flat two-dimensional surface by multiple key lock binding (see Fig. \ref%
{substrate}). \ At each location on the surface there are $m$ key-lock
bridges which may be open or closed, with a binding energy of $\epsilon $
for each key-lock pair. \ Here we have neglected the variation in $\epsilon $%
. \ In the case of the DNA-colloidal system mentioned in the introduction,
the model parameter $\epsilon $ is related to the hybridization free energy
of the DNA. \ The resulting $m$-bridge free energy plays the role of an
effective local potential for the particle \cite{statmech}:%
\begin{align}
U(m)& =-Tm\Delta  \label{potenergy} \\
\Delta & \equiv \log (1+\exp [\epsilon /T])
\end{align}%
Generically, $m$ is a Poisson distributed random number $P_{m}=\overline{m}%
^{m}\exp (-\overline{m})/m!$ where $\overline{m}$ denotes the mean of the
distribution. \ The model parameter $\overline{m}$ is a collective property
of the particle-surface system. \ For example, consider the case of
dendrimers functionalized with folic acid, which can be utilized for
targeted, cell specific chemotherapy. \ The folic acid on the dendrimer
branch ends form key-lock bridges with folate receptors in the
cell-membrane. \ In this case $\overline{m}$ will depend on the distribution
of keys (folic acids) on the dendrimer, and the surface coverage of locks
(folate receptors) in the cell membrane. \ 

At each location, the particle is attached to the surface by $m$ bridges. \
To detach from the surface the particle must break all its connections, in
which case it departs and diffuses away into solution. \ Alternatively the
particle can hop a distance $a$ to a new location characterized by a new
value of the bridge number $m$. \ By introducing the correlation length $a$,
we have coarse-grained the particle motion by the distance after which the
new value of the bridge number becomes statistically independent of the
value at the previous location. \ In the localized regime the particle
remains close to its original location until departing. \ In the diffusive
regime the particle is able to fully explore the surface through a random
walk by multiple breaking and reforming of bridges. \ 
\begin{figure}[tbp]
\includegraphics[width=4.6138in,height=3.4714in]{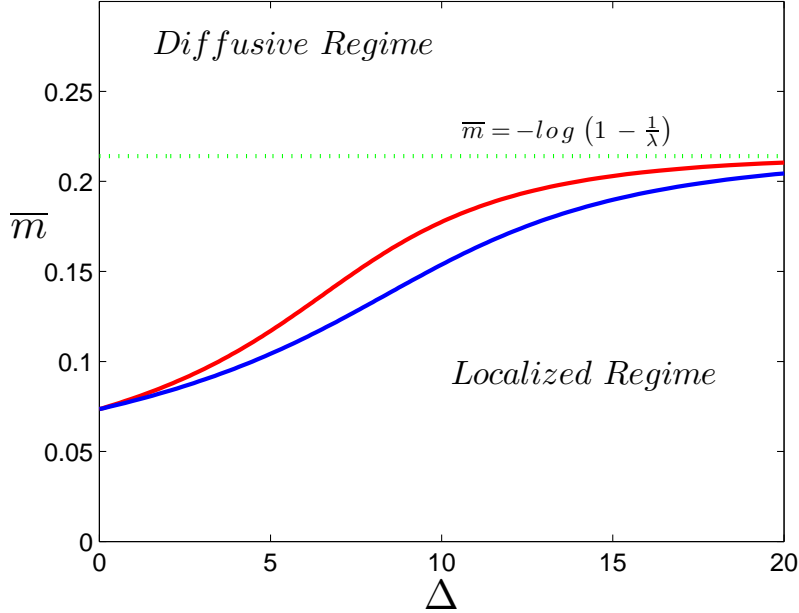}
\caption{The crossover from the localized to the diffusive regime below the
percolation threshold. \ Estimates based on the characteristic cluster size
(Eq. \protect\ref{crossnum}, red line) and confinement of the random walk
(Eq. \protect\ref{crosswalk}, blue line) give similar crossovers. \ For
large $\Delta $ the crossover condition is $\overline{m}=-\log \left( 1-%
\frac{1}{\protect\lambda }\right) $. \ }
\label{crosspoint3}
\end{figure}

\section{Crossover from Localized to Diffusive Behavior}

\qquad Naively one might expect the crossover between the two regimes to
occur at the percolation threshold, where one first encounters an infinitely
connected cluster of sites with $m>0$. \ However, the crossover from the
localized to diffusive regime occurs at smaller $\overline{m}$\ than
predicted by percolation theory. \ If $p_{c}=1/2$ denotes the critical
probability for site percolation on the triangular lattice, the percolation
transition occurs at $\overline{m}=\log \left( 2\right) $. \ There are two
alternative estimates for the crossover from the localized to the diffusive
regime. \ The first is to compare the average number of steps $n=\exp \left(
\Delta \overline{m}\right) $ the particle takes before departing (see
section 3.5) to the characteristic cluster size $s_{c}=1/\log (1/\lambda p)$
below the percolation threshold. \ Here $\lambda =5.19$ is a numerical
constant for the triangular lattice \cite{cluster}, and in the percolation
language $p=1-\exp (-\overline{m})$ is the occupancy probability. \ The
crossover condition $n=s_{c}$ can be expressed as a function of $\overline{m}
$. \ 
\begin{equation}
\Delta =-\frac{1}{\overline{m}}\log \left[ -\log \left\{ \lambda \left(
1-e^{-\overline{m}}\right) \right\} \right]  \label{crossnum}
\end{equation}%
Alternatively, in the localized regime the particles' random walk is
confined by the characteristic cluster size. \ Below percolation the radius
of gyration of the cluster is $R_{s}\sim s^{\rho }$ with $\rho =0.641$ in
two dimensions. \ Comparing the radius of gyration of the cluster to the
radius of gyration for the particles' random walk, the crossover occurs at $%
n=\left( s_{c}\right) ^{2\rho }$. \ 
\begin{equation}
\Delta =-\frac{2\rho }{\overline{m}}\log \left[ -\log \left\{ \lambda \left(
1-e^{-\overline{m}}\right) \right\} \right]  \label{crosswalk}
\end{equation}%
Since $2\rho $ differs from $1$ by less than $30\%$, both conditions give
similar crossovers (see Fig. \ref{crosspoint3}). \ The saturation at $%
\overline{m}=-\log \left( 1-\frac{1}{\lambda }\right) $ occurs for very
large $\Delta $, as a result for binding energies of a few $T$ per bridge
the crossover occurs at $\overline{m}\simeq 0.1$. \ 

\section{Localized Regime}

\qquad In the percolation language, when the occupancy probability $p=1-\exp
(-\overline{m})$ is small, particles are localized on finite clusters. \ In
this localized regime particles are able to fully explore the cluster to
which they are attached before departing. \ This thermalization of particles
with finite clusters permits an equilibrium calculation of the cluster free
energy $F=-T\log \langle Z\rangle $. \ The departure rate is given by the
Arrhenius relation $K=\frac{1}{\tau _{0}}\exp \left( F/T\right) $. \ Here $%
\tau _{0}$ is a characteristic timescale for bridge formation. \ The
probability that the particle departs between $t$ and $t+dt$ is determined
from the departure time distribution $\Phi (t)dt\simeq K\exp [-Kt]dt$. \ 

To begin we calculate the partition function for the finite clusters. \ The
cluster is defined as $s$ connected sites on the lattice, all of which are
characterized by $0<m<m^{\ast }$ bridges. \ For Poisson distributed bridge
numbers the partition function for the finite cluster is: \ 
\begin{equation}
Z(m^{\ast },s)=\sum_{i=1}^{s}\sum_{m_{i}=1}^{m^{\ast }-1}\widetilde{P}%
_{m_{i}}\exp (\Delta m_{i})=\frac{s}{\exp (\overline{m})-1}(\exp (\overline{m%
}e^{\Delta })Q(\overline{m}e^{\Delta },m^{\ast })-1)
\end{equation}%
\ Because by definition the cluster does not contain sites with $m=0$
bridges we have renormalized the probability distribution $\widetilde{P}%
_{m}=P_{m}/(1-\exp (-\overline{m}))$ so that $\sum_{m=1}^{\infty }\widetilde{%
P}_{m}=1$. \ Here $Q(x,m^{\ast })\equiv \Gamma (x,m^{\ast })/\Gamma (m^{\ast
})=\exp (-x)\sum_{k=0}^{m^{\ast }-1}x^{k}/k!$ is the regularized upper
incomplete $\Gamma $ function. \ In the language of the statistics of
extreme events, $m^{\ast }-1$ is the maximum "expected" value of $m$ in a
sample of $s$ independent realizations \cite{dispersive}. The point is that
on finite clusters we should not expect to achieve arbitrarily large values
of the bridge number. \ Hence when averaging the partition function to
obtain the cluster free energy one should only average over sites with $%
m<m^{\ast }$. \ The distribution function for $m^{\ast }$ is obtained by
noting that the probability that all $s$ values of $m$ are less than $%
m^{\ast }$ is $\left( \sum_{m=1}^{m^{\ast }-1}\widetilde{P}_{m}\right)
^{s}=\left( \frac{\exp (\overline{m})Q(\overline{m},m^{\ast })-1}{\exp (%
\overline{m})-1}\right) ^{s}$. \ By differentiating this quantity with
respect to $m^{\ast }$ we obtain the distribution function for the maximum
expected value of $m$. \ 
\begin{equation}
f_{s}(m^{\ast })=s\left( \frac{\exp (\overline{m})Q(\overline{m},m^{\ast })-1%
}{\exp (\overline{m})-1}\right) ^{s-1}\widetilde{P}_{m^{\ast }}
\end{equation}%
\ The cluster size distribution below the percolation threshold is
exponential \cite{percolationtheory} with characteristic cluster size $%
s_{c}=1/\log (1/\lambda p)$. \ 
\begin{equation}
p_{s}(p)=\frac{1-\lambda p}{\lambda }\exp \left( -\frac{s}{s_{c}}\right)
\end{equation}%
The summation over $s$ can be performed analytically, which allows the
result to be expressed as a single summation over $m^{\ast }$. \ 
\begin{equation}
\langle Z\rangle =\sum_{m^{\ast }=2}^{\infty }\sum_{s=1}^{\infty
}p_{s}(p)f_{s}(m^{\ast })Z(m^{\ast },s)=
\end{equation}%
\begin{equation}
\frac{\lambda (1-\lambda (1-\exp (-\overline{m})))}{\exp (\overline{m})-1}%
\sum_{m^{\ast }=2}^{\infty }\widetilde{P}_{m^{\ast }}(\exp (\overline{m}%
e^{\Delta })Q(\overline{m}e^{\Delta },m^{\ast })-1)\frac{1+y(m^{\ast })}{%
(1-y(m^{\ast }))^{3}}  \notag
\end{equation}%
\begin{equation}
y(m^{\ast })=\lambda (Q(\overline{m},m^{\ast })-\exp (-\overline{m}))
\end{equation}

\begin{figure}[tbp]
\includegraphics[width=4.6138in,height=3.4714in]{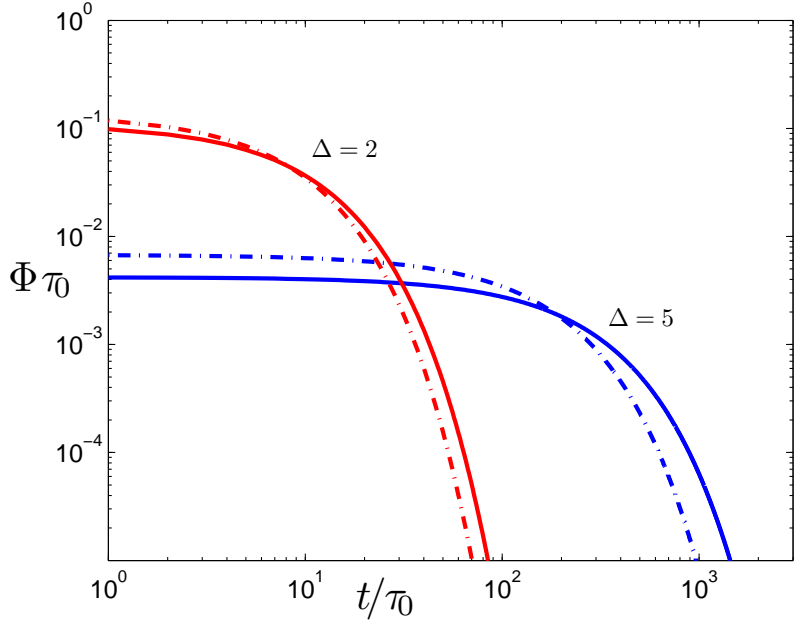}
\caption{Departure time distribution function versus time in the localized
regime with $\overline{m}=0.1$. The results of our calculation (solid lines)
are compared to single exponential relaxation with departure rate $K=\frac{%
\exp (-\Delta )}{\protect\tau _{0}}$ (dotted lines). }
\label{localreg3}
\end{figure}

For fixed $\overline{m}$, as in the plot (see Fig. \ref{localreg3}),
changing $\Delta $ is directly related to a change in the average binding
free energy. \ Increasing $\Delta $ leads to a reduction in the rate of
particle departure. \ 

\section{Diffusive Regime}

\qquad The departure time distribution changes significantly in the
diffusive regime. \ In this regime the particle can explore the surface to
find a more favorable connection site, which leads to a longer lifetime for
the bound state. \ This phenomenon is qualitatively similar to \textit{aging}
in glassy systems. \ In these systems one finds that the response to an
external field is time dependent \cite{glassybook}. \ In the magnetic
analogy this leads to a time dependence of the magnetization. \ Below the
glass temperature, the longer one waits before applying the external
magnetic field, the more time the system has to settle into deep energy
wells, and the smaller the response. \ In our case, the diffusive
exploration of the particle allows it to find a deeper energy well, which
leads to an increase in the bound state lifetime. \ 

\begin{figure}[tbp]
\includegraphics[width=4.6138in,height=3.4714in]{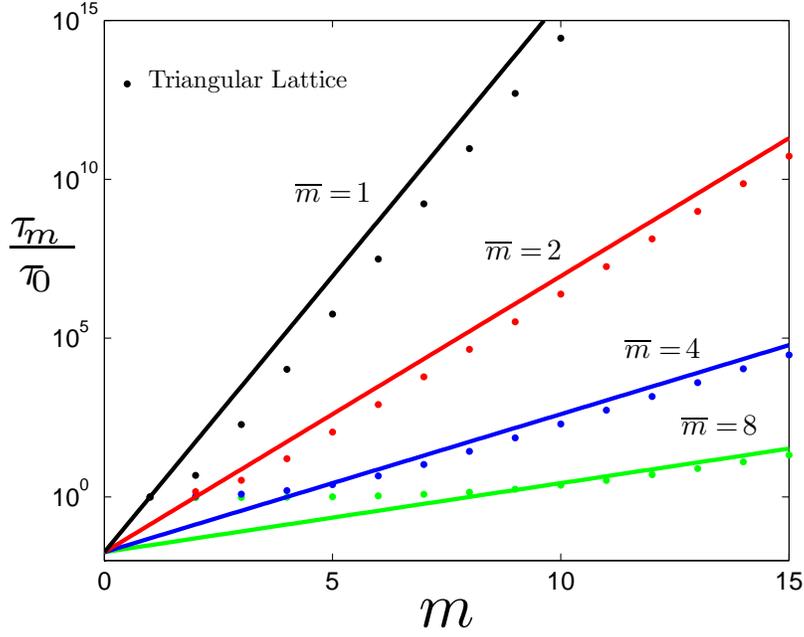}
\caption{Comparison of the ensemble averaged dwell time in a lattice model
(Eq. \protect\ref{hoplattice}) to the effective Arrhenius approximation (Eq. 
\protect\ref{arrhenius}). \ In the plot the product $\Delta \overline{m}=4$
is held constant. }
\label{tauplot6}
\end{figure}

As a result, the departure time distribution must now reflect not only
desorption, but also hopping to adjacent sites. \ The hopping rate between
neighboring sites $i$ and $j$ is given by an Arrhenius law $\kappa
_{i\rightarrow j}=\frac{1}{\tau _{0}}\exp [-\Delta (m_{i}-m_{j})\theta
(m_{i}-m_{j})]$, with $\theta (x)$ the Heaviside step function. \ In a
lattice model with coordination number $z$ the dwell time $\tau _{m}$ at a
site with $m$ bridges is calculated by averaging over the hopping rates to
the nearest neighbors (see Fig. \ref{tauplot6}). \ 

\begin{equation}
\tau_{m}=\left\langle \frac{1}{\frac{1}{z}\sum_{i=1}^{z}\kappa_{m\rightarrow
i}}\right\rangle
_{m_{1}...m_{z}}=z\sum_{m_{1}=1}^{\infty}\cdots\sum_{m_{z}=1}^{\infty}%
\widetilde{P}_{m_{1}}\cdots\widetilde{P}_{m_{z}}\frac{1}{\sum_{i=1}^{z}%
\kappa_{m\rightarrow i}}  \label{hoplattice}
\end{equation}
Fortunately, this ensemble averaging procedure can be accurately
approximated by an effective Arrhenius relation:%
\begin{equation}
\tau_{m}=\tau_{0}\exp[\Delta(m-\overline{m})]  \label{arrhenius}
\end{equation}
The validity of the approximation is most important for sites with $m\geq%
\overline{m}$ bridges, since the diffusive exploration allows the particle
to quickly cascade into these deep energy wells. \ 

The effective Arrhenius relation greatly simplifies the calculation, since
so long as $\Delta \overline{m}$ is sufficiently large, the probability of
the particle still being attached to the surface after an $n$ step random
walk is $\left( 1-K_{m}\tau _{m}\right) ^{n-1}=[1-\exp (-\Delta \overline{m}%
)]^{n-1}$. \ Here $K_{m}=\frac{1}{\tau _{0}}\exp (-\Delta m)$ is the
departure rate from a site with $m$ bridges. \ Interestingly, in this
approximation scheme the attachment probability is independent of the
particular bridge numbers $\{m_{1},...,m_{n}\}$ realized during the walk. \
Thus, the probability of departure $f_{n}$ after exactly $n$ steps is:%
\begin{align}
f_{n}& =[\exp (\gamma )-1]\exp (-\gamma n) \\
\gamma & \equiv -\log [1-\exp (-\Delta \overline{m})]
\end{align}%
The average number of steps for the random walk is $\sum\nolimits_{n=1}^{%
\infty }nf_{n}=\exp (\Delta \overline{m})$. \ To calculate the departure
time distribution $\Phi (t)$ we use $f_{n}$ to average over the departure
time distribution for walks with a given $n$, $\phi _{n}(t)$. \ \ 
\begin{equation}
\Phi (t)=\sum\limits_{n=1}^{\infty }f_{n}\phi _{n}(t)
\end{equation}%
\ 
\begin{equation}
\phi _{n}(t)=\prod\limits_{j=1}^{n}\left( \sum_{m_{j}=1}^{\infty }\widetilde{%
P}_{m_{j}}\int_{0}^{\infty }dt_{j}\left( \frac{-dS_{m_{j}}(t_{j})}{dt_{j}}%
\right) \right) \delta \left( t-\sum_{k=1}^{n}t_{k}\right)
\end{equation}

Here $S_{m}(t)$ is the survival probability at time $t$ for a site with $m$
bridges, used to determine the probability of departure between $t$ and $%
t+dt $. \ If there was only one hopping pathway with rate $\kappa$, we would
have $-\frac{dS}{dt}=\kappa\exp(-\kappa t)$. \ The generalization accounts
for the fact that the particle can hop to any of its $z$ neighbours, and the
probability of departure is not simply exponential. \ 
\begin{equation}
S_{m}(t)=\left( \sum\limits_{a=1}^{\infty}\widetilde{P}_{a}\exp\left[
-t\kappa_{m\rightarrow a}\right] \right) ^{z}
\end{equation}
It is convenient to Fourier transform $\phi_{n}(t)$ so that one can sum the
resulting geometric series for $\Phi(\omega)$. \ 
\begin{align}
\phi_{n}(\omega) & =\int_{-\infty}^{\infty}\phi_{n}(t)\exp[-i\omega t]%
dt=X(\omega)^{n}  \label{Xomega} \\
X(\omega) & \equiv\sum_{m=1}^{\infty}\widetilde{P}_{m}\sum_{m_{1}=1}^{%
\infty}\cdots\sum_{m_{z}=1}^{\infty}\widetilde{P}_{m_{1}}\cdots\widetilde {P}%
_{m_{z}}\left( \frac{\sum\limits_{i=1}^{z}\kappa_{m\rightarrow m_{i}}}{%
\sum\limits_{i=1}^{z}\kappa_{m\rightarrow m_{i}}+i\omega}\right) \\
\Phi(\omega) & =[\exp(\gamma)-1]\sum\limits_{n=1}^{\infty}\left[
\exp(-\gamma)X(\omega)\right] ^{n}=[\exp(\gamma)-1]\frac{X(\omega)}{%
\exp(\gamma)-X(\omega)}
\end{align}

To facilitate a simpler calculation, we employ a coarse-graining procedure
to dispense with the tensor indices $\{m,m_{1},...,m_{z}\}$ in the
definition of $X(\omega)$. \ In the summation there are many terms for which
the value of $\sum\limits_{i=1}^{z}\kappa_{m\rightarrow m_{i}}$ are equal,
but with different weight factors $\widetilde{P}_{m}\widetilde{P}%
_{m_{1}}\cdots\widetilde{P}_{m_{z}}$. \ To eliminate this degeneracy we
introduce a smooth function $f(\kappa)$ normalized according to $\int
f(\kappa)d\kappa=1$. \ 
\begin{equation}
X(\omega)\simeq\int f(\kappa)\frac{\kappa}{\kappa+i\omega}d\kappa
\end{equation}
\ 

The inverse Fourier transform is performed using the residue theorem to
obtain the final result. \ The contour integral is closed in the upper half
plane, with all the poles on the imaginary axis at $\omega =iz$. \ 
\begin{align}
\Phi (t)& =\frac{1}{2\pi }\int_{-\infty }^{\infty }\Phi (\omega )\exp
[i\omega t]d\omega =\frac{[\exp (\gamma )-1]}{2\pi }2\pi i\sum_{r=1}^{\infty
}res_{\omega =\omega _{r}}\left[ \frac{\exp [i\omega t]X(\omega )}{\exp
(\gamma )-X(\omega )}\right]  \label{deptimedistdif} \\
& =[\exp (\gamma )-1]i\sum_{r=1}^{\infty }res_{\omega =\omega _{r}}\left[ 
\frac{\exp [i\omega _{r}t]\left\{ X(\omega _{r})+(\omega -\omega _{r})\left( 
\frac{dX}{d\omega }\right) _{\omega =\omega _{r}}+\cdots \right\} }{\exp
(\gamma )-\left\{ X(\omega _{r})+(\omega -\omega _{r})\left( \frac{dX}{%
d\omega }\right) _{\omega =\omega _{r}}+\cdots \right\} }\right]  \notag \\
& =[\exp (\gamma )-1]i\sum_{r=1}^{\infty }\left[ \frac{-\exp [i\omega
_{r}t]X(\omega _{r})}{\left( \frac{dX}{d\omega }\right) _{\omega =\omega
_{r}}}\right]  \notag \\
& =\exp (\gamma )[\exp (\gamma )-1]\sum_{r=1}^{\infty }\frac{\exp (-z_{r}t)}{%
Y(z_{r})}  \notag \\
Y(z_{r})& \equiv \int f(\kappa )\frac{\kappa }{\left( \kappa -z_{r}\right)
^{2}}d\kappa
\end{align}%
\qquad Here $z_{r}$ labels the roots of the equation%
\begin{equation}
\exp (\gamma )-X(iz)=0
\end{equation}%
The benefit of the coarse-graining is now more transparent, as the residues
are all labeled by a single index $r$ as opposed to the tensor indices $%
\{m,m_{1},...,m_{z}\}$. \ 

In Fig. \ref{experiment2} the departure time distribution is plotted in the
diffusive regime. \ The optimal regime for fast particle departure is to
have a large number ($\overline{m}\sim 10$) of weakly bound key-lock
bridges. \ In this scenario the departure time distribution is accurately
approximated as a single exponential, $\Phi (t)=K_{\overline{m}}\exp (-K_{%
\overline{m}}t)$. \ 

\begin{figure}[h]
\includegraphics[width=4.5221in,height=3.4058in]{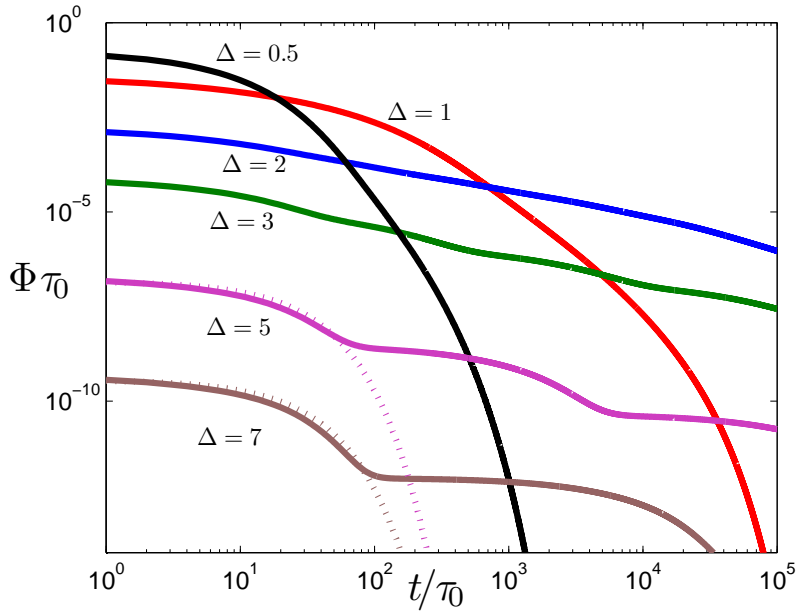}
\caption{Departure time distribution function versus time with $\overline{m}%
=3$. \ The dotted lines for $\Delta =5$ and $7$ show the exponential
approximation $\Phi (t)=K_{\overline{m}}\exp (-\protect\kappa ^{\ast }t)$. \ 
}
\label{universalbehavior}
\end{figure}

\begin{figure}[h]
\includegraphics[width=4.6836in,height=3.5274in]{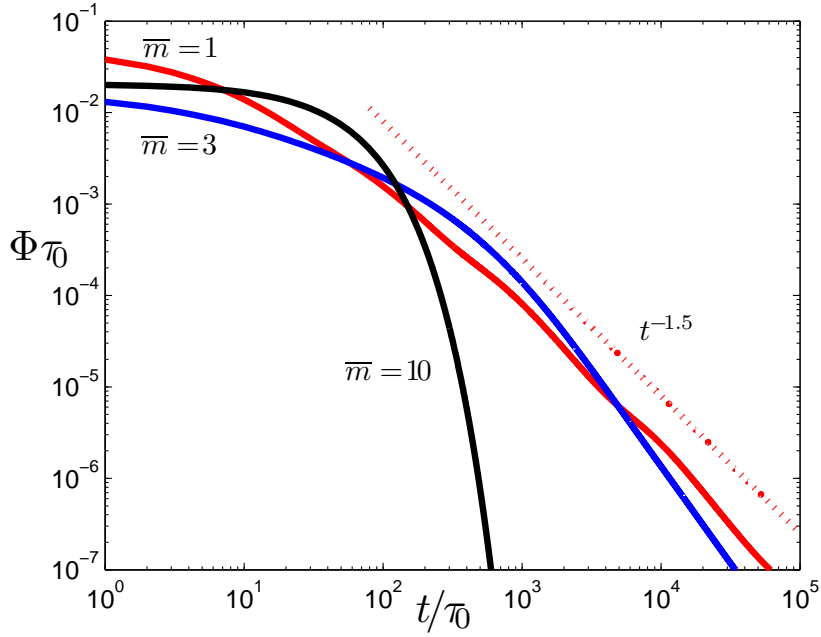}
\caption{Departure time distribution function versus time as determined by
Eq. \protect\ref{deptimedistdif} in the diffusive regime. \ In the plot the
average binding energy is held constant at $4T$. The theoretically
determined departure time distribution can be compared to an experiment with
DNA-grafted colloids which observed power law behavior with exponent $-1.5$.
\ }
\label{experiment2}
\end{figure}

We now discuss the behavior of the departure time distribution in several
regimes of interest. \ At fixed $\overline{m}$, for small $\Delta $ the
behavior is \textit{non-universal}. \ The departure time distribution
exhibits multi-stage behavior, where the initial departure and long time
behavior may both take the shape of a power law, albeit with different
exponents (see $\Delta =0.5$, $1$ curves in Fig. \ref{universalbehavior}). \ 

As the strength of the key-lock binding increases ($\Delta \gtrsim 1$) there
is a crossover from non-universal behavior to \textit{universal power law}
behavior for the first several decades in time (see $\Delta =2$, $3$ curves
in Fig. \ref{universalbehavior}). \ 
\begin{equation}
\Phi (t)\sim t^{-.7}
\end{equation}

For $\Delta \gtrsim 3$ we enter the regime of \textit{multiexponential
beating}. \ The initial departure behavior is well described as an
exponential with initial departure rate $K_{\overline{m}}=\exp (-\Delta 
\overline{m})/\tau _{0}$ and characteristic timescale $1/\kappa ^{\ast
}\simeq 15\tau _{0}$. \ 
\begin{equation}
\Phi (t)\simeq K_{\overline{m}}\exp (-\kappa ^{\ast }t)
\end{equation}%
We attribute $\kappa ^{\ast }$ to the diffusive cascade of particles from
states with $\overline{m}$ bridges into more highly connected states. \
Since this process involves particles finding a lower energy state, $\kappa
^{\ast }$ does not depend on $\Delta $. \ As indicated by the small
departure probability, the binding is nearly irreversible in this regime
(see $\Delta =5$, $7$ curves in Fig. \ref{universalbehavior}). \ \ 

We also plot the departure time distribution relevant to the experimental
situation where the average binding energy is held constant \cite{statmech}%
. \ 
\begin{equation}
\frac{\Delta \overline{m}}{1-\exp (-\overline{m})}+\ln (1-\exp (-\overline{m}%
))=const.\ 
\end{equation}%
The optimal regime for fast departure is to have a large number ($\overline{m%
}\sim 10$) of weakly bound bridges (see Fig. \ref{experiment2}). \ In this
fast departure regime the departure time distribution is well approximated
as a single exponential, $\Phi (t)=K_{\overline{m}}\exp (-K_{\overline{m}}t)$
. \ 

\section{Diffusion}

\qquad We now turn to discuss the statistics for the in-plane diffusion of
the particle. \ We first note that the in-plane trajectory of the particle
subjected to a delta-correlated random potential remains statistically
equivalent to an unbiased random walk. \ As a result, the mean squared
displacement for an $n$ step random walk remains $\left\langle
r^{2}\right\rangle =na^{2}$. \ As the particle explores the landscape it
cascades into deeper energy wells, the hopping time increases, and the
diffusion gets slower. \ In the limit $n\rightarrow \infty $ the average
hopping time can be determined from the equilibrium canonical distribution.
\ For Poisson distributed bridge numbers $m$, this corresponds to a finite
renormalization of the diffusion coefficient $D^{\ast }$ with $%
D_{0}=a^{2}/4\tau _{0}$. \ 

\begin{figure}[tbp]
\includegraphics[width=4.6138in,height=3.4714in]{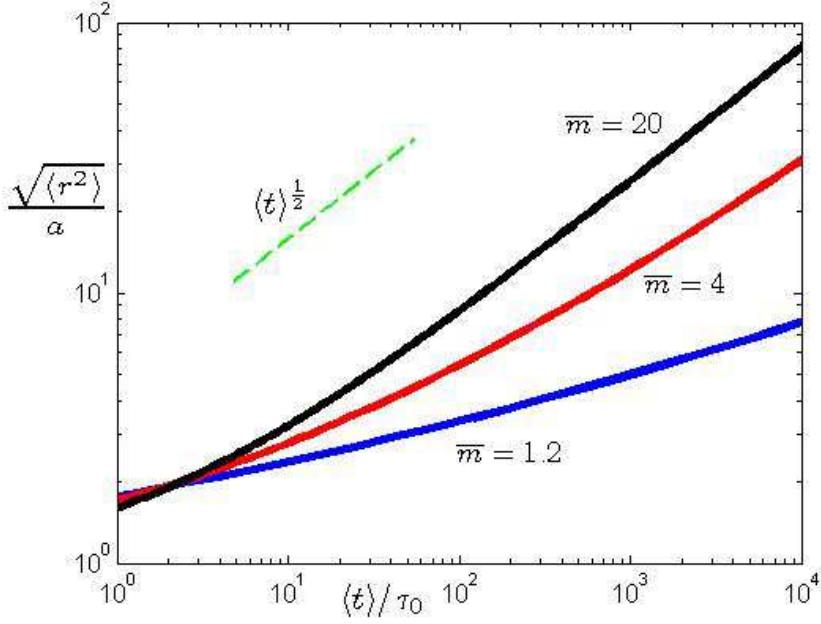}
\caption{Root mean squared displacement vs. time with $\Delta \overline{m}=4$%
. \ The curves are calculated from the parametric equations \protect\ref%
{rsquared}, \protect\ref{tav}. \ }
\label{rvtnew}
\end{figure}

\begin{figure}[tbp]
\includegraphics[width=4.6138in,height=3.4714in]{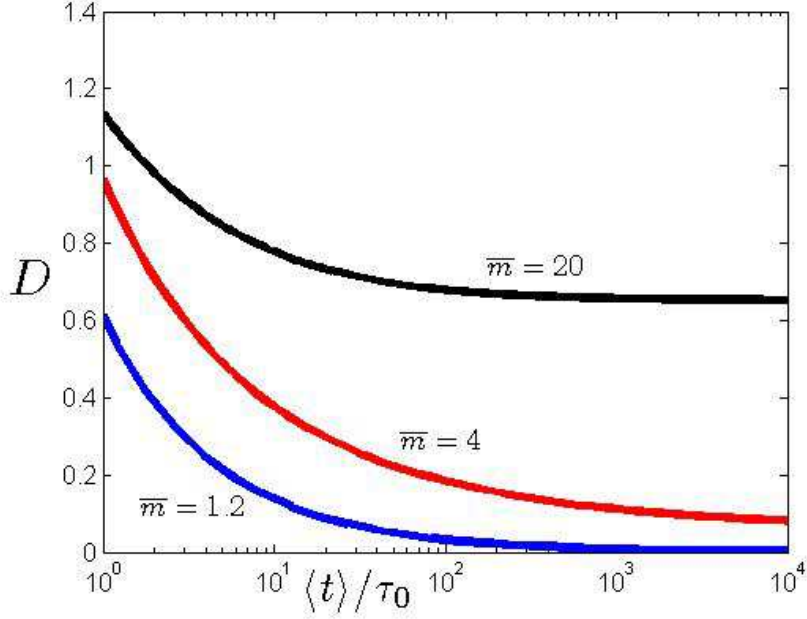}
\caption{The dimensionless diffusion coefficient $D\equiv \frac{1}{4D_{0}}%
\frac{\partial \left\langle r^{2}\right\rangle }{\partial \left\langle
t\right\rangle }$ plotted against time. \ }
\label{Dnew}
\end{figure}

\ 
\begin{equation}
\left\langle t\right\rangle =n\left\langle \tau_{m}\right\rangle =n\tau
_{0}\exp\left( -\Delta\overline{m}\right) \sum_{m=1}^{\infty}\widetilde {P}%
_{m}\exp\left( \Delta m\right) =n\tau_{0}\frac{\exp\left( \overline {m}%
e^{\Delta}\right) -1}{\exp\left( \Delta\overline{m}\right) [\exp(\overline{m}%
)-1]}
\end{equation}%
\begin{equation}
D^{\ast}\equiv\frac{1}{4}\frac{\partial\left\langle r^{2}\right\rangle }{%
\partial\left\langle t\right\rangle }=D_{0}\frac{\exp\left( \Delta \overline{%
m}\right) [\exp(\overline{m})-1]}{\exp\left( \overline{m}e^{\Delta}\right) -1%
}
\end{equation}

This "ergodic" behavior is only achieved after a very long time. \
Generally, an $n$ step random walk cannot visit sites with arbitrarily large 
$m$. \ In this transient regime one should only average over sites with $%
m<m^{\ast }$. \ In the language of the statistics of extreme events, $%
m^{\ast }-1$ is the maximum "expected" value of $m$ in a sample of $n$
independent realizations \cite{dispersive}. \ Even with this complication,
the average diffusion time $\left\langle t\right\rangle $ and the mean
squared displacement $\left\langle r^{2}\right\rangle $ can both be
expressed in terms of $m^{\ast }$, which defines their relationship in
parametric form. \ \ 
\begin{equation}
\left\langle r^{2}\right\rangle =\frac{a^{2}}{P(\overline{m},m^{\ast })}
\label{rsquared}
\end{equation}%
\begin{equation}
\left\langle t\right\rangle =\frac{\left\langle r^{2}\right\rangle }{D^{\ast
}}\left( 1-\frac{P(\overline{m}e^{\Delta },m^{\ast })}{1-\exp (-\overline{m}%
e^{\Delta })}\right)  \label{tav}
\end{equation}

Here $P(x,m^{\ast })\equiv \gamma (x,m^{\ast })/\Gamma (m^{\ast })=\exp
(-x)\sum_{k=m^{\ast }}^{\infty }x^{k}/k!$ is the regularized lower
incomplete $\Gamma $ function. \ In the limit $m^{\ast }\rightarrow \infty $
we recover the renormalized diffusion relation $\left\langle t\right\rangle
=\left\langle r^{2}\right\rangle /D^{\ast }$, although this occurs at very
long, often unrealistic times. \ In the transient regime we expect
anomalous, subdiffusive behavior. \ As indicated in Fig. \ref{rvtnew}, this
subdiffusive behavior is typical for strong enough key-lock interactions. \
Figure \ref{Dnew} is a plot of the dimensionless diffusion coefficient
versus time. \ 

\begin{figure}[h]
\includegraphics[width=4.6836in,height=3.5274in]{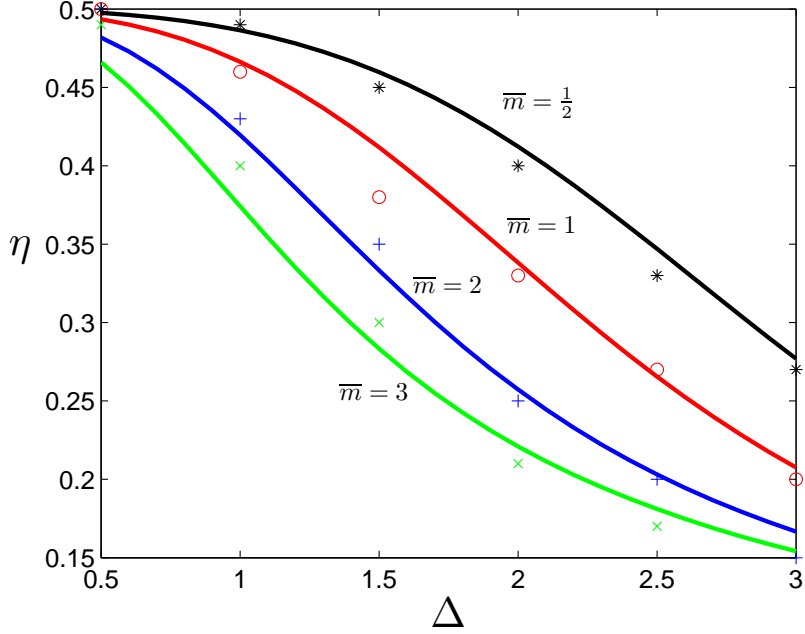}
\caption{Comparison of the power law exponent determined numerically to Eq. 
\protect\ref{diffusionexponent} in the transient regime. \ }
\label{difexppaper}
\end{figure}

By approximating the incomplete gamma functions, the transient behavior may
be well described by a power law with a single free parameter $\beta \simeq
0.15$ (see Fig. \ref{difexppaper}). \ 
\begin{equation}
\frac{\left\langle r^{2}\right\rangle ^{\frac{1}{2}}}{a}\simeq \left( \frac{%
\left\langle t\right\rangle }{\tau _{0}}\right) ^{\eta }
\end{equation}%
\begin{equation}
\eta =\frac{1}{2}-\frac{1+[\Delta -1]\exp (\Delta )}{2\Delta \lbrack \exp
(\Delta )-1]-\frac{2}{\beta \overline{m}}\ln \left[ 1-\exp (-\beta \overline{%
m})\right] }  \label{diffusionexponent}
\end{equation}%

There is an analogy between our results and dispersive transport in
amorphous materials. \ In these systems a length dependence of the effective
mobility \cite{semiconductortransport} can be interpreted within the context
of the statistics of extreme events \cite{dispersive}. \ The time required
for charge carriers to travel through the material depends on the dwell
times spent at all of the trapping centers. \ Since this transit time will
be dominated by the dwell time of the deepest trapping center, one would
like to know how the thickness of the material effects the distribution for
the largest trapping depth. \ The analogy to our result is made by replacing
the material thickness with the number of steps in the random walk $n$, and
replacing the distribution for the largest trap depth with the distribution
for $m^{\ast }$, since the bridge number is related to energy by Eq. \ref%
{potenergy}. \ 

\section{Connection to Experiments}

\qquad We now would like to make a connection between our results and recent
experiments with DNA-grafted colloids. \ The departure time distribution can
be compared to an experiment which determined the time-varying separation of
two DNA-grafted colloids in an optical trap \cite{crocker}. \ In the
experimental setup, two particles are bound by DNA bridges, and after
breaking all connections diffuse to the width of the optical trap. \ Because
the length of the DNA\ chains grafted on the particle is much shorter than
the particle radius, surface curvature effects can be neglected. \ The
interaction resembles that of a particle interacting with a patch on a $2D$
substrate. \ Experimentally the tail of the departure time distribution was
observed to be a power law $\Phi (t)\sim t^{-1.5}$. \ Qualitatively similar
behavior is predicted by the theory with $\overline{m}\sim 1$ and average
binding free energy of several $T$ (see $\overline{m}=1$ curve in Fig. \ref%
{experiment2}). \ 

\begin{figure}[tbp]
\includegraphics[width=4.6112in,height=3.4705in]{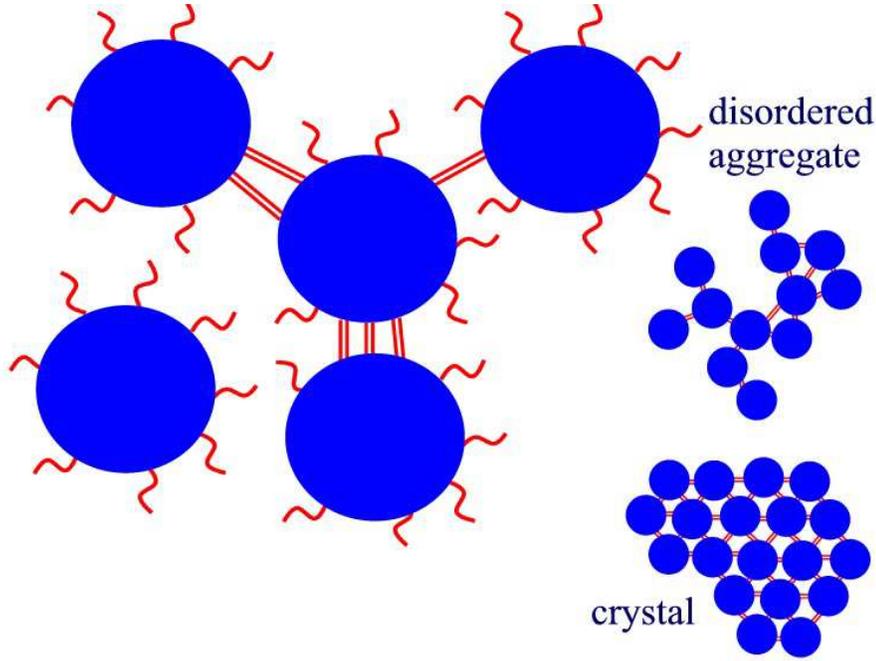}
\caption{Graphical depiction of key-lock binding between nanoparticles
functionalized with complementary ssDNA. \ The resulting structures can be
disordered, fractal-like aggregates, or crystalline. \ }
\label{crystal}
\end{figure}

In addition, our work provides insight into the slow crystallization
dynamics of key-lock binding particles (see Fig. \ref{crystal}). \ In \cite%
{chaikin}, $1\mu m$ diameter particles grafted with ssDNA formed reversible,
disordered aggregates. \ The average number of key-lock bridges between
particles was $\overline{m}\sim 2$. \ The authors of \cite{crocker} observed
random hexagonal close packed crystals by further reducing the surface
density of DNA strands on the particles. \ The crystallization process
requires that particles rapidly detach and reattach at the desired lattice
location. \ In the localized regime particle desorption is the relevant
process. \ 

\begin{figure}[tbp]
\includegraphics[width=4.6138in,height=3.4714in]{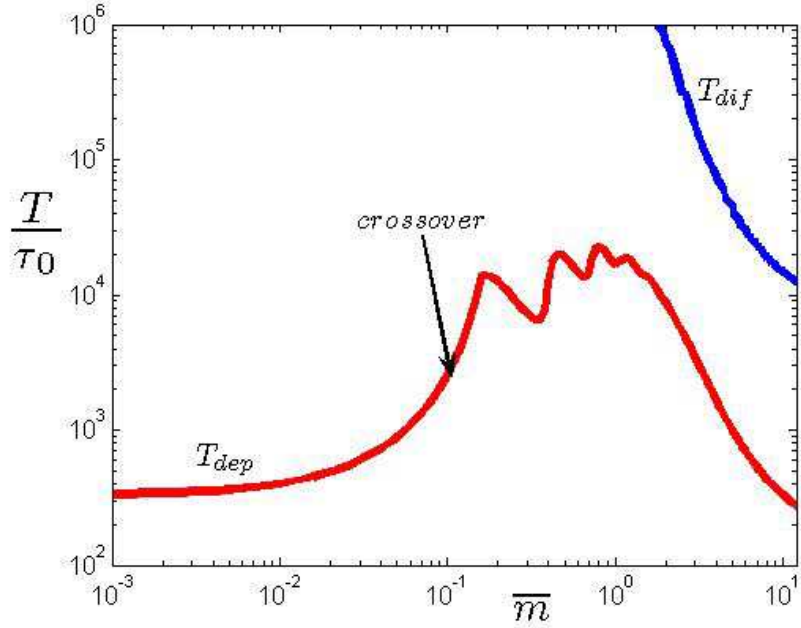}
\caption{Plot of the characteristic times $T_{dep}$ and $T_{dif}$ versus $%
\overline{m}$ at constant binding energy ($const=4$ in Eq. \protect\ref%
{constbinding}). \ }
\label{timecompare}
\end{figure}

In the diffusive regime surface diffusion also plays a role in the
rearrangement of particles into the desired crystalline structure. \ To
determine which process is more important for particle rearrangement, we can
compare the departure time with the time required for a particle to
diffusively explore the surface of a particle to which it is bound. \ The
time $T_{dep}$ required for $90\%$ of the particles to depart is: \ \ 
\begin{equation}
0.1=\int_{T_{dep}}^{\infty }\Phi (t)dt  \label{Tdep}
\end{equation}

To estimate the time required for diffusive rearrangement $T_{dif}$ we use
the parametric equations \ref{rsquared}, \ref{tav}. \ In \cite{chaikin}
particles of radius $R=.5\mu m$ were grafted with DNA chains of length $%
l\sim 20nm$. \ Assuming the correlation length $a\sim l$ we have $\frac{%
\left\langle r^{2}\right\rangle }{a^{2}}\sim \left( \frac{\pi R}{l}\right)
^{2}\simeq 10^{3}$. \ Figure \ref{timecompare} shows a comparison of $%
T_{dif} $ and $T_{dep}$ at constant binding free energy. \ 
\begin{equation}
\frac{\Delta \overline{m}}{1-\exp (-\overline{m})}+\log (1-\exp (-\overline{m%
}))=const  \label{constbinding}
\end{equation}

This expression for the binding energy takes into account the entropy
reduction associated with the non-ergodic degrees of freedom. \ For a
detailed discussion of this topic see reference \cite{statmech}. \ Since $%
T_{dif}>T_{dep}$, colloidal desorption and reattachment is the dominant
mechanism by which particles rearrange. \ 

\ As the figure indicates, the optimal regime of fast departure is to have a
large number ($\overline{m}\gtrsim 10$) of weakly bound key-lock bridges. \
We predict a localized regime below the crossover where particle departure
is relatively fast. \ Just beyond the crossover there is a relative maximum
in $T_{dep}$ before it decreases at large $\overline{m}$. \ The increase in
departure time at the onset of diffusive behavior is indicative of a regime
where the system ages. \ In this regime the interplay of diffusion an
desorption leads to longer bound state lifetimes, and an increase in the
departure time. \ 

\begin{figure}[tbp]
\includegraphics[width=4.6112in,height=3.4705in]{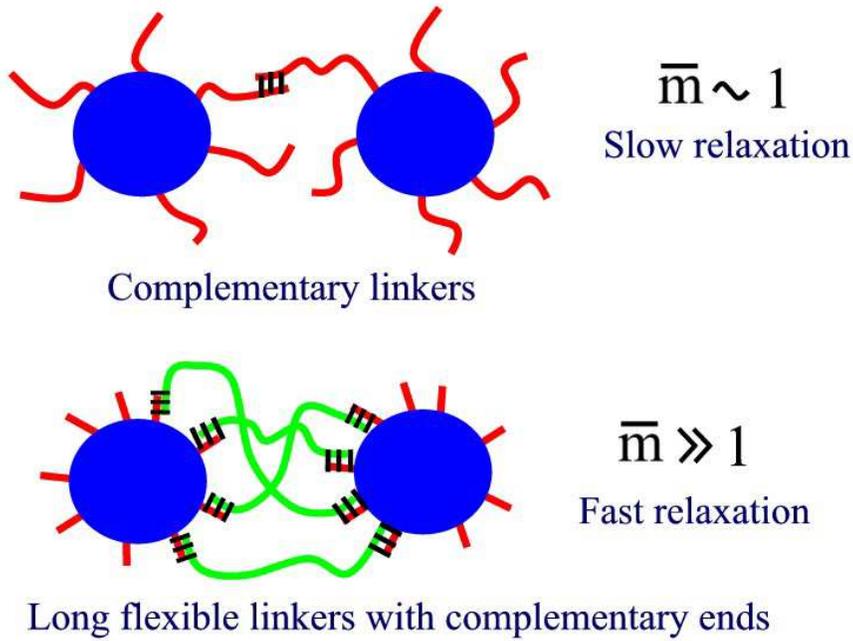}
\caption{Graphical depiction comparing recent experiments with DNA-grafted
colloids to a future implementation with long flexible linkers. \ Increasing
the number of key-lock bridges between particle pairs potentially decreases
the time required for crystallization. \ }
\label{longlinkers}
\end{figure}

We now turn to the question of designing a future experiment which will
facilitate fast particle departure and hence colloidal crystallization. \
The essential goal is to increase the average number of key-lock bridges
between particle pairs. \ Increasing the surface density of DNA strands
alone results in the formation of a brush, which decreases the effective
cross section for the interaction between complementary DNA. \ Instead we
propose the introduction of long, flexible DNA\ linkers between particles
with a high coverage of short ssDNA (see Fig. \ref{longlinkers}). \ This
system has the potential to realize more key-lock bridges between particle
pairs as compared to previous experiments, and therefore substantially
reduce the time required for crystallization. \ 

\section{Summary}

\qquad In this chapter we studied the dynamics of particles which interact
through the reversible formation of multiple key-lock bridges. \ Well before
the percolation threshold is reached there is a crossover from a localized
regime to a diffusive regime. \ In the localized regime the particles remain
close to their original attachment site until departing. \ In this regime
particles are attached to finite clusters, and the system exhibits an
exponential distribution of departure times. \ Once the radius of gyration
of the cluster exceeds the characteristic radius for the particles' random
walk, the finite clusters behave effectively as infinite clusters. \
Diffusion allows the particles to cascade into deeper energy wells, which
leads to a decrease in hopping rate. \ The diffusion slows and the bound
state lifetime increases, a phenomenon qualitatively similar to \textit{aging%
} in glassy systems. \ In the diffusive regime we discussed the statistics
for the particles' in-plane diffusion. \ Weak key-lock interactions give
rise to a finite renormalization of the diffusion coefficient. \ However, as
the strength of the interaction increases (larger $\Delta $), the system
exhibits anomalous, subdiffusive behavior. \ This situation is analogous to
dispersive transport in disordered semiconductors. \ We then made the
connection between our calculation of the departure time distribution and
recent experiments with DNA-coated colloids. \ The findings indicate that
the optimal regime for colloidal crystallization is to have a large number
of weakly bound key-lock bridges. \ A modified experimental setup was
proposed which has the potential to realize this regime of fast particle
departure. \ 

\chapter{Self-Assembly of DNA-Coded Clusters}

\section{Motivation and Problem Description}

\qquad Over the past decade, a number of proposals have identified potential
applications of DNA for self-assembly of micro- and nanostructures \cite%
{nucleic}, \cite{falls}, \cite{template}, \cite{angstrom}, \cite{nanotube}.
\ Among these proposals, one common theme is finding a way to utilize the
high degree of selectivity present in DNA-mediated interactions. \ An
exciting and potentially promising application of these ideas is to use
DNA-mediated interactions to programmable self-assemble nanoparticle
structures \cite{rational}, \cite{natreview}, \cite{designcrystals}, \cite%
{nanocrystals}. \ Generically, these schemes utilize colloidal particles
functionalized with specially designed ssDNA (markers), whose sequence
defines the particle type. \ Selective, type-dependent interactions can then
be introduced either by making the markers complementary to each other, or
by using linker-DNA chains whose ends are complementary to particular maker
sequences. \ Independent of these studies, there are numerous proposals to
make sophisticated nano-blocks which can be used for hierarchical
self-assembly. One recent advance in the self-assembly of anisotropic
clusters is the work of Manoharan et. al \cite{packing}. \ They devised a
scheme to produce stable clusters of $n$ polystyrene microspheres. \ The
clusters were assembled in a colloidal system consisting of evaporating oil
droplets suspended in water, with the microspheres attached to the droplet
interface. \ The resulting clusters, unique for each $n$, are optimal in the
sense that they minimize the second moment of the mass distribution $%
M_{2}=\sum_{i=1}^{n}(\mathbf{r}_{i}-\mathbf{r}_{cm})^{2}$. \ 

In this chapter \cite{nanoclusters}, we present a theoretical discussion of
a method which essentially merges the two approaches. We propose to utilize
DNA to self-assemble colloidal clusters, somewhat similar to those in Ref. 
\cite{packing}. \ An important new aspect of the scheme is that the clusters
are "decorated": each particle in the resulting cluster is distinguished by
a unique DNA marker sequence. As a result, the clusters have additional
degrees of freedom associated with particle permutation, and potentially may
have more selective and sophisticated inter-cluster interactions essential
for hierarchic self-assembly. \ In addition,\ the formation of such clusters
would be an important step towards programmable self-assembly of micro- and
nanostructures of an arbitrary shape, as suggested\ in Ref. \cite{licata}. \
\ 

\begin{figure}[h]
\includegraphics[width=4.77in,height=3.5831in]{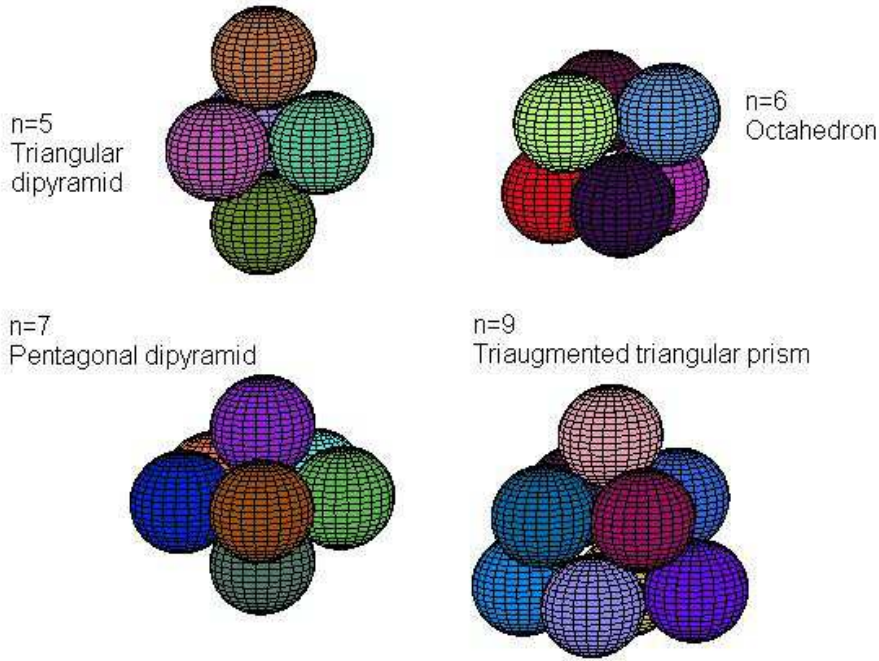}
\caption{The minimal second moment clusters for $n=5,6,7,$ and $9$. \
Pictures of all the clusters from $n=4$ to $15$ are available in 
\protect\cite{packing}.}
\label{packingspic}
\end{figure}

\begin{table}[h]
\caption{Minimal Second Moment Clusters}
\label{table:polyhedra}\centering        
\begin{tabular}{ccc}
\hline\hline
$n$ & Polyhedra Name & Sch\"{o}nflies Point Group \\[0.5ex] \hline
$4$ & Tetrahedron & $T_{d}$ \\ 
$5$ & Triangular dipyramid & $D_{3h}$ \\ 
$6$ & Octahedron & $O_{h}$ \\ 
$7$ & Pentagonal dipyramid & $D_{5h} $ \\ 
$8$ & Snub disphenoid & $D_{2d}$ \\ 
$9$ & Triaugmented triangular prism & $D_{3h}$ \\ 
$10$ & Gyroelongated square dipyramid & $D_{4d}$ \\ 
$11$ & (non-convex) & $C_{S}$ \\[1ex] \hline
\end{tabular}%
\end{table}

We begin with octopus-like particles functionalized with dsDNA, with each
strand terminated by a short ssDNA marker sequence. \ We assume that each
particle $i$ has a unique code, i.e. the maker sequence $s_{i}$ of ssDNA
attached to it. \ We then introduce anchor DNA to the system, ssDNA with
sequence $\overline{s}_{A}\overline{s}_{B}...\overline{s}_{n}$, with $%
\overline{s}_{i}$ the sequence complementary to the marker sequence $s_{i}$.
\ The anchor is designed to hybridize with one particle of each type. \
Consider a cluster of $n$ particles attached to a single anchor. \ 

\begin{figure}[h]
\includegraphics[width=4.77in,height=3.5831in]{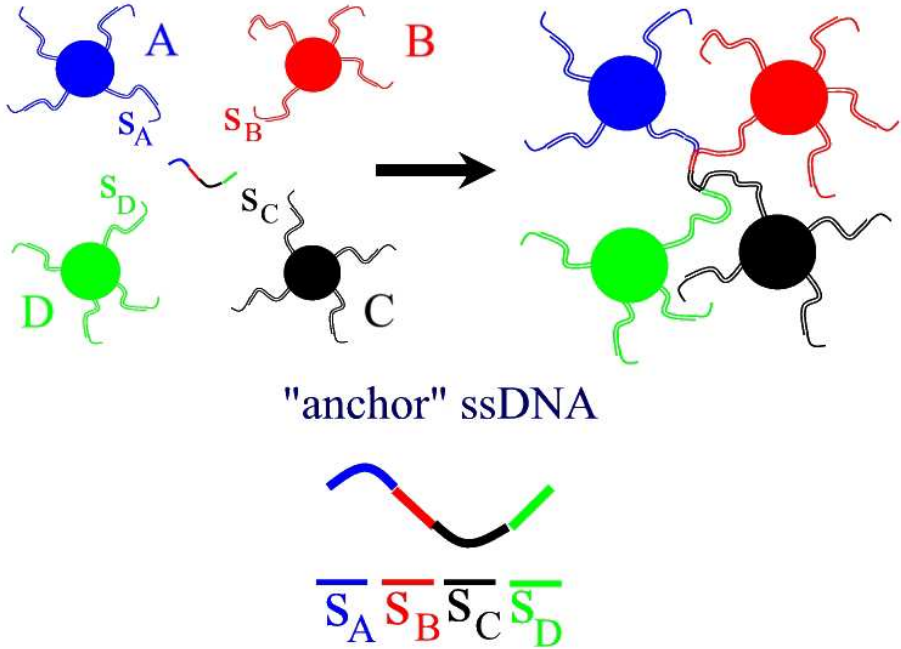}
\caption{Schematic representation of the method for constructing decorated
colloidal clusters using ssDNA "anchors". \ }
\label{anchornew}
\end{figure}

If we treat the DNA\ which link the particles to the anchor as Gaussian
chains, there is an entropic contribution to the cluster free energy which
can be expressed in terms of the particle configuration $\{\mathbf{r}%
_{1},...,\mathbf{r}_{n}\}$ as follows. \ Here $R_{g}$ is the radius of
gyration of the octopus-like DNA arms. \ \ \ $\ $%
\begin{equation}
F=\frac{3T}{2R_{g}^{2}}\sum_{i=1}^{n}(\mathbf{r}_{i}-\mathbf{r}_{anchor})^{2}
\end{equation}%
This approximation of the DNA arms as Gaussian chains is acceptable provided
their length $L$ exceeds the persistence length $l_{p}\simeq 50$ $nm$ and
the probability of self-crossing is small \cite{morphology}. \ The physical
mechanism which determines the final particle configuration in our system is
quite different from the capillary forces of Manoharan et al. \ However,
because the functional form of the free energy is equivalent to the second
moment of the mass distribution, the ground state of the cluster should
correspond to the same optimal configuration. \ 

\section{Equilibrium Treatment}

\qquad Consider a system with $n$ particle species and an anchor of type $%
\overline{s}_{A}\overline{s}_{B}...\overline{s}_{n}$. \ The clusters we
would like to build contain $n$ distinct particles (each particle in the
cluster carries a different DNA marker sequence) attached to a single
anchor. \ Let $C_{n}$ denote the molar concentration of the desired one
anchor cluster. \ Because there are many DNA attached to each particle,
multiple anchor structures can also form. \ The question is whether the
experiment can be performed in a regime where the desired one anchor
structure dominates, avoiding gelation. \ 

We consider the stability of type $C_{n}$ with respect to alternative two
anchor structures. \ To do so we determine the concentration $C_{n+1}$ of $%
n+1$ particle structures which are maximally connected, but do not have a $%
1:1:\cdots :1$ composition. \ In particular, these structures contain more
than one particle of each type, which could cause problems in our
self-assembly scheme \cite{licata}. \ There are also $n$ particle structures 
$\widetilde{C}_{n}$ with the correct composition, but which contain two
anchors. \ We would like to avoid the formation of these structures as well,
as their presence decreases the overall yield of type $C_{n}$. \ Figure \ref%
{clusternew} enumerates the various structures for an $n=3$ species system.
\ If the experiment can be performed as hoped, we will find a regime where
the ratios $\frac{C_{n+1}}{C_{n}}$ and $\frac{\widetilde{C}_{n}}{C_{n}}$\
are small. \ To this end, the equilibrium concentrations $C_{n}$, $C_{n+1}$,
and $\widetilde{C}_{n}$ are determined by equilibrating the chemical
potential of the clusters with their constituents. \ \ 

\begin{figure}[h]
\includegraphics[width=4.7607in,height=3.2967in]{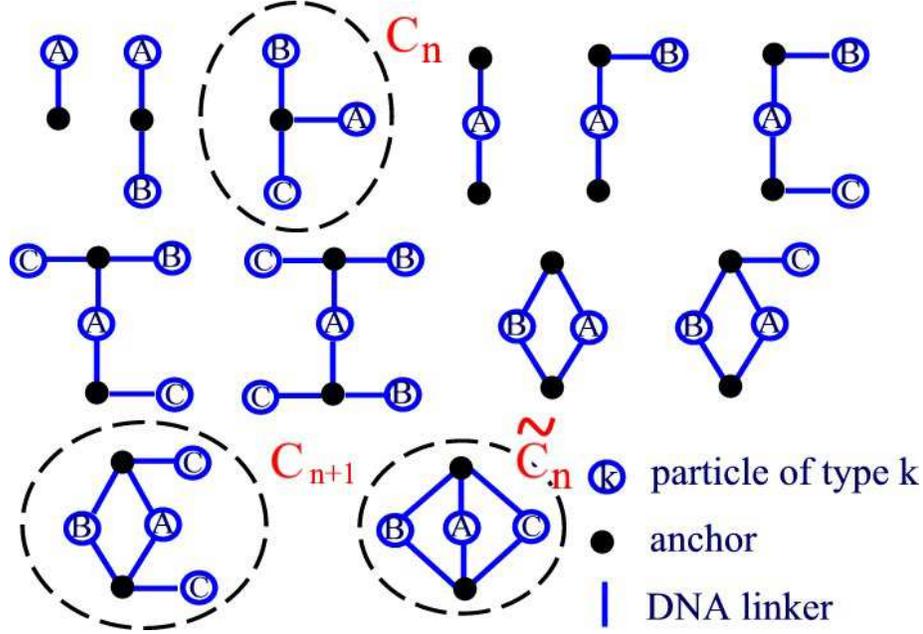}
\caption{The topologically distinct one and two anchor structures for an
anchor ssDNA with sequence $\overline{s}_{A}\overline{s}_{B}\overline{s}_{C}$%
. \ Different structure varieites may be obtained by relabeling the particle
indices subject to the constraint that no more than one particle of each
type is attached to a given anchor. \ }
\label{clusternew}
\end{figure}

First we determine the molar concentration $C_{n}$ of the desired one anchor
structure, composed of one anchor and $n$ particles. \ Let $c_{i}$ denote
the molar concentration of species $i$, and $c_{a}$ the molar anchor
concentration. \ Here $c_{o}=1M$ is a reference concentration. \ \ 
\begin{gather}
T\log \left( \frac{c_{a}\tprod\limits_{i=1}^{n}c_{i}}{c_{o}^{n+1}}\right)
=T\log \left( \frac{C_{n}}{c_{o}}\right) +\epsilon _{n} \\
\epsilon _{n}=\tsum\limits_{i=1}^{n}\Delta G_{i}-T\log \left(
N^{n}v^{n}c_{o}^{n}\right) \\
C_{n}=N^{n}v^{n}c_{a}\tprod\limits_{i=1}^{n}c_{i}\exp \left[ -\frac{\Delta
G_{i}}{T}\right]
\end{gather}

Here $\epsilon _{n}$ is the binding free energy of the cluster, which has a
contribution from the hybridization free energy $\Delta G_{i}$ associated
with attaching particle $i$ to the anchor, and an entropic contribution from
the number of ways to construct the cluster(since each particle has $N$
hybridizable DNA arms). \ In addition we must take into account the entropy
for the internal degrees of freedom in the structure stemming from the
flexibility of the DNA attachments to the anchor. \ In the Gaussian
approximation, neglecting the excluded volume between particles this
localization volume $v\sim R_{g}^{3}$ can be calculated exactly. \ 
\begin{eqnarray}
F(\mathbf{r}_{1},...,\mathbf{r}_{n}) &=&\frac{3T}{2R_{g}^{2}}%
\sum_{i=1}^{n}r_{i}^{2} \\
v^{n} &=&\int d^{3}r_{1}...d^{3}r_{n}\exp \left[ -\frac{F(\mathbf{r}_{1},...,%
\mathbf{r}_{n})}{T}\right] \\
v &=&R_{g}^{3}\left( \frac{2\pi }{3}\right) ^{\frac{3}{2}}
\end{eqnarray}

We now consider the competing two anchor structure $C_{n+1}$. \ To determine
the equilibrium concentration it is instructive to consider the reaction in
which two clusters of type $C_{n}$\ combine to form a single cluster of type 
$C_{n+1}$ and release $n-1$ particles into solution. \ Since there are many
DNA attached to each particle, in what follows we omit factors of $\frac{N-1%
}{N}$. \ 
\begin{gather}
2T\log \left( \frac{C_{n}}{c_{o}}\right) +2\epsilon _{n}=T\log \left( \frac{%
C_{n+1}}{c_{o}}\right) +\epsilon _{n+1}+\tsum\limits_{i=1}^{n-1}T\log \left( 
\frac{c_{i}}{c_{o}}\right) \\
\epsilon _{n+1}=2\tsum\limits_{i=1}^{n}\Delta G_{i}-T\log \left(
N^{n+1}(N-1)^{n-1}v_{2}^{n+2}c_{o}^{n+2}\right) \\
C_{n+1}=\frac{v_{2}^{n+2}}{v^{2n}}\frac{C_{n}^{2}}{\tprod%
\limits_{i=1}^{n-1}c_{i}}
\end{gather}%
The localization volume $v_{2}$ can be calculated in a similar fashion,
fixing one particle at the origin and integrating over the $n$ remaining
particle positions $\{\mathbf{r}_{1},...,\mathbf{r}_{n}\}$ and the position
of the two anchors $\mathbf{r}_{a1}$ and $\mathbf{r}_{a2}$. \ We make use of
the following formula for multivariate Gaussian integrals. \ 
\begin{equation}
\didotsint_{-\infty }^{\infty }\exp \left[ -\frac{q^{T}Mq}{2}\right]
dq_{1}...dq_{3(n+2)}=\frac{\left( 2\pi \right) ^{\frac{3(n+2)}{2}}}{\sqrt{%
\det (M)}}
\end{equation}%
$\ $ \ \ 
\begin{gather}
F(\mathbf{r}_{1},...,\mathbf{r}_{n,}\mathbf{r}_{a1},\mathbf{r}_{a2})=\frac{3T%
}{2R_{g}^{2}}[r_{a1}^{2}+(\mathbf{r}_{n}-\mathbf{r}_{a2})^{2}+\tsum%
\limits_{i=1}^{n-1}\left\{ (\mathbf{r}_{i}-\mathbf{r}_{a1})^{2}+(\mathbf{r}%
_{a2}-\mathbf{r}_{i})^{2}\right\} ] \\
v_{2}^{n+2}=\int d^{3}r_{1}...d^{3}r_{n}d^{3}r_{a1}d^{3}r_{a2}\exp \left[ -%
\frac{F}{T}\right] \\
v_{2}=R_{g}^{3}\left( \frac{2\pi }{3}\right) ^{\frac{3}{2}}n^{\frac{-3}{n+2}%
}2^{\frac{-3(n-1)}{2(n+2)}}=n^{\frac{-3}{n+2}}2^{\frac{-3(n-1)}{2(n+2)}}v
\end{gather}%
Similarly one can obtain the cluster concentration for the two anchor
structure $\widetilde{C}_{n}$. \ The localization volume $v_{3}$ can be
calculated in a similar fashion to $v_{2}$. \ 
\begin{equation}
v_{3}=R_{g}^{3}\left( \frac{2\pi }{3}\right) ^{\frac{3}{2}}n^{\frac{-3}{%
2(n+1)}}2^{\frac{-3n}{2(n+1)}}=n^{\frac{-3}{2(n+1)}}2^{\frac{-3n}{2(n+1)}}v
\end{equation}

We consider the symmetrical case $\Delta G_{i}=\Delta G$ and equal initial
particle concentrations $c_{i}^{(o)}=c^{(o)}$ for all species $i$. \ In this
case we have $c_{A}=c_{B}=...=c_{n}\equiv c$. \ We can express the
concentration of the competing two anchor structures $C_{n+1}$ and $%
\widetilde{C}_{n}$ in terms of the concentration of the desired one anchor
cluster $C_{n}$. \ Since there are many DNA attached to each particle, we
omit factors of $\frac{N-1}{N}$. \ 
\begin{equation}
C_{n}=c_{a}\left( Nvc\ \exp \left[ -\frac{\Delta G}{k_{B}T}\right] \right)
^{n}
\end{equation}%
\begin{equation}
C_{n+1}\simeq \frac{v_{2}^{n+2}}{v^{2n}}\frac{C_{n}^{2}}{c^{n-1}}
\end{equation}%
\ 
\begin{equation}
\widetilde{C}_{n}\simeq \frac{v_{3}^{n+1}}{v^{2n}}\frac{C_{n}^{2}}{c^{n}}
\end{equation}%
The concentration of free anchors $c_{a}$ can be determined from the
equation for anchor conservation. \ \ 
\begin{equation}
c_{a}^{(o)}=c_{a}+C_{n}+2\widetilde{C}_{n}+2nC_{n+1}
\label{anchorconservation}
\end{equation}

\begin{figure}[h]
\includegraphics[width=4.7527in,height=3.5765in]{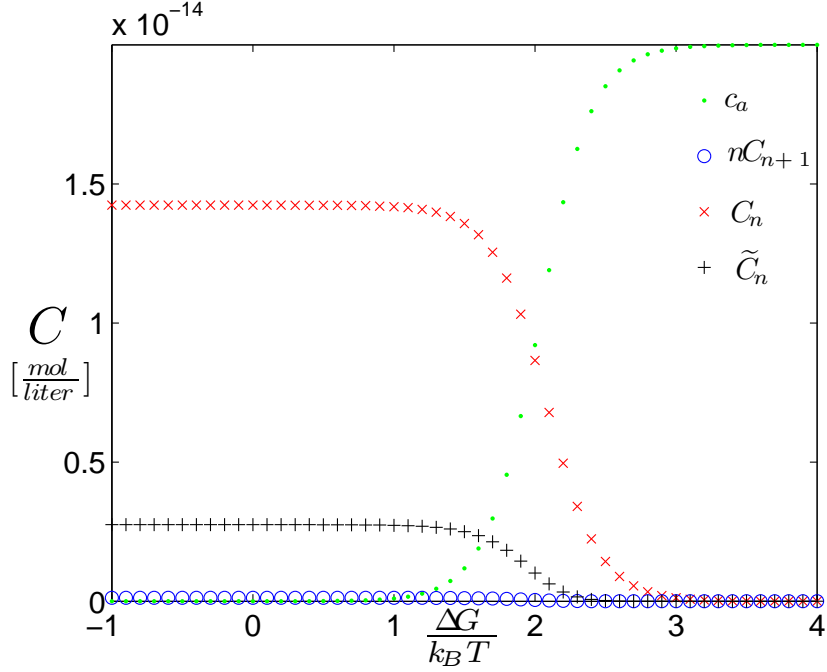}
\caption{The molar concentrations $c_{a}$, $C_{n}$, $nC_{n+1}$, and $%
\widetilde{C}_{n}$ in the symmetrical case for a system with $n=5$ particle
species. \ \ The total particle volume fraction $n\protect\phi \approx .25$
and $\frac{c_{a}^{(o)}}{c^{(o)}}=10^{-3}$. \ }
\label{science3new}
\end{figure}

\ \ We are interested in the low temperature regime where there are no free
anchors in solution. \ We determine the saturation values for the ratios of
interest by noting that the Boltzmann factor $\delta \equiv \exp \left[ -%
\frac{\Delta G}{T}\right] \gg 1$ in this regime. \ 
\begin{equation}
\frac{C_{n+1}}{C_{n}}\simeq \frac{n^{-3}2^{-3n+\frac{9}{2}}}{%
(R_{g}^{3}c^{(o)})^{n-2}}\frac{c_{a}^{(o)}}{c^{(o)}}+O\left( \frac{1}{\delta
^{n}}\right)
\end{equation}

\begin{equation}
\frac{\widetilde{C}_{n}}{C_{n}}\simeq \frac{n^{-3/2}2^{-3n+\frac{3}{2}}}{%
(R_{g}^{3}c^{(o)})^{n-1}}\frac{c_{a}^{(o)}}{c^{(o)}}+O\left( \frac{1}{\delta
^{n}}\right)  \label{cntilde}
\end{equation}

Since $\frac{C_{n+1}}{\widetilde{C}_{n}}\ll 1$, eq. \ref{cntilde} provides
the experimental constraint for suppressing the two anchor structures. \
Taking the radius of the hard spheres $R\sim R_{g}$, it can be interpreted
as a criterion for choosing the initial anchor concentration $c_{a}^{(o)}$
for an $n$ species system with $\phi =\frac{4\pi }{3}R_{g}^{3}c^{(o)}$ the
particle volume fraction for an individual species. \ 
\begin{equation}
\frac{c_{a}^{(o)}}{c^{(o)}}\lesssim n^{\frac{3}{2}}(2\phi )^{n-1}
\label{criterion}
\end{equation}%
The condition gives the maximum anchor concentration for the two anchor
structures to be suppressed. \ Since $\phi \leq \frac{1}{n}$ the theoretical
limits are $\frac{c_{a}^{(o)}}{c^{(o)}}\lesssim 1$, $.29$, $.06$, and $.01$
for $n=4$, $5$, $6$, and $7$ respectively. \ In figure \ref{science3new} we
plot the solution for the concentrations. \ There is a large temperature
regime($\frac{\Delta G}{k_{B}T}\lesssim 2$) where the two anchor structures
are suppressed in favor of the desired one anchor structures. \ 

\section{Irreversible Binding}

\qquad In the previous section we performed an equilibrium calculation to
determine the yield of the cluster $C_{n}$ \cite{nanoclusters}. \ The
results of that study indicated that the concentration of anchors must be
kept very small to prevent the aggregation of larger clusters (see Eq. \ref%
{criterion}). \ From an experimental perspective this result is somewhat
disappointing, since the overall yield of the cluster $C_{n}$ is
proportional to the anchor concentration. \ The situation is considerably
improved in the regime of irreversible binding of particles to anchors. \ In
what follows we present a calculation for the yield of the desired one
anchor cluster far from equilibrium. \ To distinguish between the results of
the previous section and the regime of irreversible binding we will change
the notation slightly. \ This change reflects the fact that the role of the
DNA anchor could also be played by a patchy colloidal particle as discussed
below. \ Henceforth we will refer to the DNA anchors as DNA\ scaffolds. \ In
the new notation the desired one anchor cluster $C_{n}$ is called the star
cluster. \ 

The plan for this section is the following. \ The goal is to maximize the
yield for the star cluster. \ We analytically calculate the yield of the
star cluster in the regime of irreversible binding. \ The analytical results
are compared to the numerical results for the full aggregation equations. \
From an experimental perspective, the most important result is the
determination of an optimal concentration ratio for experiments (see Eq. \ref%
{popt}). \ To conclude we discuss the experimental feasibility of the
self-assembly proposal. \ 

\begin{figure}[h]
\includegraphics[width=4.711in,height=3.5475in]{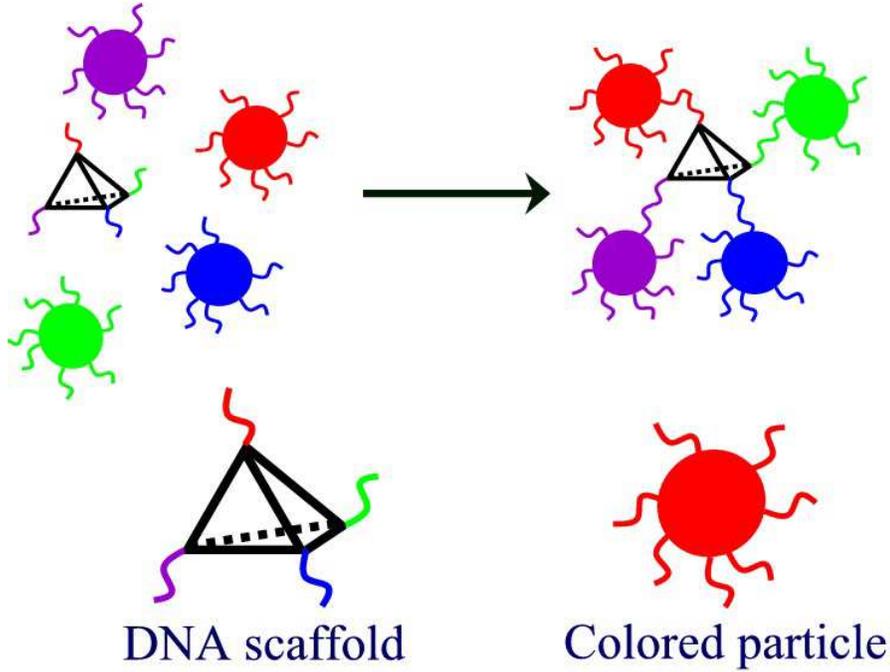}
\caption{A graphical depiction of the scheme for self-assembling star
clusters using DNA scaffolds. \ In the diagram (not drawn to scale) the
scaffold funcionality $n=4$. \ }
\label{dnascaffold}
\end{figure}

The basic idea behind the procedure is as follows (see Fig. \ref{dnascaffold}%
). \ Particles are functionalized with single-stranded DNA (ssDNA) markers
which determine the particle color. \ There may be many DNA attached to each
particle, but on any given particle the marker sequence is identical. \ One
then introduces DNA\ scaffolds to the system. \ The scaffold is a structure
with $n$ ssDNA\ markers, each marker complementary to one of the particle
colors. \ Hybridization of the ssDNA markers on the particles to those on
the scaffold results in the formation of colored particle clusters. \
Because there are many DNA\ attached to each particle, clusters can form
which contain more than one scaffold. \ The essential goal of the procedure
is to maximize the concentration of a particular type of cluster which we
denote the star cluster. \ The star cluster contains one and only one
scaffold to which $n$ particles are attached, each particle having a
distinct color. \ 

We should note that the role of the scaffold could also be played by a
patchy particle (\cite{patchy},\cite{patchy2},\cite{patchy3}). \ For
example, these patches are regions on the particle surface where one can
graft ssDNA markers. \ In this case there may be several DNA connections
between a patch and colored particle. \ Our conclusions will still be valid,
provided the patch size is chosen so that a patch interacts with at most one
particle. \ 

To understand the basic physics behind the aggregation process we consider
the mobility mismatch between the particles and the scaffolds. \ In
solution, a particle with radius $R\simeq 1\mu m$ has a diffusion
coefficient given by the Stokes-Einstein relation $D=k_{B}T/6\pi \eta R$. \
On the other hand, the size of the scaffold $a\simeq 10nm$. \ As a result
the scaffolds diffuse $R/a\simeq 100$ times faster than the particles. \ To
first approximation the resulting aggregation is a two stage process. \ In
the first stage the particles recruit different numbers of scaffolds via the
fast scaffold diffusion and subsequent DNA\ hybridization. \ Since we
consider the regime of strong binding where these bonds are irreversible,
the result is a Poisson distribution over the concentration of particles
with $m$ scaffolds attached. \ Let $C_{i}$ denote the concentration of
particles with color $i$, and $c$ denote the total concentration of
scaffolds. \ The total particle concentration $C_{tot}=\sum_{i=1}^{n}C_{i}$.
\ The concentration $C_{i}^{(m)}$ of particles of color $i$ with $m$
scaffolds attached is%
\begin{eqnarray}
C_{i}^{(m)} &=&C_{i}\frac{p^{m}\exp (-p)}{m!}  \label{pdef} \\
p &=&\frac{c}{C_{tot}}
\end{eqnarray}

In the second stage there are no free scaffolds left in solution, and these
particles decorated with scaffolds aggregate to form the final clusters. \
The seed to build a star cluster is a particle of any color with exactly one
scaffold attached. \ This seed must aggregate with $n-1$ particles of
different colors, each of which has no scaffolds. \ We now calculate the
concentration of the star cluster $C_{\ast }$. \ The yield of the desired
star cluster is quantified in terms of the star mass fraction $M_{\ast
}=(nC_{\ast })/C_{tot}$. \ 
\begin{equation}
M_{\ast }=\frac{n}{C_{tot}}\sum\limits_{i=1}^{n}C_{i}^{(1)}\prod\limits 
_{\substack{ j=1  \\ j\neq i}}^{n}\frac{C_{j}^{(0)}}{C_{j}}=x\exp (-x)
\label{cstar}
\end{equation}%
Here $x=np$ is the scaffold functionality $n$ multiplied by the
concentration ratio $p$. \ By choosing $p=1/n$ the mass fraction attains a
maximum of $\exp (-1)\simeq 0.37!$ \ This result indicates that by selecting
the appropriate scaffold concentration, in the nonequilibrium regime up to $%
37\%$ of the particles will aggregate to form star clusters. \ This is a
significant improvement over the situation in the equilibrium regime. \ 

This treatment of the problem captures the physics of star cluster
formation, but it does not account for the loss of star clusters due to
aggregation. \ In particular, as long as there are scaffolds with markers
available for hybridization, when these scaffolds encounter a star cluster
they can aggregate to form a larger cluster. \ We now estimate how this
aggregation effects the final concentration of star clusters. \ 

Consider the beginning of the second stage in our aggregation process. \
There are no longer any free scaffolds in solution, but a scaffold can have
up to $n-1$ DNA markers still available for hybridization. \ We would like
to determine how the star cluster mass fraction $M_{\ast }(y)$ changes as a
function of the fraction of saturated scaffolds $y$. \ Here a saturated
scaffold has particles hybridized to all $n$ of its DNA markers, and is
therefore unreactive. \ If $s$ is the expectation that a slot on the
scaffold is filled, then the fraction of saturated scaffolds is $y=s^{n-1}$.
\ The average number of open slots on a scaffold is $(n-1)(1-s)$. \ Consider
filling an open slot on the scaffold. \ The probability that the particle
which filled the slot was part of a star cluster is $M_{\ast }(y)$. \ The
average rate $r(y)$ at which star clusters are lost to aggregation is then%
\begin{equation}
r(y)=-M_{\ast }(y)\frac{d}{dy}\left[ (n-1)(1-s)\right] =M_{\ast
}(y)y^{-\alpha }\text{.}
\end{equation}%
Here the exponent $\alpha =(n-2)/(n-1)$. \ We can then construct a
differential equation for $M_{\ast }$ taking into account this loss due to
aggregation. \ 
\begin{equation}
\frac{dM_{\ast }}{dy}=\frac{dM_{\ast }^{(o)}}{dy}-xr(y)
\end{equation}%
In the absence of this loss term the result of the calculation should
recover our previous result Eq. \ref{cstar}. \ This zeroth order
approximation is just $M_{\ast }^{(o)}(y)=xy\exp (-xy)$ which gives the
correct star cluster concentration once all of the scaffolds are saturated ($%
y=1$). \ To simplify the analysis a bit we take $\alpha =1$ which is an
excellent approximation in the limit of large scaffold functionality $n$. \
This is an inhomogeneous first order differential equation which can be
solved by introducing an integrating factor $u(y)=y^{x}$. $\ $The initial
condition which must be satisfied is $M_{\ast }(0)=0$. \ We are interested
in the final star mass fraction $M_{\ast }$, which is $M_{\ast }(y=1)$. \
The result is%
\begin{eqnarray}
M_{\ast } &=&x\sum_{k=0}^{\infty }\frac{(-x)^{k}}{k!}\left[ \frac{1}{x+k+1}-%
\frac{x}{x+k+2}\right]  \label{massfractheory} \\
&=&x\exp (-x)+x^{2}E_{-x}(x)-x^{1-x}\Gamma (1+x)  \notag
\end{eqnarray}%
Here $\Gamma (x)$ is the gamma function and $E_{\nu
}(x)=\int\limits_{1}^{\infty }t^{-\nu }\exp (-xt)dt$ is the exponential
integral of order $\nu $. \ 

We can perform a similar type of analysis in the case when there is only one
particle color. \ In this case the $n$ ssDNA markers on the scaffold all
have identical sequences complementary to this color. \ It turns out that
the result for the mass fraction is the same. \ Because the mass fraction is
the same in both cases, we can gain insight into the behavior of the system
with many colors by analyzing the much simpler one color system. \ To test
our predictions, we numerically solved a system of differential equations
which models the irreversible aggregation between particles (one color) and
scaffolds. \ 
\begin{equation}
\frac{dC_{IJ}}{dt}=\frac{1}{2}\sum\limits_{\substack{ i+i^{\prime }=I  \\ %
j+j^{\prime }=J}}K_{iji^{\prime }j^{\prime }}C_{ij}C_{i^{\prime }j^{\prime
}}-C_{IJ}\sum\limits_{i,j}K_{ijIJ}C_{ij}
\end{equation}

This equation is the Smoluchowski coagulation equation \cite{Smoluchowski}
adapted to our system. \ $C_{ij}$ is the concentration of the cluster with $%
i $ scaffolds and $j$ particles. $\ K_{iji^{\prime }j^{\prime }}$ is the
rate constant for the irreversible reaction $C_{ij}+C_{i^{\prime }j^{\prime
}}\rightarrow C_{i+i^{\prime }j+j^{\prime }}$. \ We assume that the rates
are diffusion limited in which case we can estimate the rate for any pair of
clusters by $K_{iji^{\prime }j^{\prime }}=4\pi D_{s}R_{l}$. \ The larger
cluster with hydrodynamic radius $R_{l}\sim n_{l}^{1/3}$ plays the role of a
sink. \ Here $n_{l}$ is the number of particles in the larger cluster and $%
D_{s}=k_{B}T/6\pi \eta R_{s}$ is the diffusion constant for the smaller
cluster. \ To simply matters we only consider tree like structures, i.e. we
do not consider the formation of clusters with internal loops. \ We have
truncated the set of equations by considering clusters with a maximum of $10$
scaffolds. \ 

\begin{figure}[h]
\includegraphics[width=4.7528in,height=3.5772in]{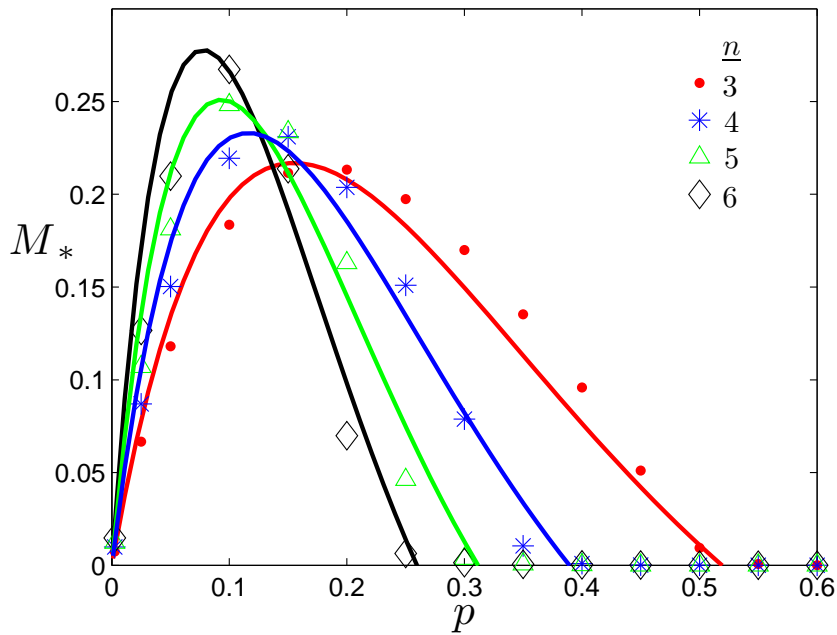}
\caption{The mass fraction $M_{\ast }$ as a function of $p$ for scaffolds
with functionality $n=3$ (red), $4$ (blue), $5$ (green), and $6$ (black). \
The results determined numerically from the full solution of the
Smoluchowski coagulation equation (markers) can be compared to results of
the anlaytical calculation (lines) Eq. \protect\ref{massfractheory}. \ The
resulting agreement is good up to an overall normalization factor $\protect%
\gamma _{n}$ in the range $1.2$ to $1.5$ which normalizes the analytical
curves. \ }
\label{massfracstar}
\end{figure}

By solving these equations we can determine the concentration of stars $%
C_{\ast }=C_{1n}$ in this notation and test the validity of our two stage
ansatz. \ As indicated in Fig. \ref{massfracstar}, the result of our
analytical calculation matches the results of the full numerical calculation
up to an overall normalization factor of order unity. \ Several points are
in order. \ 

The optimal concentration ratio $p$ for experiments is easily determined
from $\frac{dM_{\ast }}{dx}=0$. \ The result is $x_{\max }\simeq 0.47$. \
For scaffolds of functionality $n$ the concentration ratio should be chosen
as:%
\begin{equation}
p=\frac{0.47}{n}\text{.}  \label{popt}
\end{equation}%
Note that the maximum attainable star cluster yield $M_{\ast }(x_{\max
})\simeq 1/4$ does not decrease with increasing $n$. \ In fact, the
numerical results predict a slight increase in star cluster yield for larger 
$n$. \ Solving the aggregation equations becomes computationally expensive,
but it can still be done by reducing the maximum number of scaffolds in a
cluster. \ For example, considering clusters with up to $5$ scaffolds for $%
n=10$ gives $M_{\ast }(x_{\max })\simeq 0.3$. \ These results are important
from the perspective of experimental feasibility for the self-assembly
method. \ This is to be contrasted with the earlier equilibrium treatment. \
There the condition to suppress the aggregation of larger clusters imposed a
fairly strict constraint\cite{nanoclusters} on the concentration ratio $%
p\lesssim n^{1/2}\left( \frac{2}{n}\right) ^{n-1}$. \ From the perspective
of self-assembling stars with large $n$ this renders the regime of
irreversible binding far more appealing than the equilibrium regime. \ 

If an experiment is performed with the optimal concentration ratio, the
clusters which self-assemble are easily separated by density gradient
centrifugation\cite{centrifugation}. \ In this regime most of the particles
are monomers, in star clusters, or in saturated two scaffold clusters. \
These clusters contain, $1$, $n$, and $2n-1$ particles respectively. \ The
disparity in hydrodynamic radius and sedimentation velocity of these
clusters makes the separation procedure experimentally feasible. \ 

In this section we considered a DNA\ scaffold method for self-assembling
star clusters of $n$ colored particles. \ By taking advantage of the
mobility mismatch between particles and scaffolds, we were able to formulate
a nonequilibrium calculation of the star mass fraction. \ The results of the
calculation were compared to the numerical results of the full Smoluchowski
coagulation equation for the system. \ Good agreement is established between
the analytical calculation and the numerics. \ In the regime of irreversible
binding the yield of the desired star cluster is drastically improved in
comparison to earlier equilibrium estimates. \ In nonequilibrium we find an
experimentally feasible regime for the self-assembly of star clusters with a
maximum mass fraction $\simeq 1/4$. \ We determined the optimal
concentration ratio for an experimental implementation of our proposal. \
The additional color degrees of freedom associated with particle permutation
in these clusters makes them ideal candidates as building blocks in a future
hierarchical self-assembly scheme. \ In addition, these clusters can serve
as the starting point to self-assemble structures of arbitrary geometry\cite%
{licata}. \ The experimental realization of self-assembling star clusters
using DNA\ scaffolds would constitute an important step towards realizing
the full potential of DNA\ mediated interactions in nanoscience. \ 

\section{Cluster Degeneracy}

\qquad Building these decorated colloidal clusters is the first major
experimental step in a new self-assembly proposal which will be discussed in
the next chapter. \ However, in order to utilize the resulting clusters as
building blocks, an additional ordering is necessary. \ The problem is that
the decoration introduces degeneracy in the ground state configuration. \
This degeneracy was not present in \cite{packing} since all the polystyrene
spheres were identical. \ Namely, in the colloidal clusters self-assembled
by our method, permuting the particle labels in a cluster does not change
the second moment of the mass distribution (see Figure \ref%
{degeneracypicture}). \ We need a method to select a single "isomer" out of
the many present after self-assembly. \ In the DNA-colloidal system
considered here, this isomer selection can be facilitated by "linker" ssDNA.
\ These are short ssDNA with sequence $\overline{s}_{A}\overline{s}_{B}$ to
connect particles $A$ and $B$. \ We first construct a list of nearest
neighbors for the chosen isomer, and introduce linker DNA for each nearest
neighbor pair. \ The octopus-like DNA arms of the given particles will
hybridize to the linkers, resulting in a sping-like attraction between the
selected particle pairs. \ Note that the length $L$ of the DNA\ arms must be
on the order of the linear dimension of the original cluster. \ Otherwise
the interparticle links cannot form upon introduction of linker DNA to the
system. \ It should be noted that although this method breaks the
permutation degeneracy of a cluster, the right-left degeneracy will still be
present. \ \ 
\begin{figure}[h]
\includegraphics[width=4.7342in,height=3.5572in]{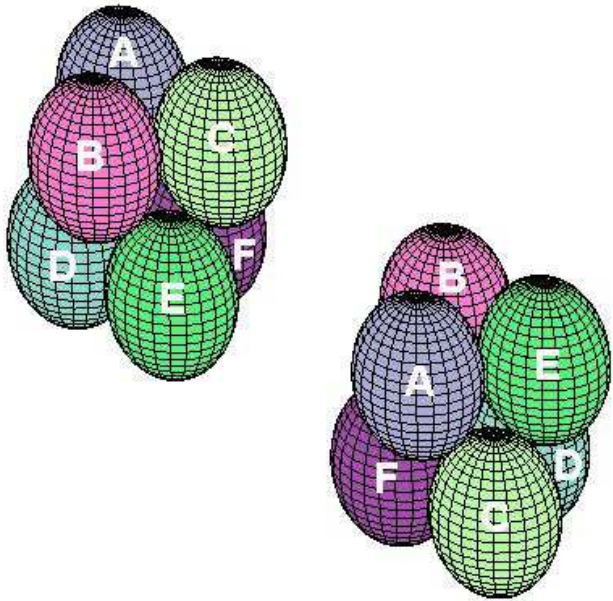}
\caption{An illustration of degeneracy in DNA-coded nanoclusters. \ Two
different $n=6$ isomers are pictured, both with the same minimal second
moment configuration, the octahedron. \ }
\label{degeneracypicture}
\end{figure}

\chapter{Programmable Self-Assembly}

\section{Introduction}

\qquad Over recent years, significant attention has been attracted to the
possibility of nanotechnological applications of DNA \cite{nucleic}, \cite%
{falls}, \cite{template}, \cite{angstrom}, \cite{nanotube}. \ Among the
various proposals, one of the most interesting directions is the use of
DNA--mediated interactions for \textit{programmable self--assembly} of
nanoparticle structures \cite{rational}, \cite{natreview}, \cite%
{designcrystals}, \cite{nanocrystals}. Several schemes of such self-assembly
have been studied both experimentally and theoretically. Their common theme
is the use of colloidal particles functionalized with specially designed
ssDNA (markers), whose sequence defines\ the particle type. In such systems,
selective type-dependent interactions can be introduced either by making the
markers complementary to each other, or by using linker-DNA chains whose
ends are complementary to particular marker sequences.

Recent theoretical studies have addressed the expected phase behavior \cite%
{phasebehav}, melting properties \cite{meltingprop}, and morphological
diversity \cite{morphology} of DNA--colloidal assemblies. \ In particular,
there are indications that these techniques can be utilized for fabrication
of photonic band gap materials \cite{photonic}, \cite{wiremesh}. \ Despite
significant experimental progress, the long-term potential of DNA--based
self--assembly is far from being realized. \ For instance, most of the
experimental studies of DNA--colloidal systems report random aggregation of
the particles \cite{colloidalgold}. \ Some degree of structural control in
these systems has been achieved, mostly by varying the relative sizes of
particles, rather than by tuning the interactions \cite{micelle}, \cite%
{blocks}.

In the present chapter \cite{licata}, we take a broader view of programmable
self-assembly. \ While this theoretical study is strongly motivated by the
prospects of DNA-colloidal systems, our main objective is to address a more
general question: how well can a desired structure be encoded by tunable
interactions between its constituents? The particular model system on which
we focus consists of distinguishable particles with individually controlled
interactions between any pair of them. In the first stage of our study, we
analyze a simplified yet generic version of such a system. \ All the
particles have the same repulsive potential, while the attraction is
introduced only between selected pairs of particles in the form of a
spring-like quadratic potential. \ This potential mimics the effect of a
stretchable DNA molecule whose ends can selectively adsorb to the particular
pair of particles. \ We then introduce a number of additional features which
make the model a more realistic description of an actual DNA-colloidal
system.

Our major result is that a combination of stretchable interparticle linkers
(e.g. sufficiently long DNA), and a soft repulsive potential greatly reduces
(or totally eliminates) the probability of self-assembling an undesired
structure. \ The experimental prototype is a system of particles in a
mixture of two types of DNA\ molecules which can selectively adsorb to the
particle surface. \ The first type are DNA molecules with two "sticky" ends,
i.e. both end sequences of the DNA are complementary to the particle marker
sequence. \ With one end adsorbed to the particle surface, the remaining
sticky end makes it possible to introduce an attractive interparticle
potential between selected particle pairs. \ The second type are DNA\
molecules with one sticky end which adsorbs to the particle surface. \ These
DNA strands give rise to a soft repulsive potential of entropic origin
between all particle pairs. \ 

There is a natural analogy between our problem and the folding of proteins
where interactions between amino acids encode the overall structure. \
However, it should be emphasized that the task of programmable self-assembly
in colloidal systems is even more demanding than protein folding: any
intermediate metastable configuration has a much longer lifetime and
therefore means a misfolding event. \ Because of this, we were looking for a
self--assembly scenario which does not require thermally activated escape
from a metastable configuration. This makes the problem additionally
interesting from the theoretical perspective of "jamming", a phenomenon
actively studied in the context of granular and colloidal systems. Our
results can be interpreted in terms of a jamming-unjamming transition
controlled by the interaction parameters.

\ It should be emphasized that the goal of this work is primarily
conceptual, as opposed to providing a manual for the immediate experimental
realization of ordered colloidal structures. \ Nevertheless, future
experimental schemes will be forced to overcome obstacles presented by
colloidal jamming. \ With this in mind, one of the most salient features of
our model is the ability to smooth the energy landscape by tuning the
interactions between particles. \ 

The plan for the chapter is as follows. \ In section 2, we address the
problem within a simplified generic model which mimics the nanoparticle
system with stretchable DNA connections. An unexpected and very encouraging
result of this study is that the misfolding (or jamming) in the model system
can be completely avoided for a certain set of parameters. \ In section 3
the original model is adapted to a more realistic situation which
incorporates the random character of the DNA-mediated interactions. \ In
section 4 we discuss the prospects for the future experimental
implementation of our scheme. \ In section 5 we summarize the major results.
\ \ 

\section{Beads and Springs Model}

\qquad Consider an isolated group of repulsive particles linked via a
polymer spring to their desired nearest neighbours. \ We assume that the
DNA\ marker sequence is the same for any two markers attached to the same
particle, but different particles have different marker sequences (i.e. each
particle has a unique code). \ In this case, the attraction between any two
particles can be effectively switched on by adding DNA "linkers" whose ends
are complementary to the corresponding marker sequences of the particles. \
As a first approach to the problem, we introduce a generic "Beads \&
Springs" model which incorporates essential features of the
DNA--nanoparticle system. \ The model system contains $N$ particles with
pairwise (type-independent) repulsive potential $U(r)$. \ In general, this
repulsion may have a hard-core or soft-core behavior, or be a combination of
the two. In order to model the DNA--induced type--dependent attraction, we
introduce a harmonic potential (linear springs) which acts only between
selected pairs of particles \cite{scmbook}. \ Thus, the model Hamiltonian
has the following form: 
\begin{equation}
H=\frac{1}{2}\sum\limits_{\alpha ,\beta }\kappa J_{\alpha \beta }\left\vert 
\mathbf{r}_{\alpha }-\mathbf{r}_{\beta }\right\vert ^{2}+U(|\mathbf{r}%
_{\alpha }-\mathbf{r}_{\beta }|)
\end{equation}%
Here $\alpha ,\beta $ are the particle indices, $\mathbf{r}_{\alpha }$ are
their current positions, and $\kappa $ is the spring constant. \ The
connectivity matrix element $J_{\alpha \beta }$ may be either 1 or 0,
depending on whether the two particles are connected by a spring, or not. \
Our goal is to program\ the desired spatial configuration by choosing an
appropriate connectivity matrix $J_{\alpha \beta }$. A natural construction
is to \textit{put }$J_{\alpha \beta }=1$\textit{\ for any pair of particles
which must be nearest neighbors} in the desired cluster, and not to connect
the particles otherwise (i.e. put $J_{\alpha \beta }=0$). \ This
construction assures that the target configuration is the ground state of
the system. \ 

Note that our problem is somewhat similar to that of heteropolymer folding.
\ In that case, the selective interactions between monomers (e.g. amino
acids in protein) are responsible for the coding of the spacial structure of
a globule. Our major concern is whether the kinetics of the system will
allow\ it to reach the ground state within a reasonable time. \ \
Unfortunately, since the Brownian motion of a typical nanoparticle is
relatively slow (compared to molecular time scales), it is unrealistic to
expect that our system will be able to find the target configuration by
"hopping" between various metastable states, as in the case of protein
folding. \ However, our case is different because the attractive force grows
with distance, as opposed to the short--range nature of heteropolymer
self--interactions. \ As we shall see below, this difference is essential,
making it possible for the system to reach the ground state without stopping
at any metastable configuration.

\begin{figure}[h]
\includegraphics[width=4.48in,height=4.858in]{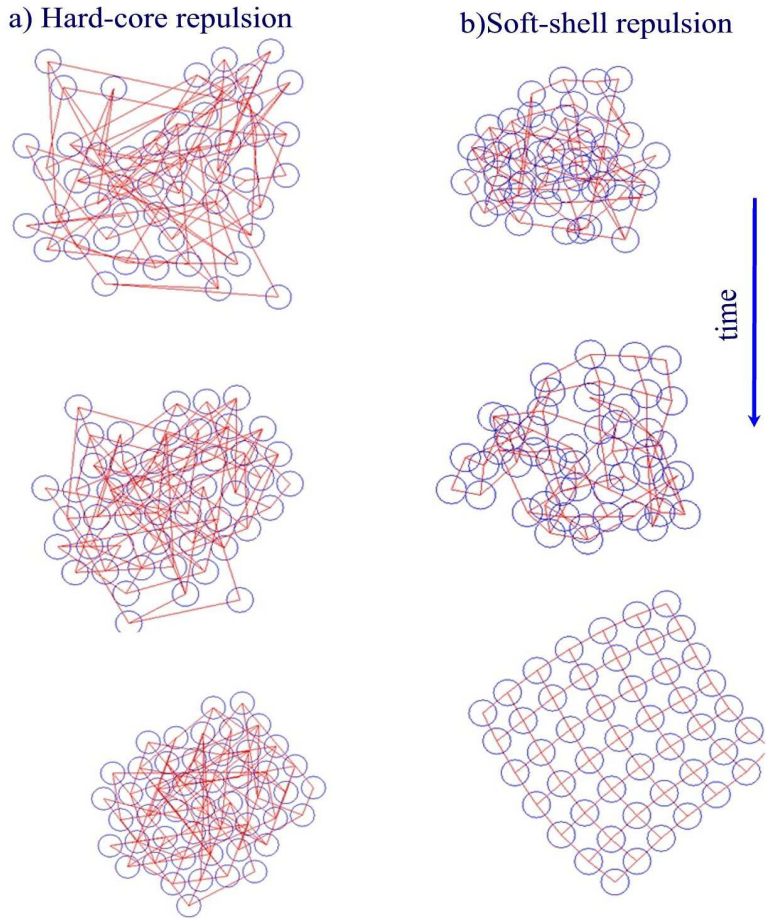}
\caption{Programmable self--assembly in 2D, studied within the Beads \&
Springs model. 49 particles are all distinct and connected with linear
springs to encode the desired configuration (7x7 square). The jamming
behavior is observed for the case of hard spheres (a). However, the assembly
of the target structure can be achieved if the repulsion is sufficiently
"soft" (b). }
\label{2D}
\end{figure}

We have performed a molecular dynamics simulation of the above model by
numerically integrating its Langevin equation:

\begin{equation}
b^{-1}\mathbf{\dot{r}}_{\alpha }=-\mathbf{\nabla }_{\alpha }H+\mathbf{\eta }%
_{\alpha }
\end{equation}
Here $b$ is the particle mobility. \ The thermal noise has been artificially
suppressed in this study (i.e. $\eta =0$ ). \ In other words, we have
assumed the worst case scenario: once the system is trapped in a local
energy minimum, it stays there indefinitely. The equations of motion were
solved numerically by a first order Runge-Kutta method. \ First, we studied
a system of $N=49$ distinguishable particles in 2D, whose native
configuration was a $7\times 7$ square cluster (see Figure \ref{2D}). \
Their initial positions were random, and the connectivity matrix was
constructed according to the above nearest--neighbor rule. \ 

First, we studied the case when $U(r)$ is a hard--core potential. More
precisely, the repulsive force was given by a semi--linear form: 
\begin{equation}
f_{hc}\left( r\right) =-\frac{\partial U_{hc}\left( r\right) }{\partial r}%
=\kappa _{0}(d-r)\Theta (d-r)
\end{equation}%
Here $\Theta $\ is the unit step function, and $d$ is the diameter of the
hard sphere. \ The parameter\ $\kappa _{0}$ determines the strength of the
hard-core repulsion, and it does not affect the results, as long as $\kappa
_{0}\gg \kappa $. \ In our simulations, we found that the hard sphere system
eventually stops in a configuration definitely different from the desired
one, a behavior which is well known in the context of granular and colloidal
systems as "jamming" \cite{geltransition}. \ Remarkably, \textit{the jamming
can be avoided when the hard-core repulsion is replaced by a soft--core
potential}: 
\begin{equation}
U_{sc}(r)=U_{0}\exp (-r/\lambda ).
\end{equation}%
Here the decay length $\lambda $ is of the order of the equilibrium
interparticle distance $r_{0}$. \ This indicates that the energy landscape
can be made smooth by a combination of long-range selective attraction and
soft-core repulsion. \ 

The result is surprising and remarkably robust. \ In particular, in order to
expand our finding to the 3D case, we studied the self-assembly of particles
into tetrahedra of various sizes ($N=10,20,35$). \ This time, the hard core
interaction potential was superimposed with\ a soft shell repulsion, which
makes the model more relevant for an actual DNA-colloidal system: 
\begin{equation}
U\left( r\right) =U_{hc}\left( r\right) +U_{sc}\left( r\right) .
\end{equation}%
After the system has fully relaxed, a geometric measure of the folding
success is determined by comparing particle separations of the desired final
state to those generated from a set of random initial conditions. Figure \ref%
{jammed} shows the "jamming phase diagram" for these systems. \ To assign a
point on the diagram to the correct folding regime, we required 100
consecutive successful folds. \ While this criterion can only give an upper
bound on the jamming probability (which is approximately $1\%$), an
additional analysis gives strong evidence that the correct folding region of
the diagram corresponds to zero probability of jamming. \ 

\begin{figure}[h]
\includegraphics[width=4.5886in, height=3.4579in]{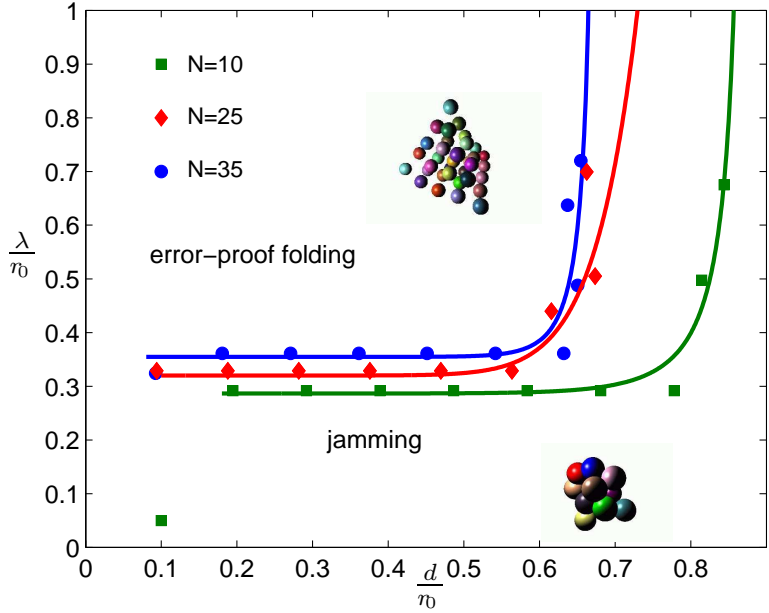}
\caption{"Jamming phase diagram" obtained for programmable self-assembly of
tetrahedral clusters, within the Beads \& Springs model. The control
parameters depend on the equilibrium interparticle distance $r_{o}$, the
diameter of the hard sphere $d$, and the range of the soft-shell repulsion $%
\protect\lambda $.}
\label{jammed}
\end{figure}

\section{Self-Assembly in DNA Colloidal Systems}

\qquad As we have seen, the introduction of a soft-core repulsive potential $%
U_{sc}$ is crucial to a successful self-assembly proposal. \ In a real
system, this repulsion can be generated e.g. by DNA\ or another water
soluble polymer adsorbed to the particle surface. \ The mechanism is quite
independent of the monomer chemistry, but for the sake of concreteness we
will speak of the repulsion generated by DNA. Namely, we assume that a
certain fraction of the DNA "arms" of the "octopus-like" particles are not
terminated by a sticky end, and only play the role of a repulsive "buffer".
\ When the polymer coverage is sufficiently low, the interparticle repulsion
is primarily due to entropy loss of a chain squeezed between two particles.
\ The characteristic length scale of this interaction is given by the radius
of gyration of the "buffer" chain, $R_{g}$. \ The corresponding repulsive
force $f_{sc}$ can be calculated exactly in the limit of relatively short
buffer chains, $R_{g}\ll d$. The result of this calculation \cite{morphology}
can be adequately expressed in the following compact form: \ 
\begin{equation}
f_{sc}\left( r\right) \approx \frac{4NR_{g}T}{d\left( r-d\right) }\exp
\left( -\frac{\left( r-d\right) ^{2}}{2R_{g}^{2}}\right)
\end{equation}%
Here $N$ is the total number of buffer chains per particle. Note that even
though this result is only valid for $R_{g}\ll d$, it correctly captures the
Gaussian decay of the repulsive force expected for longer chains as well.
Therefore, we expect the results to be at least qualitatively correct beyond
the regime of short buffer chains.\ 

In addition to the modified soft potential, we have to take into account the
random character of the realistic DNA-mediated attraction. \ It originates
from the fact that (1) the number of DNA "arms" of the original octopus-like
particles will typically be determined by a random adsorption process, and
(2) the fraction of the DNA chains recruited for linking a particular pair
of particles is also random. \ In terms of our original model, this means
that the "springs" will not have the same spring constant. \ If the
individual linkers are modelled by Gaussian chains \cite{statprop}, the
overall spring constant for a \ particular pair of connected particles is
given by $\kappa _{\alpha \beta }=Tm_{\alpha \beta }/2R_{g}^{^{\prime }2}$ ,
where $R_{g}^{^{\prime }}$ is the\ radius of gyration of a single linker,
and $m_{\alpha \beta }$ is the number of individual chains connecting the
particles. \ We assume that this number obeys the generic Poisson
distribution: \ $P(m)=\overline{m}^{m}e^{-\overline{m}}/m!$. \ As
formulated, the model is cast as a system of coupled differential equations:
\ 
\begin{equation}
\mathbf{\dot{r}}_{\alpha }=b\sum\limits_{\beta }\left[ -\frac{T}{%
2R_{g}^{^{\prime }2}}J_{\alpha \beta }m_{\alpha \beta }r_{\alpha \beta
}+f_{hc}\left( r_{\alpha \beta }\right) +f_{sc}\left( r_{\alpha \beta
}\right) \right] \mathbf{\hat{n}}_{\alpha \beta }
\end{equation}%
Here $r_{\alpha \beta }=\left\vert \mathbf{r}_{\alpha }-\mathbf{r}_{\beta
}\right\vert $, $\mathbf{\hat{n}}_{\alpha \beta }=\left( \mathbf{r}_{\alpha
}-\mathbf{r}_{\beta }\right) /r_{\alpha \beta }$.

We have studied the behavior of the system as a function of two
dimensionless parameters, one of which is the ratio of the buffer radius of
gyration to the particle diameter, $R_{g}/d$ . The other parameter
characterizes the relative strength of the attractive and repulsive forces: 
\begin{equation}
\alpha =\frac{\overline{m}}{N}\left( \frac{R_{g}}{R_{g}^{^{\prime }}}\right)
\end{equation}%
Including the above modifications of the model, the essential result of the
study is that the jamming probability can be drastically suppressed,
similarly to the previous case. However, the jamming in this system cannot
be eliminated completely. \ Instead of an actual "phase boundary", we have
observed a sharp crossover to the regime of predominantly good folding, in
which the error probability is suppressed to a modest level $\sim 10-20\%$.
Interestingly, the behavior is nearly independent of the energy parameter $%
\alpha $, which makes $R_{g}/d$ the only major control parameter. Figure \ref%
{error} shows the error probability as a function of this geometric
parameter for tetrahedral clusters of different sizes. As this plot
indicates, the misfolding behavior gets suppressed as $R_{g}/d$ exceeds $1$.

\begin{figure}[h]
\includegraphics[width=4.65in, height=3.41in]{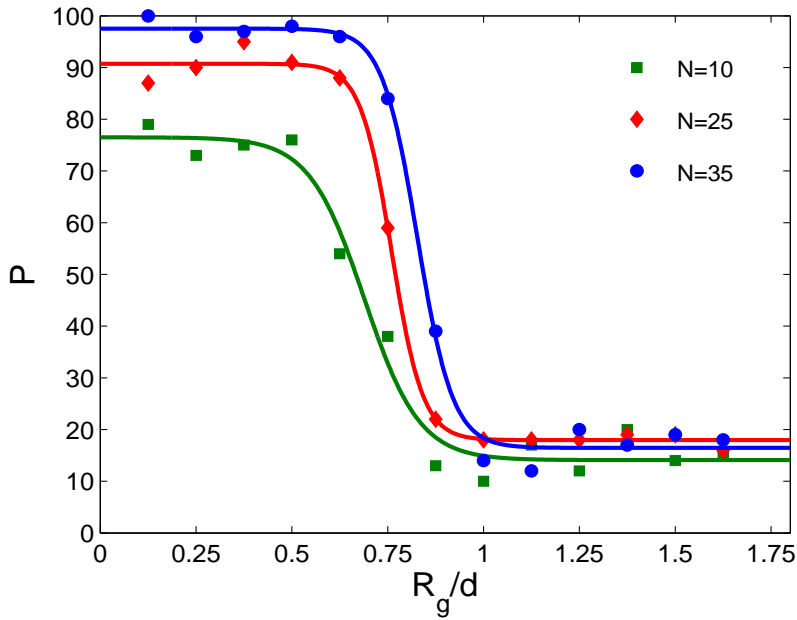}
\caption{Error probability $P$ as a function of aspect ratio $R_{g}/d$ for
tetrahedral clusters with modified soft-potential and realistic DNA-mediated
attraction. Each data point on the misfolding profile represents 100 trials.
\ }
\label{error}
\end{figure}
\ 

\begin{figure}[h]
\includegraphics[width=4.65in,height=3.41in]{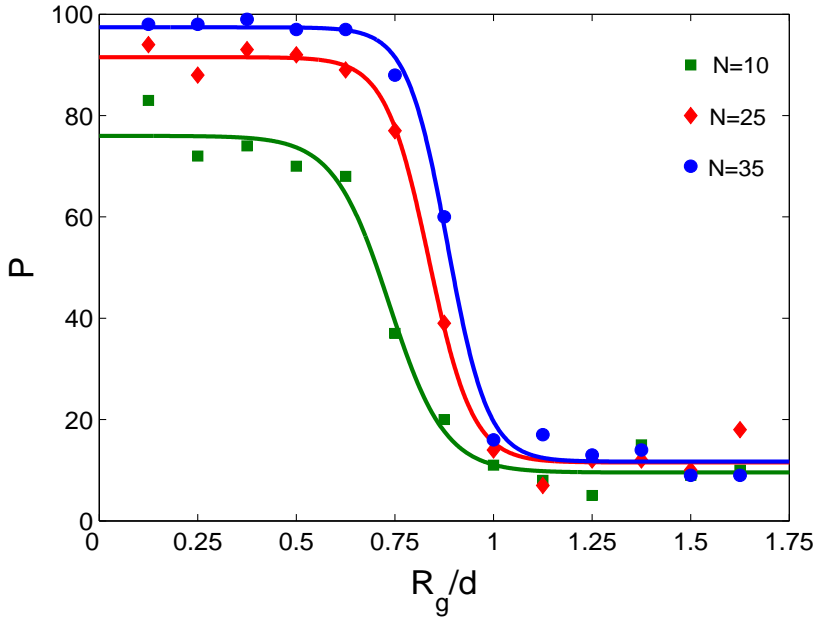}
\caption{Error probability $P$ as a function of aspect ratio $R_{g}/d$ for
tetrahedral clusters in the Pincus regime. \ Each data point on the
misfolding profile represents 100 trials. \ }
\label{pincus}
\end{figure}

As a further test of the robustness of the model, we consider a modified
attractive potential which deviates from the linear Hooke's law. \ For
larger forces we enter the Pincus regime \cite{Pincus}, where the end-to-end
extension of the polymer chain $r$ is related to the external tension $f\sim 
$ $r^{\frac{3}{2}}$. \ This tension law incorporates the excluded volume
interaction of individual linker DNA with themselves, which was not
previously considered. \ Remarkably, the major result for error suppression
carries over, as illustrated in figure \ref{pincus}. \ \ 

Our description has several limitations which are to be addressed in future
work. \ \ In particular, our discussion is only applicable to the limit of
modest coverage of particles with buffer chains. This case of weakly
overlapping adsorbed chains (known as the "mushroom regime") is drastically
different from the high--coverage "polymer brush" behavior \cite{scaling}. \
\ Nevertheless, our major conclusions appear to be rather robust. \ The
condition for error suppression($R_{g}/d\gtrsim 1$) is the same in both the
harmonic and Pincus regimes. \ 

\begin{figure}[h]
\includegraphics[width=4.7041in,height=3.5414in]{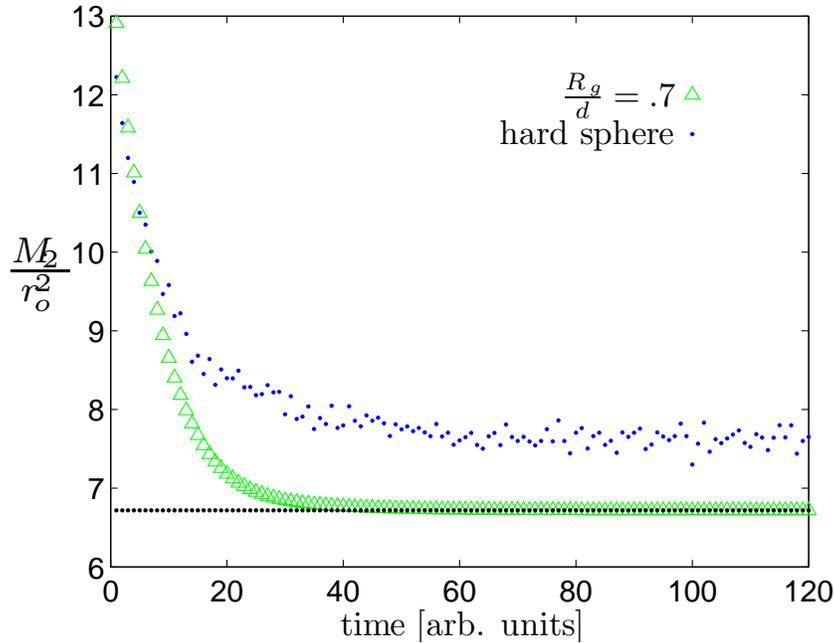}
\caption{Plot of the dimensionless second moment $\frac{M_{2}}{r_{o}^{2}}$
as function of time for $n=9$ particles. \ Results are shown for the case of
hard spheres and also for a system with a soft-core repulsion with geometric
parameter $\frac{R_{g}}{d}=.7$. \ The dashed line is the theoretical moment
for the triaugmented triangular prism, which is the minimal $n=9$ structure.
\ }
\label{jam}
\end{figure}

In the previous chapter we presented a detailed discussion of a method to
self-assemble DNA-coded clusters. \ Within the context of the beads and
springs model presented here we can discuss the role that jamming plays in
preventing the one anchor structures from assuming the minimal second moment
configuration. \ We performed simulations of the assembly of optimal
colloidal clusters up to $n=9$ particles by numerically integrating the
particles' Langevin equations. \ As indicated in Figure \ref{jam}, the hard
sphere system gets trapped in a configuration with a larger $M_{2}$ than the
optimal cluster, whereas the soft-core system is able to fully relax. \ The
jamming behavior is largely determined by the single control parameter $%
\frac{R_{g}}{d}$, with $d$ the diameter of the hard sphere. \ Beyond the
critical value $\frac{R_{g}}{d}\gtrsim .5$ the jamming behavior is either
completely eliminated, or greatly reduced in the case of larger clusters. \ 

\section{Discussion}

\qquad Previous studies have demonstrated an experimental implementation of
self--assembly in a DNA--colloidal system \cite{chaikin}, \cite{crocker}. \
The approach in these studies differs from our vision of controllable
self-assembly in two major ways: (1) because the "linker" DNA\ chains($\sim
20nm$) used are much shorter than the particle diameter($\sim 1\mu m$), the
particles behave as sticky spheres, and (2) there is little diversity in
particle type, where the structures result from interactions of a one or two
component system. \ We would like to discuss issues related to a future
experimental implementation similar to the modified Beads \& Springs model.
\ 

In our molecular dynamics simulations we assumed that (1) the desired group
of particles has already been localized in a small region of space, and that
(2) the interparticle connections have already been made. \ There are a
number of experimental challenges associated with implementing the
self-assembly proposal of our simulations. \ In another manuscript \cite%
{nanoclusters} we provide a detailed discussion of the localization problem,
which is the first major experimental intermediary. \ 

After localization, the next step is to make the desired connections between
particles within the cluster. \ To do so one can add short ssDNA with
sequences $\bar{s}_{A}\bar{s}_{B}$ to link particles $A$ and $B$. \ The DNA
marker sequence $s_{A}$ for particle $A$ is a sequence of nucleotides
complementary to the $\bar{s}_{A}$ portion of the linker sequence $\bar{s}%
_{A}\bar{s}_{B}$. \ The hydrogen bonding of complementary nucleotides forms
base pairs which join both marker strands to the linker, creating a DNA\
bridge between the two particles. \ After the interparticle links are
formed, they should be made permanent by ligation. \ Since the spring
constants of the above dsDNA chains are too small to drive the self-assembly
of a desired cluster, we propose to melt them either by changing the
temperature or pH. As a result, the dsDNA links will be turned into ssDNA
with a much higher effective spring constant (due to the shorter persistence
length). \ This will trigger the self--assembly scenario similar to the one
discussed within the Beads \& Springs model. \ Note that DNA entanglements
may be effectively eliminated if the procedure is done in the presence of
DNA Topoisomerases. \ 

\section{Summary}

\qquad We presented a model of DNA-colloidal self-assembly which exhibits a
tunable jamming-unjamming transition. \ The combination of a soft-core
repulsion with a type-dependent long range attraction provides a natural
funneling of the energy landscape to the ground state configuration. \ This
is to be contrasted with the case of protein folding, where under
physiological conditions the interactions between amino acids are screened
to several angstroms. \ Because this lengthscale is much shorter than the
spatial extent of the native structure, large regions of the energy
landscape are flat, which prohibits formation of the native state on the
basis of funneling alone. \ As a result the folding rate is necessarily
limited by the diffusion of amino acid segments looking for their desired
nearest neighbors \cite{proteinlandscape}, \cite{proteinfolding}. \ The fact
that the potential in the DNA-colloidal system is long ranged is essential,
allowing us to avoid the pitfall of slow particle diffusion. \ 

Within the Beads and Springs Model, we obtained the jamming phase diagram
for several modest sized tetrahedral clusters. \ We identified a regime of
parameter space with error-proof folding, and demonstrated the importance of
introducing a soft-core repulsion. \ The original model was then adapted to
include several features of realistic DNA-mediated interactions. \ Although
the jamming cannot be completely eliminated in the modified system, we
identified a regime of predominantly good folding, and calculated the error
probability for tetrahedral clusters. \ The jamming behavior is determined
by a single geometric parameter $R_{g}/d$. \ We concluded by discussing
prospects for an experimental implementation of our self-assembly scheme. \ 

\chapter{Cooperativity Based Drug Delivery Systems}

\section{Introduction}

\qquad Nanoparticle based drug delivery systems have attracted substantial
attention for their potential applications in cancer treatment \cite{target1}%
, \cite{target2}, \cite{target3}, \cite{target4}, \cite{nanocluster}. \ It
is hoped that by selectively targeting cancer cells with chemotherapeutic
agents one can reduce side effects and improve treatment outcomes relative
to other drug delivery systems which do not discriminate between normal and
cancerous cells. \ For example, many epithelial cancer cells are known to
overexpress the folate receptor \cite{ovarian}, \cite{folate1}, \cite%
{folate2}, \cite{folate3}, \cite{folate4}. \ A nanoparticle with many folic
acid ligands will preferentially bind to cancerous cells. \ A recent study 
\cite{pamam} of a potential drug delivery platform consisting of generation
5 PAMAM\ dendrimers with different numbers of folic acid found that
multivalent interactions have a pronounced effect on the dissociation
constant $K_{D}$. \ This enhancement is the signature for cooperativity of
the binding, which should lead to a greater specificity to cancerous cells
in vivo. \ 

\begin{figure}[tbp]
\includegraphics[width=4.6683in,height=3.5182in]{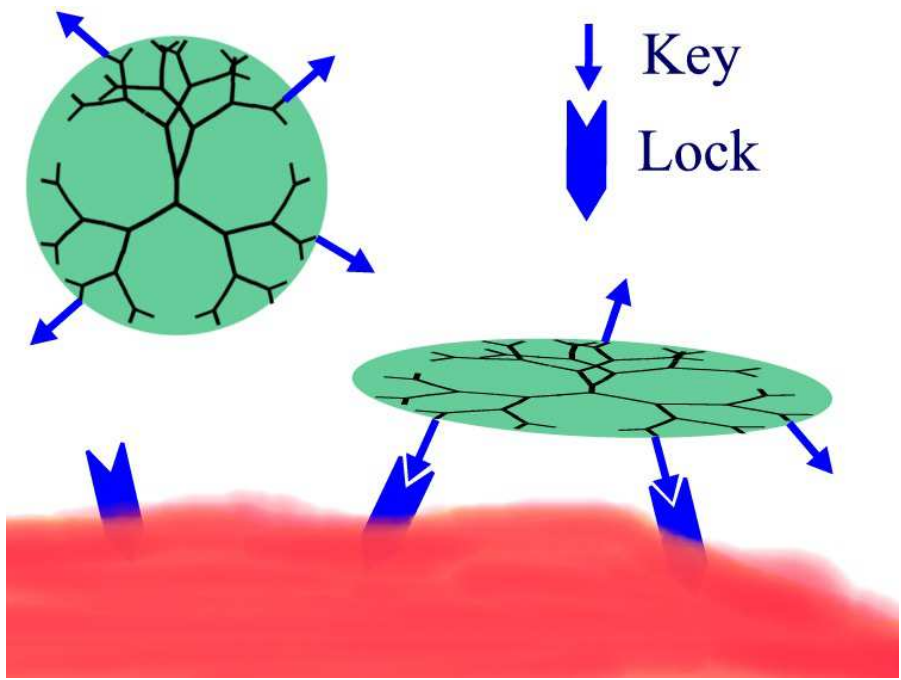}
\caption{A picture of the dendrimer "key-lock" binding to the cell membrane
surface. \ }
\label{dendrimernew}
\end{figure}

In this chapter \cite{cancerpaper} we present a theoretical study of these
key-locking nanodevices (see Fig. \ref{dendrimernew}). \ We introduce the
idea that there are kinetic limitations to cooperativity-based drug delivery
systems. \ In vivo the finite timescale for endocytosis prevents arbitrarily
high cooperativity in the drug delivery system. \ To begin we provide a
detailed analysis of the in vitro experiments \cite{pamam}. \ Although
enhancement of the association is the signature of greater cooperativity, in
this case it is due mostly to non-specific binding of the dendrimers to the
surface. \ Due to the finite time window of the experiments, only indirect
support can be offered to the notion of enhanced cooperativity. \ In the
latter half of the chapter we expand the notion of kinetically limited
cooperativity to the system in vivo. \ The equilibrium coverage of
nanodevices on the cells is related to the concentration of folate-binding
proteins and the strength of the key-lock binding. \ We quantify the
preferential adsorption of nanodevices to the cancerous cells, and discuss
how kinetic effects prohibit arbitrarily high cooperativity in the drug
delivery system. \ The implications of the work for designing new drug
delivery vehicles with enhanced specificity to cancerous cells are
discussed. \ 

\section{Key-Lock Model}

\qquad We now consider a simple model of the nanodevice system. \ A
dendrimer with a maximum of $M$ keys (e.g. folic acids) interacts with locks
(e.g. folate-binding proteins) in the cell membrane surface. \ A simple
order of magnitude estimate for $M\simeq 30$ can be obtained from the ratio
of the surface area of the dendrimer to the surface area of the folic acid.
\ In this way we implicitly take into account the excluded volume effect
between the keys. \ The free energy for the dendrimer connected to the
surface by $m$ key-lock bridges is \cite{statmech} $\ $%
\begin{equation}
F_{m}=-Tm\Delta \text{.}
\end{equation}%
The dimensionless energy parameter $\Delta $ contains information about the
binding energy of a single key-lock pair, and the entropy loss associated
with localizing a dendrimer on the cell-membrane surface. \ An estimate of $%
\Delta \simeq 17.5$ can be obtained from the dissociation constant of free
folic acid $K_{D}^{(o)}$ using the equilibrium relation between the
dissociation constant and the free energy change for the formation of a
single key-lock bridge, $K_{D}^{(o)}=\frac{1}{\xi ^{3}}\exp (-\Delta )$. \
Here $\xi ^{3}$ is the localization volume of an "unbound" key. \ Below we
determine the value $\xi \simeq 0.2nm$ from analysis of the in vitro
experiments, which was used to determine $\Delta $. \ 

\begin{figure}[tbp]
\includegraphics[width=4.7528in,height=3.5772in]{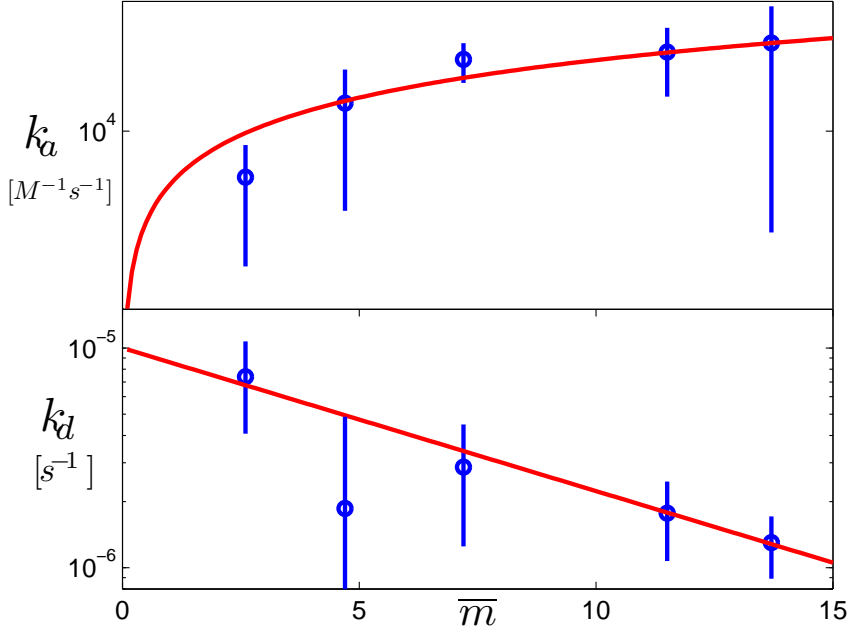}
\caption{Top: Plot of the association rate constant (Eq. \protect\ref{kaeqn}%
) $k_{a}[M^{-1}s^{-1}]$ versus $\overline{m}$. \ Bottom: Plot of the
effective dissociation rate constant (Eq. \protect\ref{kdeffeqn}) $%
k_{d}[s^{-1}]$ versus $\overline{m}$. \ In the fit $%
k_{d}^{(o)}=10^{-5}[s^{-1}]$ and $\protect\alpha =0.15$. \ The experimental
data points are taken from Figure 5 in \protect\cite{pamam}.}
\label{kplot5}
\end{figure}

The measured association rate constant $k_{a}$ of the dendrimer with folic
acid is a factor of $10^{3}$ times greater than $k_{a}^{(o)}$ of free folic
acid. \ Only a factor of $\overline{m}$ can be attributed to the dendrimer
having many folic acids attached to it. \ Here $\overline{m}$ is the average
number of keys attached to the dendrimer. \ This pronounced enhancement of $%
k_{a}$ is the primary evidence for non-specific attraction between the
dendrimer and the surface. \ 
\begin{equation}
k_{a}=\overline{m}k_{a}^{(o)}\exp \left( \frac{-\epsilon _{0}}{T}\right)
\label{kaeqn}
\end{equation}%
The non specific attraction $\epsilon _{0}$ accounts for the Van der Waals
attraction to the surface and hydrophobic enhancement. \ The experimentally
measured $k_{a}$ values are reproduced by a reasonable energy scale $%
-\epsilon _{0}\simeq 7T$ (see Fig. \ref{kplot5}). \ 

We provide a simple explanation for the experimentally observed dependence
of the dissociation rate constant $k_{d}$ on $\overline{m}$. \ The
dissociation rate constant of free folic acid $k_{d}^{(o)}\sim 10^{-5}\left[
s^{-1}\right] $ provides a characteristic departure time of $%
1/k_{d}^{(o)}\simeq 30$ $hours$ for those dendrimers attached by a single
key-lock bridge. \ Moreover, the departure time for multiple bridge states
increases exponentially in $\Delta $, for two bridges it is $\exp (\Delta
)/k_{d}^{(o)}\simeq 10^{9}$ $hours$. \ Strictly speaking the relaxation is
multiexponential, with time constants for each bridge number. \ However, the
experimental $k_{d}$ values are well fit by a single exponential. \ On the
timescale of the experiment, we will only see the departure of dendrimers
attached by a single bridge. \ 

The experiment measures the departure rate of dendrimers which are connected
to the surface by a single bridge, but are unable to form an additional
connection. \ Consider a dendrimer attached to the surface by one key-lock
bridge. \ If the dendrimer has a total of $j$ keys, the probability that
none of the remaining $j-1$ keys can form bridges is $(1-\alpha )^{j-1}$. \
We now compute the probability $\alpha $ that a remaining key is available
to form a bridge. \ In the vicinity of the surface the dendrimer is a
disclike structure \cite{mecke} with radius $a\simeq 4.8nm$. \ By rotation
of the dendrimer about the first bridge, a key located at position $\rho $
searches the annulus of area $2\pi \rho \xi $ to find a lock. \ The
probability of encountering a lock in this region is $2\pi \rho \xi \sigma
_{o}$, where the surface density of the locks $\sigma _{o}\simeq \frac{16}{%
100nm^{2}}$. \ By averaging over the key location we obtain the final result%
\begin{equation}
\alpha =\frac{1}{a}\int_{0}^{a}2\pi \rho \xi \sigma _{o}d\rho \simeq \xi
a\sigma _{o}.  \label{alpha}
\end{equation}%
Assuming that during dendrimer preparation the attachment of folic acid to
the dendrimer is a Poisson process, the probability of a dendrimer having
exactly $j$ keys is $P_{j}(\overline{m})=\exp (-\overline{m})\overline{m}%
^{j}/j!$. \ The final result is obtained by averaging the probability that
no additional bridges can form over this distribution. \ The factor of $j$
counts the number of ways to make the first connection. \ 
\begin{equation}
k_{d}=k_{d}^{(o)}\frac{\sum\limits_{j=1}^{\infty }(1-\alpha )^{j-1}jP_{j}(%
\overline{m})}{\sum\limits_{j=1}^{\infty }jP_{j}(\overline{m})}%
=k_{d}^{(o)}\exp (-\alpha \overline{m})  \label{kdeffeqn}
\end{equation}

The formula predicts an exponential decay of the effective dissociation rate
constant with the average number of folic acids on the dendrimer, which
allows for a quantitative comparison to the experiment (see Fig. \ref{kplot5}%
). \ Using $\alpha \simeq 0.15$, we can determine the localization length $%
\xi \simeq 0.2nm$ for locks in the experiment from Eq. \ref{alpha}. \ This
estimate for $\xi $ is physically reasonable, and comparable to the bond
length of the terminal group on the dendrimer. \ 

Similar to the finite timescale of the experiments in vitro \cite{pamam}, in
vivo the endocytosis time provides kinetic limitations to cooperative
binding. \ In equilibrium the concentration of dendrimers on the cell
surface $n$ is related to the concentration of dendrimers in solution $%
c_{sol}$ through the association constant $K_{A}=n/(\sigma _{o}c_{sol})$. \
Although it is tempting to use our in vitro results to define the
association constant as $K_{A}=k_{a}/k_{d}$, this approach is only valid
provided there is a single rate for both association and dissociation. \
Because the dendrimer can form multiple bridges, there are many different
rate constants. \ We present a partition function method which accounts for
the multiple rate constants in the problem, and for the possibility that in
vivo there is surface diffusion of locks. \ 

\section{\protect\bigskip Calculation of the Association Constant $K_{A}$}

\qquad In this section we consider the reaction $\sigma
_{o}+c_{sol}\rightleftharpoons n$ for which the association constant $K_{A}$
is defined as%
\begin{equation}
K_{A}=\frac{n}{\sigma _{o}c_{sol}}.
\end{equation}%
Here $\sigma _{o}$ is the concentration of locks in the cell membrane, $%
c_{sol}$ is the concentration of dendrimers in solution, and $n$ is the
concentration of dendrimers attached to the surface of the cell membrane. \
To proceed we construct a vector $\mathbf{s}$ of length $M$, which is a list
of the possible sites folic acid can attach to the dendrimer. \ If a folic
acid is present at site $i$ we have $s_{i}=1$, and otherwise $s_{i}=0$. \
The concentration of dendrimers on the cell surface $n$ is proportional to
the partition function of the system. \ 
\begin{gather}
n=\frac{c_{sol}\xi ^{3}}{A}\sum\limits_{m=1}^{m_{\max }}\int \frac{d^{2}%
\mathbf{r}_{1}\cdots d^{2}\mathbf{r}_{m}}{m!}\sum_{i\neq j\neq \cdots \neq
p}s_{i}\cdots s_{p} \\
\times \sigma (\mathbf{r}_{1})\cdots \sigma (\mathbf{r}_{m})\exp \left[
m\Delta -\frac{\epsilon _{0}+\varepsilon _{ij\cdots p}(\mathbf{r}_{1},\cdots
,\mathbf{r}_{m})}{T}\right]  \notag
\end{gather}%
Here $\sigma (\mathbf{r})$ is the surface density of locks on the cell
membrane at position $\mathbf{r}$, and $A$ denotes the total area of the
cell membrane. \ The energy $\varepsilon _{ij\cdots p}(\mathbf{r}_{1},\cdots
,\mathbf{r}_{m})$ that appears in the Boltzmann weight is the elastic energy
penalty required to form multiple bridges. \ The point is that in solution
the dendrimer is roughly spherical, but must flatten to a pancake like shape
to form multiple connections with the cell surface \cite{mecke}. \ 

The ensemble averaging is performed by assuming that during nanodevice
preparation the attachment of folic acid to the dendrimer is a Poisson
process. $\ $In this case $\left\langle s_{i}\right\rangle =\frac{\overline{m%
}}{M}$ is given by the success probability that a folic acid attaches to the
dendrimer, and the$\ m$ point correlator $\left\langle s_{i}s_{j}\cdots
s_{p}\right\rangle =\left( \frac{\overline{m}}{M}\right) ^{m}$. \ In other
words, the probability of attachment of a given folic acid to a terminal
group on the dendrimer is unaffected by the presence of other folic acids up
to an exclusion rule which has already been taken into account. \ If the
interaction potential between locks in the cell membrane is $V(\mathbf{r}%
_{1,}\cdots ,\mathbf{r}_{m})$ we have $\left\langle \sigma (\mathbf{r}%
_{1})\cdots \sigma (\mathbf{r}_{m})\right\rangle =\left( \sigma _{o}\right)
^{m}\exp \left[ -V(\mathbf{r}_{1,}\cdots ,\mathbf{r}_{m})/T\right] $. \ By
performing the ensemble averaging we arrive at the result for the
equilibrium coverage $n_{m}^{eq}$ of dendrimers connected to the cell
surface by $m$ bridges. \ Since the lock which forms the first bridge can be
anywhere on the cell membrane, without loss of generality we place this lock
at $\mathbf{r}_{1}=\mathbf{0}$. \ The integrand is then independent of $%
\mathbf{r}_{1}$, and the first areal integration gives a factor of the cell
surface area $A$. \ 
\begin{eqnarray}
\left\langle n\right\rangle &=&\frac{c_{sol}\xi ^{3}}{A}A\sum_{m=1}^{m_{\max
}}\int \frac{d^{2}\mathbf{r}_{2}\cdots d^{2}\mathbf{r}_{m}}{m!}\left\langle
\sigma \mathbf{(0)}\sigma \mathbf{\mathbf{(r}}_{2}\mathbf{\mathbf{)}\cdots }%
\sigma \mathbf{(r}_{m}\mathbf{)}\right\rangle \times \\
&&\sum_{i\neq j\neq \cdots \neq p}\left\langle s_{i}\cdots
s_{p}\right\rangle \exp \left( m\Delta -\frac{\epsilon _{0}+\varepsilon
_{ij\cdots p}(\mathbf{0},\cdots ,\mathbf{r}_{m})}{T}\right)  \notag
\end{eqnarray}%
\begin{eqnarray}
\left\langle n\right\rangle &=&c_{sol}\xi ^{3}\sum_{m=1}^{m_{\max }}\int 
\frac{d^{2}\mathbf{r}_{2}\cdots d^{2}\mathbf{r}_{m}}{m!}\left( \sigma
_{o}\right) ^{m}\exp \left[ \frac{-V(\mathbf{0},\mathbf{\mathbf{r}}%
_{2},\cdots ,\mathbf{r}_{m})}{T}\right] \times \\
&&\sum_{i\neq j\neq \cdots \neq p}\left( \frac{\overline{m}}{M}\right)
^{m}\exp \left( m\Delta -\frac{\epsilon _{0}+\varepsilon _{ij\cdots p}(%
\mathbf{0},\mathbf{r}_{2},\cdots ,\mathbf{r}_{m})}{T}\right)  \notag
\end{eqnarray}%
\begin{eqnarray}
\left\langle n\right\rangle &=&c_{sol}\xi ^{3}\exp \left( \frac{-\epsilon
_{0}}{T}\right) \sum_{m=1}^{m_{\max }}\frac{\left( \overline{m}\sigma
_{o}\exp (\Delta )\right) ^{m}}{m!}\frac{\xi ^{2(m-1)}}{\xi ^{2(m-1)}}\frac{1%
}{M^{m}}\times \\
&&\sum_{i\neq j\neq \cdots \neq p}\int d^{2}\mathbf{r}_{2}\cdots d^{2}%
\mathbf{r}_{m}\exp \left[ -\frac{(V(\mathbf{0},\mathbf{\mathbf{r}}%
_{2},\cdots ,\mathbf{r}_{m})+\varepsilon _{ij\cdots p}(\mathbf{0},\mathbf{r}%
_{2},\cdots ,\mathbf{r}_{m}))}{T}\right]  \notag
\end{eqnarray}%
Here $K_{A}^{(o)}=1/K_{D}^{(o)}$ $=\xi ^{3}\exp (\Delta )$ is the
association constant of free folic acid which has been measured
experimentally. \ 
\begin{eqnarray}
\left\langle n\right\rangle &=&\frac{c_{sol}\xi ^{3}}{\xi ^{2}}%
\sum_{m=1}^{m_{\max }}\frac{\left( \overline{m}\sigma _{o}\xi ^{2}\exp
(\Delta )\right) ^{m}}{m!}\exp \left[ \frac{-\left( \epsilon _{0}+\epsilon
_{el}^{(m)}\right) }{T}\right] = \\
&&c_{sol}\xi \sum_{m=1}^{m_{\max }}\frac{1}{m!}\left( \frac{\overline{m}%
\sigma _{o}K_{A}^{(o)}}{\xi }\right) ^{m}\exp \left[ \frac{-\left( \epsilon
_{0}+\epsilon _{el}^{(m)}\right) }{T}\right]  \notag
\end{eqnarray}%
This gives the final result for the ensemble averaged concentration of
dendrimers on the surface $\left\langle n\right\rangle $:%
\begin{eqnarray}
\left\langle n\right\rangle &=&\sum_{m=1}^{m_{\max }}n_{m}^{eq} \\
n_{m}^{eq} &=&\frac{c_{sol}\xi }{m!}\left( \frac{\overline{m}\sigma
_{o}K_{A}^{(o)}}{\xi }\right) ^{m}\exp \left[ \frac{-\left( \epsilon
_{0}+\epsilon _{el}^{(m)}\right) }{k_{B}T}\right]
\end{eqnarray}%
Here the elastic energy $\epsilon _{el}^{(m)}$ is defined by:%
\begin{eqnarray}
\exp \left( \frac{-\epsilon _{el}^{(m)}}{T}\right) &\equiv &\frac{1}{M^{m}}%
\int \frac{d^{2}\mathbf{r}_{2}\cdots d^{2}\mathbf{r}_{m}}{\xi ^{2(m-1)}} \\
&&\times \sum_{i\neq j\neq \cdots \neq p}\exp \left[ -\frac{(V(\mathbf{0},%
\mathbf{\mathbf{r}}_{2},\cdots ,\mathbf{r}_{m})+\varepsilon _{ij\cdots p}(%
\mathbf{0},\mathbf{r}_{2},\cdots ,\mathbf{r}_{m}))}{T}\right]  \notag
\end{eqnarray}%
\ \ \ \ \ \ Defined in this manner, $\exp (-\epsilon _{el}^{(m)}/T)$ has a
physical interpretation as the Boltzmann weight for the elastic energy of
the optimal $m$ bridge configuration. \ The membrane surface can only
accommodate a finite number of locks in the vicinity where the dendrimer is
attached \cite{pamam}. \ As a result $n_{m}^{eq}=0$ for $m>m_{\max }$ since
forming additional key-lock pairs would require deforming the dendrimer into
configurations prohibited by elastic stress and steric hindrance. \ 

The calculation of the equilibrium coverage above is applicable with and
without diffusion of locks in the cell membrane. \ In the regime of fast
diffusion the locks are free to diffusively explore the surface. \ Their
positions are ergodic variables, and the overall ensemble averaged
equilibrium coverage counts the Boltzmann weights for different lock
configurations. \ In the regime of slow diffusion, locks are immobilized in
the cell membrane. \ This is the relevant situation when the locks have
phase separated into protein rich (lipid rafts) and protein poor phases. \ 

\section{Kinetically Limited Cooperativity}

\qquad When kinetic effects are taken into account, these two regimes are
drastically different. \ When locks are diffusing, the dendrimer is able to
attain the maximum cooperativity $m_{\max }$. \ After the dendrimer makes
the first connection, it simply waits for locks to diffuse in the vicinity
of available keys to make additional connections. \ In the absence of
diffusion, the optimal configuration can only be obtained by multiple
binding and unbinding events, the timescale for which is prohibitively long.
\ This is the case for lipid rafts where the locks are immobilized similar
to the experiments in vitro \cite{pamam}, and the dendrimer is unable to
attain the maximum cooperativity. \ This is the kinetic origin of limited
cooperativity in the drug delivery system. \ 

We now quantify the preferential attachment of nanodevices to the cancerous
cells, taking into account kinetic effects. \ Let $n_{m}$ denote the
concentration of dendrimers attached to the cell by $m$ bridges. \ We can
construct a differential equation for $n_{m}$ by considering linear response
to the deviation from thermal equilibrium $n_{m}^{eq}$. \ 
\begin{equation}
\frac{dn_{m}}{dt}=k_{d}^{(m)}(n_{m}^{eq}-n_{m})-\gamma n_{m}
\end{equation}%
Here $\gamma $ is the rate for endocytosis \cite{endocytosis}. \ The
dissociation rate constant $k_{d}^{(m)}$ for breaking all $m$ bridges is: \ 
\begin{equation}
k_{d}^{(m)}=m\frac{k_{a}^{(o)}}{\xi ^{3}}\exp \left( \frac{\epsilon
_{el}^{(m)}}{T}\right) \exp (-m\Delta )
\end{equation}%
The steady state concentration $n_{m}^{ss}$ is the solution to $\frac{dn_{m}%
}{dt}=0$. \ As a result we obtain the total coverage $n$ of dendrimers on
the cell surface in the following form:%
\begin{equation}
n=\sum_{m=1}^{m_{\max }}n_{m}^{ss}=\sum_{m=1}^{m_{\max }}\frac{n_{m}^{eq}}{%
1+\gamma /k_{d}^{(m)}}
\end{equation}

\begin{figure}[tbp]
\includegraphics[width=4.7946in,height=3.6101in]{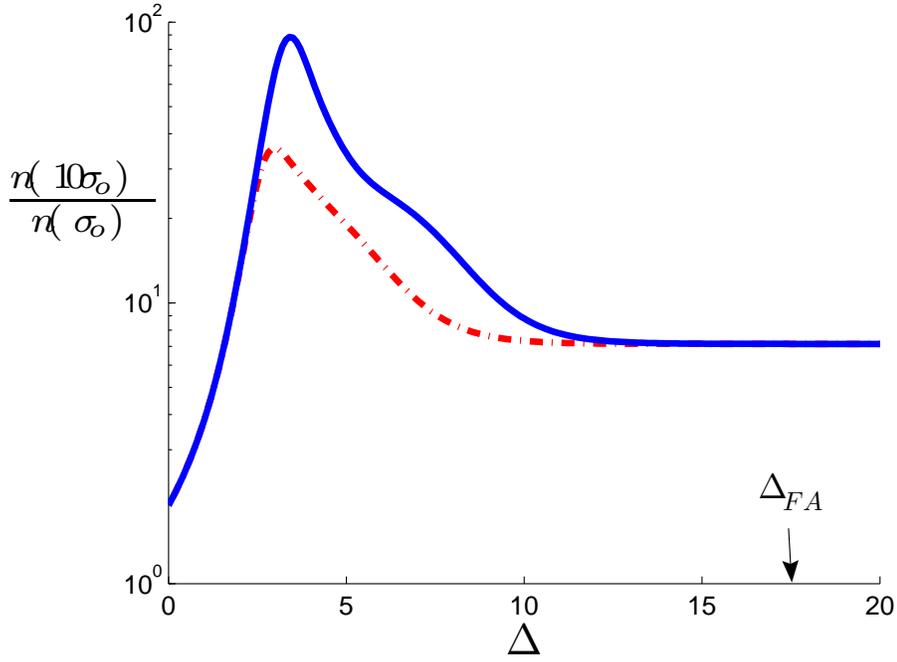}
\caption{The ratio of surface concentrations of dendrimers on cancerous to
normal cells $\frac{n(10\protect\sigma _{o})}{n(\protect\sigma _{o})}$ as a
function of $\Delta $ with $r=10$. \ The dotted line corresponds to an
endocytosis time $1/\protect\gamma =1\left[ hr\right] $ and the solid line
is $1/\protect\gamma =10\left[ hr\right] $. \ Here $\overline{m}=15$, $%
m_{\max }=4$, $\protect\xi =3[nm]$, and $\protect\sigma _{o}=2\times
10^{-3}[nm^{-2}]$. \ $\protect\varepsilon _{el}^{(m)}=3T$ for $m\geq 3$
bridges. \ }
\label{logratio}
\end{figure}

We now have a means to discuss the preferential attachment of dendrimers to
the cancerous cell. \ The folate binding proteins on the cancerous cell are
overexpressed, i.e. if their concentration on the normal cell is $\sigma
_{o} $, their concentration on the cancerous cell is $r\sigma _{o}$ with $%
r>1 $. \ The value of $r$ is determined by the biology, and cannot be
changed by the experimenter. \ To quantify the preferential binding of the
dendrimer to the cancerous cells we calculate the ratio of coverage on
cancerous to normal cells $\frac{n(r\sigma _{o})}{n(\sigma _{o})}$. \ Values
of this ratio greater than $r$ indicate the nature of cooperative dendrimer
binding (see Fig. \ref{logratio}).

\begin{figure}[tbp]
\includegraphics[width=4.6858in,height=3.5182in]{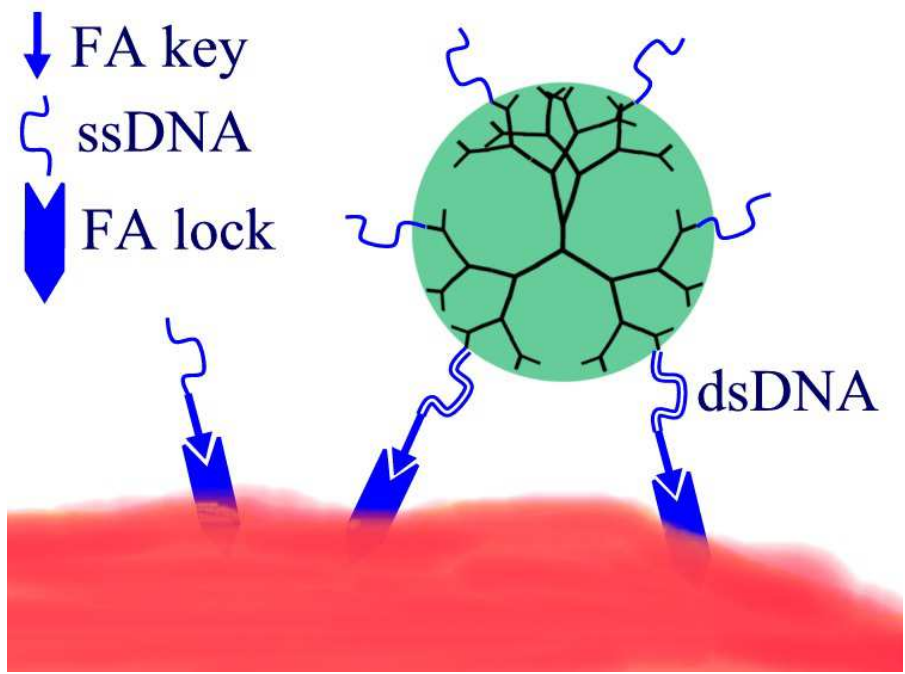}
\caption{Single-stranded DNA (ssDNA) on the dendrimer hybridize to the ssDNA
attached to the folic acid (FA) key. \ }
\label{dendrimerscheme}
\end{figure}

\section{Design of an Improved Drug-Delivery Platform}

\qquad The current experimental scheme uses direct targeting with folic acid
($\Delta _{FA}\simeq 17.5$), which does not optimize the coverage on
cancerous cells. \ By decreasing $\Delta $ the drug delivery can be tuned to
the favorable regime. \ To do so, consider binding to the cell through an
intermediary, perhaps single-stranded DNA (ssDNA). \ Instead of folic acid,
attach many identical sequences of ssDNA to the dendrimer. \ Then, one also
constructs a folic acid-ssDNA complex with the ssDNA sequence complementary
to that of the ssDNA attached to the dendrimer. \ The folic acid will bind
very strongly to the folic acid receptors on the cell membrane, leaving the
unhybridized ssDNA as a receptor (see Fig. \ref{dendrimerscheme}). \
Effectively one has replaced $\Delta _{FA}$ with a new value $\Delta _{DNA}$
which can be tuned very precisely by controlling the length and sequence of
the DNA. \ Due to the large degree of overexpression \cite{ovarian}, this
change substantially increases the ratio of dendrimers on cancerous to
normal cells. \ As indicated in Fig. \ref{logratio}, with $r\simeq 10$ there
is a $5$ fold improvement over direct targeting with folic acid!

\section{Summary}

\qquad In this chapter we presented a theoretical study of a cell-specific,
targeted drug delivery system. \ A simple "key-lock" model was proposed to
determine the effective dissociation rate and association rate constants of
the dendrimers as a function of the average number of folic acids, which
permits a direct comparison to the experimental results. \ The equilibrium
coverage of dendrimers on the cell surface was calculated, and the
differences between in vitro experiments and in vivo studies were discussed.
\ The degree of cooperativity of the drug delivery system is kinetically
limited. \ We quantified the notion of preferential selection of dendrimers
to cancerous cells, and demonstrated that the selectivity can be enhanced by
decreasing the strength of individual bonds. \ A particular implementation
of this idea using ssDNA was discussed. \ 

\chapter{Conclusions and Perspectives}

\section{The Relationship Between Crystallization and Jamming}

\qquad After the publication of \cite{licata} two recent studies reported
the self-assembly of colloidal crystals using DNA mediated interactions \cite%
{colloidalcrystal}, \cite{mirkincrystal}, \cite{crockerdnaletter}. \ Up
until this point most of the studies reported aggregation of colloids into
disordered, amorphous aggregates \cite{chaikin} or close packed crystals 
\cite{crocker}. \ In these earlier studies polystyrene spheres of diameter $%
d\sim 1\mu m$ were grafted with dsDNA of length $L\simeq 20nm$. \ Because $%
L<l_{p}\simeq 50nm$ for dsDNA the DNA linkers behaved effectively as rigid
rods. \ The essential change in the new studies was to work with flexible ($%
L\gg l_{p}$) DNA with contour length $L\gtrsim d$. \ For this purpose Gang
et al. \cite{colloidalcrystal} grafted $g\simeq 60$ ssDNA onto gold
nanoparticles of diameter $d\simeq 11nm$. \ Interestingly, they found that
as one varies the length $L\simeq Na$ of the DNA ($a\simeq .43nm$ for ssDNA 
\cite{persistencelength}), the system goes from a disordered to crystalline
configuration. \ Increasing the lengthscale for the repulsive interactions
between particles provides a means to smooth the energy landscape to the
equilibrium body-centered cubic configuration. \ 

It seems that an important control parameter in such systems is the ratio $%
\xi /d$ where $\xi $ is the lengthscale for the repulsion and $d$ is the
hard sphere diameter. \ In fact it has been theoretically predicted that one
can obtain different crystalline morphologies by varying this control
parameter \cite{morphology}. \ This control parameter is something we have
already encountered in chapter 5. \ There $\xi \simeq R_{g}$ was the radius
of gyration of the buffer DNA. \ Of course a direct analogy between the two
systems is not possible, as the self-assembly of a modest number of
particles into colloidal clusters is quite different than the
crystallization problem. \ Motivated by their similarities, it is at least
plausible that the explanation for which systems are able to attain their
ground state configuration and which remain "jammed" in metastable
configurations is qualitatively similar. \ 

To proceed with such a comparison we have to determine an appropriate
definition of the control parameter. \ Here I propose that by defining $\xi
\simeq R$ one might be able to obtain a qualitative understanding of which
systems crystallize and which remain disordered in terms of a critical value
of the control parameter $R/d$. \ Here $R$ is the so-called "coronal
thickness" for a polymer brush attached to a small colloidal particle $d\ll
R $. \ To determine $R$ we consider a simple scaling argument of Daoud and
Cotton \cite{Daoud}, \cite{grafted}. \ By balancing the osmotic pressure
against the polymer elasticity we obtain%
\begin{equation}
\frac{1}{2}c^{2}a^{3}T\simeq \frac{kRg}{4\pi R^{2}}\text{.}
\end{equation}%
Assuming that the monomer concentration takes its average value $c=(Ng)/V$
with $V=(4\pi R^{3})/3$ yields the result for the coronal thickness%
\begin{equation}
R\simeq g^{\frac{1}{5}}N^{\frac{3}{5}}a\text{.}
\end{equation}%
We are now in a position to compare the value of the control parameter for
the different systems studied experimentally \cite{colloidalcrystal}. \ 

\begin{table}[h]
\caption{Control Parameters for Equilibrium Colloidal Crystallization}
\label{table:jam}\centering        
\begin{tabular}{ccc}
\hline\hline
$N$ & $R/d$ & Aggregate? \\[0.5ex] \hline
$18$ & $0.48$ & amorphous \\ 
$30$ & $0.66$ & amorphous \\ 
$40$ & $0.78$ & amorphous \\ 
$50$ & $0.89$ & crystalline \\ 
$65$ & $0.95$ & crystalline \\[1ex] \hline
\end{tabular}%
\end{table}

It is interest to note this approximation gives a critical value of the
control parameter $\frac{R}{d}\gtrsim 0.9$, which is very close to the
critical value given for the self-assembly of colloidal clusters in chapter
5. \ The value $N=40$ is the mean for a system prepared with particles of
type $A$ for which $N_{A}=50$ and particles of type $B$ for which $N_{B}=30$%
. \ All other systems were prepared in a homogeneous fashion with $%
N_{A}=N_{B}=N$. \ This analogy seems to indicate that the system with
shorter ssDNA is trapped in a metastable configuration analogous to the
jammed configuration for the colloidal clusters, and the introduction of a
long-range repulsive potential provides a kinetically feasible pathway to
the equilibrium configuration. \ One topic for future research would be a
more detailed consideration of the jamming analogy in colloidal crystals
beyond the crude scaling argument given here. \ Comparison of the results to
the experiments of \cite{colloidalcrystal}\ is potentially problematic,
since they increase the length of the DNA but keep the grafting density
constant for all systems studied. \ As a result the properties of the
polymer brush that forms on the particle surface likely vary for different
linker lengths. \ 

\section{Building Up Complexity}

\qquad Studies on DNA-mediated colloidal crystallization have emphasized the
importance of thermal annealing as a means to facilitate crystallization. \
Here we explore another idea which might assist in the equilibrium
DNA-mediated self-assembly of colloidal crystals. \ The approach, which to
date has received little attention in experimental studies, is to
self-assemble crystalline structures in a hierarchical fashion (see Fig. \ref%
{honeycombthesis}). \ Consider a binary system of particles with two colors $%
A$ and $B$. \ The DNA marker sequences are chosen so that $AB$ bonds form as
a result of DNA hybridization, but not $AA$ or $BB$. \ To take advantage of
this cohesive energy, the nearest neighbours in the resulting crystalline
morphologies should be of different color \cite{morphology}. \ These types
of crystal structures are commonly described as a lattice with a basis \cite%
{Ashcroft}. \ For example, one can consider placing $A$ and $B$ type
particles at the points of a body-centered cubic lattice so that each
particle has $8$ nearest neighbours of the opposite type. \ This structure,
sometimes called the cesium chloride structure, is obtained by taking a
simple cubic lattice with a basis consisting of an $A$ type particle at the
origin and a $B$ type particle at the center of the cube $\left( a/2\right) (%
\mathbf{x}+\mathbf{y}+\mathbf{z})$. \ Here $a$ is the lattice constant and
the primitive lattice vectors are $\mathbf{x=(}%
\begin{array}{ccc}
1 & 0 & 0%
\end{array}%
)$, $\mathbf{y=(}%
\begin{array}{ccc}
0 & 1 & 0%
\end{array}%
)$, and $\mathbf{z=(}%
\begin{array}{ccc}
0 & 0 & 1%
\end{array}%
)$. \ Similarly one may obtain the sodium chloride structure where the
particles are located at the vertices of a simple cubic lattice and each
particle has $6$ nearest neighbours of the opposite type. \ This crystal
structure can be described as a face centered cubic lattice with a basis. \ 

In fact, the same is true of more exotic crystal structures, for example the
diamond structure. \ This structure, the same as that adopted by gallium
arsenide, can be described as a face-centered cubic lattice with a two
particle basis. \ The diamond structure is of particular interest for its
potential applications in the self-assembly of photonic band gap materials 
\cite{photonic}. \ Since these crystal structures can be built up by as $AB$
pairs, one could attempt to assemble these crystals where the fundamental
units are not individual particles but dimers. \ The dimers could be
assembled first and then used as the fundamental components. \ 

\begin{figure}[tbp]
\includegraphics[width=4.6858in,height=3.5182in]{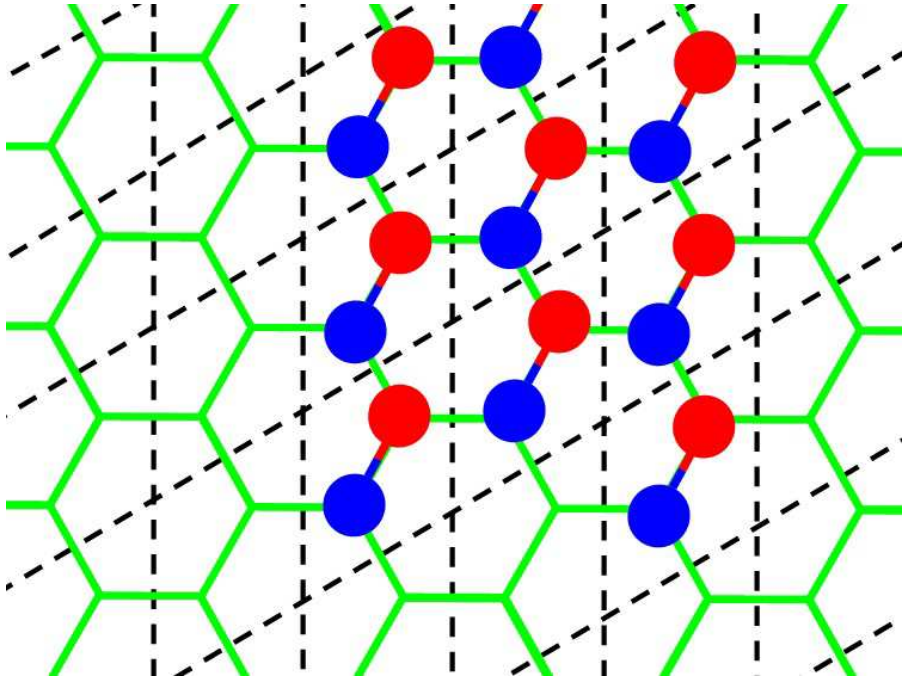}
\caption{Building up a honeycomb structure using dimers. \ The honeycomb can
be viewed as a two dimensional triangular lattice with a two particle basis.
\ }
\label{honeycombthesis}
\end{figure}

Alternatively one could consider grafting two different types of DNA onto
the $A$ and $B$ type particles. \ DNA with sequence $A_{1}$ is chosen
complementary to DNA with sequence $B_{1}$ and $A_{2}$ is chosen
complementary to $B_{2}$. \ By varying the length of the recognition
sequence, one can tune the melting temperatures $T_{1}$ and $T_{2}$ for the $%
A_{1}B_{1}$ and $A_{2}B_{2}$ complexes. \ In fact, the experimenter can
control the relative grafting density of the strands on the particles \cite%
{gangDNAcontrol}. \ By choosing a low grafting density for the $A_{2}$ and $%
B_{2}$ strands, the system will behave in an interesting fashion. \ For
temperatures $T>T_{2}$ the particles will be dispersed as monomers. \ As the
temperature is lowered $T_{1}<T<T_{2}$ we expect to see the formation of
predominantly dimers. \ Finally, for $T<T_{1}$ we expect aggregation of
large clusters, and under appropriate conditions crystallization. \ The hope
is that this intermediate stage in the assembly might facilitate faster
crystallization, i.e. provide a kinetically feasible pathway to the
equilibrium ground state. \ 

This is just one example of how complexity can be incorporated into the
DNA-colloidal system. \ Other examples have already been considered earlier
in the thesis, for example the self-assembly of DNA coded clusters. \ There
we saw the potential advantages of working with a multiple colored system,
as opposed to most of the experimental studies to date which have worked
with at most two colors. \ These ideas naturally lead us to the next topic,
the inverse problem in self-assembly. \ 

\section{The Inverse Problem in Self-Assembly}

\qquad A natural extension of these ideas is to consider the so-called
inverse problem in self-assembly. \ Presume that you have a particular
target structure $S$ in mind. \ For example, this structure might be a
particular crystalline morphology, or clusters with prescribed geometrical
features. \ Given $S$, the goal is to determine the set of components $C$
who self-assemble into $S$. \ The DNA\ grafted colloids considered in this
thesis, or combinations thereof, constitute a particularly promising
component class due to the diversity and complexity of the potential
interactions. \ This procedure requires knowledge of the interaction
potential which governs the dynamics between components. \ Some progress has
been made on the inverse problem for isotropic colloidal systems which
interact through a pair potential \cite{Torquato}, \cite{Torquato2}, \cite%
{Torquato3}, \cite{Torquato4}. \ This work determined a class of designer
potentials which give rise to a number of crystal structures, in both two
and three dimensions. \ The drawback to this approach is that the designer
potentials are generally quite complex. \ They contain a large number of
numerical fitting constants, so it is not known how the designer potentials
could be realized experimentally. \ 

A first step towards understanding the inverse problem in the DNA-colloidal
system is to determine the pair potential for two colloids grafted with DNA.
\ Some steps in this direction include \cite{crocker}, \cite{gangDNAcontrol}%
, \cite{polyelectrolyte}. \ In general one would have to consider the
effects of anisotropy. \ The major goal is to understand how the
experimental variables (grafting density of DNA, DNA flexibility and length,
hybridization free energy, etc.) affect the interaction potential. \ By
tuning these variables one might be able to construct potentials which mimic
the designer potentials determined through Monte Carlo simulations. \ Of
course knowledge of the potential which gives rise to a particular
equilibrium structure is only part of the problem. \ The free energy
landscape is generally quite rugged, so one has to take care to provide a
feasible pathway to reach the ground state configuration on experimental
timescales. \ Future studies which combine insights drawn from experiment,
theory, and modeling will hopefully lead to a realization of the incredible
technological potential of DNA-colloidal systems. \ 

\appendix

\chapter{Bridging Probability for Scheme B}

In this appendix we calculate $c_{eff}$ for scenario B (see Fig. \ref%
{hybridschemes}). \ Taking the $z$ direction normal to the planar surfaces
separated by a distance $2h$, an eigenfunction expansion \cite{edwards}
yields the following expression for the probability distribution function. \
Here $\beta \equiv \frac{\pi ^{2}R_{g}^{2}}{24h^{2}}$, and $N$ is the
normalization (unimportant for our purposes). \ 
\begin{equation}
P(\mathbf{r},\mathbf{r}^{\prime })=NP_{X}(x,x^{\prime })P_{Y}(y,y^{\prime
})P_{Z}(z,z^{\prime })  \label{pdf}
\end{equation}%
\begin{eqnarray}
P_{X}(x,x^{\prime }) &=&\exp \left( -\frac{3}{2R_{g}^{2}}(x-x^{\prime
})^{2}\right) \\
P_{Y}(y,y^{\prime }) &=&\exp \left( -\frac{3}{2R_{g}^{2}}(y-y^{\prime
})^{2}\right) \\
P_{Z}(z,z^{\prime }) &=&\sum_{n=1}^{\infty }\sin \left( \frac{n\pi z}{2h}%
\right) \sin \left( \frac{n\pi z^{\prime }}{2h}\right) \exp \left[ -\beta
n^{2}\right]
\end{eqnarray}%
To circumvent the Dirichlet boundary conditions, we start the chains at a
small distance $\lambda $ from the planes. \ We then have $\mathbf{r}%
_{1}^{\prime }=(x_{1}^{^{\prime }},y_{1}^{^{\prime }},\lambda )$ and $%
\mathbf{r}_{2}^{\prime }=(x_{2}^{^{\prime }},y_{2}^{^{\prime }},2h-\lambda )$%
. \ We first expand the distribution functions to first order in $\lambda $
and then perform the integration over all the possible hybridized
conformations $\mathbf{r}_{1}=\mathbf{r}_{2}$. \ 
\begin{eqnarray}
P_{1} &\equiv &P(\mathbf{r}_{1},\mathbf{r}_{1}^{\prime }=(x_{1}^{^{\prime
}},y_{1}^{^{\prime }},\lambda ))=\frac{N\pi \lambda }{2h}%
P_{X}(x_{1},x_{1}^{^{\prime }})P_{Y}(y_{1},y_{1}^{^{\prime }})\times \\
&&\sum_{n=1}^{\infty }n\sin \left( \frac{n\pi z_{1}}{2h}\right) \exp \left[
-\beta n^{2}\right]  \notag \\
P_{2} &\equiv &P(\mathbf{r}_{2},\mathbf{r}_{2}^{\prime }=(x_{2}^{^{\prime
}},y_{2}^{^{\prime }},2h-\lambda ))=-\frac{N\pi \lambda }{2h}%
P_{X}(x_{2},x_{2}^{^{\prime }})P_{Y}(y_{2},y_{2}^{^{\prime }})\times \\
&&\sum_{m=1}^{\infty }(-1)^{m}m\sin \left( \frac{m\pi z_{2}}{2h}\right) \exp %
\left[ -\beta m^{2}\right]  \notag
\end{eqnarray}%
The integration over the $x_{1}$ and $y_{1}$ components are Gaussian, which
leaves the integration in the direction normal to the plane. \ 
\begin{align}
\int P_{1}P_{2}\delta ^{3}(\mathbf{r}_{1}-\mathbf{r}_{2})d^{3}\mathbf{r}%
_{1}d^{3}\mathbf{r}_{2}& =-\frac{N^{2}\pi ^{3}\lambda ^{2}R_{g}^{2}}{12h^{2}}%
\sum_{n=1}^{\infty }\sum_{m=1}^{\infty }(-1)^{m}nmI_{nm}(h)\exp \left[
-\beta (n^{2}+m^{2})\right] \\
I_{nm}(h)& \equiv \int_{0}^{2h}\sin \left( \frac{n\pi z_{1}}{2h}\right) \sin
\left( \frac{m\pi z_{1}}{2h}\right) dz_{1}=h\delta _{n,m}
\end{align}%
The second line follows since $m$ and $n$ are integers, using $\lim_{\delta
\rightarrow 0}\frac{\sin \delta }{\delta }=1$. \ Defining $\mathbf{\Delta r}%
^{^{\prime }}\equiv \mathbf{r}_{1}^{\prime }-\mathbf{r}_{2}^{\prime }$ the
numerator in the expression for $c_{eff}$ is:%
\begin{eqnarray}
&&\int P(\mathbf{r}_{1},\mathbf{r}_{1}^{\prime })P(\mathbf{r}_{2},\mathbf{r}%
_{2}^{\prime })\delta ^{3}(\mathbf{r}_{1}-\mathbf{r}_{2})d^{3}\mathbf{r}%
_{1}d^{3}\mathbf{r}_{2} \\
&=&-\frac{N^{2}\pi ^{3}\lambda ^{2}R_{g}^{2}}{12h}\exp \left[ -\frac{3}{4}%
\left( \frac{\mathbf{\Delta r}^{^{\prime }}}{R_{g}}\right) ^{2}\right]
\sum\limits_{n=1}^{\infty }(-1)^{n}n^{2}\exp [-2\beta n^{2}]  \notag
\end{eqnarray}%
We also make use of the following result. \ 
\begin{equation}
\int P_{1}d^{3}\mathbf{r}_{1}=\frac{4N\pi \lambda R_{g}^{2}}{3}\sum_{k=1%
\text{, }k\text{\ }odd}^{\infty }\exp \left[ -\beta k^{2}\right]
\end{equation}%
This leads us to the overlap density for complementary, flexible linkers
with $x=\frac{h}{R_{g}}$. \ 
\begin{eqnarray}
c_{eff} &=&\frac{c(x)}{R_{g}^{3}}\exp \left[ -\frac{3}{4}\left( \frac{%
\mathbf{\Delta r}^{^{\prime }}}{R_{g}}\right) ^{2}\right]  \label{dimconc} \\
c(x) &=&\frac{-3\pi }{64x}\frac{\sum\limits_{n=1}^{\infty }(-1)^{n}n^{2}\exp %
\left[ -2\beta n^{2}\right] }{\left( \sum\limits_{k=1\text{, }k\text{\ }%
odd}^{\infty }\exp \left[ -\beta k^{2}\right] \right) ^{2}}
\end{eqnarray}%
When the chains are strongly compressed, $\beta \sim \frac{1}{x^{2}}\gg 1$,
the asymptotic behavior of $c(x)$ is easily determined as the sums converge
rapidly. \ The more interesting physical regime is as particles approach at
separations large compared to the radius of gyration of the marker strands,
which is $\beta \ll 1$. \ To extract the asymptotics in this regime we can
massage the summation as follows. \ 
\begin{eqnarray}
\sum\limits_{n=1}^{\infty }(-1)^{n}n^{2}\exp [-2\beta n^{2}] &=&\frac{-1}{2}%
\frac{\partial }{\partial \beta }\sum\limits_{n=1}^{\infty }(-1)^{n}\exp
[-2\beta n^{2}]= \\
&&\frac{-1}{2}\frac{\partial }{\partial \beta }\left[ 2\sum\limits_{n=1}^{%
\infty }\exp [-8\beta n^{2}]-\sum\limits_{n=1}^{\infty }\exp [-2\beta n^{2}]%
\right]  \notag \\
\sum\limits_{k=1\text{, }k\text{\ }odd}^{\infty }\exp \left[ -\beta k^{2}%
\right] &=&\sum\limits_{k=1}^{\infty }\exp \left[ -\beta k^{2}\right]
-\sum\limits_{k=1}^{\infty }\exp \left[ -4\beta k^{2}\right]
\end{eqnarray}%
Then note that for any even function of n, $f(n)$ we have:%
\begin{equation}
\sum_{n=1}^{\infty }f(n)=\frac{1}{2}\left( \sum_{n=-\infty }^{\infty
}f(n)-f(0)\right)
\end{equation}%
Finally we use an identity derived from properties of the theta function 
\cite{theta}. \ 
\begin{equation}
\sum_{n=-\infty }^{\infty }\exp \left[ -\beta n^{2}\right] =\sqrt{\frac{\pi 
}{\beta }}\sum_{n=-\infty }^{\infty }\exp \left[ \frac{-\pi ^{2}n^{2}}{\beta 
}\right]
\end{equation}%
Expanding the sums using these identities, the result of some
straightforward but tedious algebra gives the asymptotic behavior of the
binding probability. \ 
\begin{eqnarray}
\Delta \widetilde{G}_{B} &\simeq &\Delta G_{B}+T\left[ 
\begin{array}{c}
\frac{3}{4}\left( \frac{\Delta r^{^{\prime }}}{R_{g}}\right) ^{2}+\log
\left( R_{g}^{3}c_{o}\right) +\log \left( \frac{32}{\pi ^{2}}\right) \\ 
+3\log (x)+\frac{\pi ^{2}}{12x^{2}}%
\end{array}%
\right] \text{ \ for }x\ll 1  \notag \\
\Delta \widetilde{G}_{B} &\simeq &\Delta G_{B}+T\left[ 
\begin{array}{c}
\frac{3}{4}\left( \frac{\Delta r^{^{\prime }}}{R_{g}}\right) ^{2}+\log
\left( R_{g}^{3}c_{o}\right) -\log \left( \frac{9}{4}\sqrt{\frac{3}{\pi }}%
\right) \\ 
-2\log (x)+3x^{2}%
\end{array}%
\right] \text{ \ for }x\gg 1  \notag
\end{eqnarray}

\chapter{Bridging Probability for Scheme C}

In this appendix we calculate $c_{eff}$ for scenario C (see Fig. \ref%
{hybridschemes}). \ If the method of images is used to construct the
probability distribution function for the flexible linker DNA, as opposed to
the eigenfunction expansion used in Appendix A, we arrive at the following
expression \cite{edwards}. \ 
\begin{eqnarray}
P(\mathbf{r},\mathbf{r}^{\prime }) &=&NP_{X}(x,x^{\prime })P_{Y}(y,y^{\prime
})P_{Z}(z,z^{\prime }) \\
P_{Z}(z,z^{\prime }) &=&\sum_{n=-\infty }^{\infty }\left\{ \exp \left[ \frac{%
-3}{2R_{g}^{2}}(z-z^{\prime }-4nh)^{2}\right] -\exp \left[ \frac{-3}{%
2R_{g}^{2}}(z+z^{\prime }-4nh)^{2}\right] \right\}  \notag \\
&\simeq &\exp \left[ \frac{-3}{2R_{g}^{2}}(z-z^{\prime })^{2}\right] -\exp %
\left[ \frac{-3}{2R_{g}^{2}}(z+z^{\prime })^{2}\right]
\end{eqnarray}%
\qquad Since $\zeta \equiv \frac{L}{R_{g}}\gg 1$ we only need the $n=0$ term
in the expression for $P_{Z}(z,z^{\prime })$. \ $P_{X}(x,x^{\prime })$ and $%
P_{Y}(y,y^{\prime })$ are the same as in Appendix A(equation \ref{pdf}). \
Once again we start the chains at a small distance $\lambda $ from the
planes. \ The majority of the hybridized conformations will have the planar
surfaces separated by approximately the linker length $L$. \ To simplify the
discussion we take $\mathbf{\Delta r}^{^{\prime }}=0$ from the beginning of
the calculation, which corresponds to orientations of the rigid linker with
a small component parallel to the surface. \ We then have $\mathbf{r}%
_{1}^{\prime }=(0,0,\lambda )$ and $\mathbf{r}_{2}^{\prime }=(0,0,L+\Delta
-\lambda )$. \ We first expand the distribution functions to first order in $%
\lambda $ and then perform the integration over all the possible hybridized
conformations $|\mathbf{r}_{1}-\mathbf{r}_{2}|=L$. \ 
\begin{eqnarray}
P_{1} &\equiv &P(\mathbf{r}_{1},\mathbf{r}_{1}^{\prime }=(0,0,\lambda ))=%
\frac{6\lambda N}{R_{g}^{2}}P_{X}(x_{1},0)P_{Y}(y_{1},0)z_{1}\exp \left[ 
\frac{-3z_{1}^{2}}{2R_{g}^{2}}\right] \\
P_{2} &\equiv &P(\mathbf{r}_{2},\mathbf{r}_{2}^{\prime }=(0,0,L+\Delta
-\lambda )=\frac{6\lambda N}{R_{g}^{2}}P_{X}(x_{2},0)P_{Y}(y_{2},0)\times \\
&&(L+\Delta -z_{2})\exp \left[ \frac{-3(L+\Delta -z_{2})^{2}}{2R_{g}^{2}}%
\right]  \notag
\end{eqnarray}

To impose the delta function constraint we write:%
\begin{eqnarray}
z_{2} &=&z_{1}+L\cos \theta \\
y_{2} &=&y_{1}+L\sin \theta \sin \phi \\
x_{2} &=&x_{1}+L\sin \theta \cos \phi \\
\int \delta (|\mathbf{r}_{1}-\mathbf{r}_{2}|-L)d^{3}\mathbf{r}_{2}
&\Rightarrow &L^{2}\int \sin \theta d\theta d\phi
\end{eqnarray}%
The integrations over $x_{1}$ and $y_{1}$ are Gaussian. \ There is no
azimuthal dependence, so the $\phi $ integration gives a factor of $2\pi $.
\ We also need the following integral:%
\begin{equation}
\int P_{1}d^{3}\mathbf{r}_{1}=\frac{4N\pi \lambda R_{g}^{2}}{3}
\end{equation}%
We define $\epsilon \equiv \frac{\Delta }{L}$ and $z\equiv \frac{z_{1}}{L}$.
\ Since we consider DNA bridges with the rigid linker aligned with a small
component parallel to the planar surfaces, the upper bound for the polar
integration is given by $\theta _{\max }\simeq \frac{1}{\zeta }$. \ 
\begin{eqnarray}
c_{eff}(\mathbf{\Delta r}^{^{\prime }} &=&0)\simeq \frac{27\zeta ^{3}}{8\pi
R_{g}^{3}}\int\limits_{0}^{\frac{1}{\zeta }}\sin \theta \exp \left[ -\frac{3%
}{4}\zeta ^{2}\sin ^{2}\theta \right] I_{z}(\theta ,\epsilon ,\zeta )d\theta
\\
I_{z}(\theta ,\epsilon ,\zeta ) &=&\int\limits_{0}^{1+\epsilon }dz(1-\cos
\theta +\epsilon -z)z \\
&&\times \exp \left[ -\frac{3}{2}\zeta ^{2}\left\{ z^{2}+(1-\cos \theta
+\epsilon -z)^{2}\right\} \right]  \notag
\end{eqnarray}%
We first calculate the $z$ integration $I_{z}(\theta ,\epsilon ,\zeta )$. \
Since we consider all of the possible conformations of the short linkers
between planes, we can see that the upper bound for the $z$ integration is $%
z_{\max }=1+\epsilon $. \ However, since $\zeta \gg 1$ the gaussian decay
allows us to extend the integration to $\infty $. \ We define $f\equiv z-%
\frac{\beta }{2}$ and the small parameter $\beta \equiv 1-\cos \theta
+\epsilon $. \ Completing the square gives:%
\begin{equation}
I_{z}(\theta ,\epsilon ,\zeta )\simeq \exp \left[ -\frac{3}{4}\zeta
^{2}\beta ^{2}\right] \int\limits_{-\frac{\beta }{2}}^{\infty }\left( \frac{%
\beta ^{2}}{4}-f^{2}\right) \exp \left[ -3\zeta ^{2}f^{2}\right] df
\end{equation}%
Since $\beta $ is small we take the lower bound for the integration to $0$.
\ 
\begin{equation}
I_{z}(\theta ,\epsilon ,\zeta )\simeq \exp \left[ -\frac{3}{4}\zeta
^{2}\beta ^{2}\right] \frac{1}{4\zeta }\sqrt{\frac{\pi }{3}}\left( \frac{%
\beta ^{2}}{2}-\frac{1}{3\zeta ^{2}}\right)
\end{equation}%
We now expand the $\theta $ integrand in terms of the small parameters $%
\theta \sim \epsilon \sim \frac{1}{\zeta }$. \ 
\begin{eqnarray}
c_{eff}(\mathbf{\Delta r}^{^{\prime }} &=&0)\simeq \frac{27}{32\sqrt{3\pi }%
R_{g}^{3}}\exp \left[ -\frac{3}{4}\zeta ^{2}\beta ^{2}\right] \left( \frac{%
\zeta ^{2}\epsilon ^{2}}{2}-\frac{1}{3}\right) \\
&&\times \int\limits_{0}^{\frac{1}{\zeta }}\theta \exp \left[ -\frac{3}{4}%
\zeta ^{2}(1+\epsilon )\theta ^{2}\right] d\theta  \notag \\
&=&\frac{9\left( 1-e^{-\frac{3}{4}}\right) }{32\sqrt{3\pi }R_{g}^{3}}\exp %
\left[ -\frac{3}{4}\zeta ^{2}\epsilon ^{2}\right] \left( \epsilon ^{2}-\frac{%
2}{3\zeta ^{2}}\right)
\end{eqnarray}%
\begin{equation}
\Delta \widetilde{G}_{C}(\mathbf{\Delta r}^{^{\prime }}=0)\simeq \Delta
G_{C}+T\left[ 
\begin{array}{c}
\log \left( R_{g}^{3}c_{o}\right) +\log \left( \frac{32\sqrt{3\pi }}{9\left(
1-e^{-\frac{3}{4}}\right) }\right) +\frac{3}{4}\zeta ^{2}\epsilon ^{2} \\ 
-\log \left( \epsilon ^{2}-\frac{2}{3\zeta ^{2}}\right)%
\end{array}%
\right]
\end{equation}%
We are interested in the minimum free energy with respect to the separation
between planar surfaces, determined by $\frac{\partial \Delta \widetilde{G}%
_{C}(\mathbf{\Delta r}^{^{\prime }}=0)}{\partial \epsilon }=0$. \ Performing
the differentiation we find $\epsilon ^{\ast }=\frac{\sqrt{2}}{\varsigma }$.
\ 
\begin{eqnarray}
\Delta \widetilde{G}_{C}(\mathbf{\Delta r}^{^{\prime }} &=&0,\epsilon
=\epsilon ^{\ast })\simeq \Delta G_{C}+T\log \left[ \frac{8\sqrt{\frac{\pi }{%
3}}L^{2}R_{g}c_{o}}{\left( e^{-\frac{3}{2}}-e^{-\frac{9}{4}}\right) }\right]
\\
&=&\Delta G_{C}+4.24T+T\log \left[ L^{2}R_{g}c_{o}\right]
\end{eqnarray}

\backmatter

\newpage \bgroup \null\vfil  \centering  {\Large \bfseries \MakeUppercase%
\bibname}

{\Large \addcontentsline{toc}{chapter}{\MakeUppercase\bibname} \vfil\newpage %
\egroup}

\bibliographystyle{achemso}
\bibliography{acompat,dna}

\end{document}